\documentclass[aoas,preprint]{imsart}

\RequirePackage{amsthm,amsmath,amsfonts,amssymb}
\RequirePackage[authoryear]{natbib}
\RequirePackage[colorlinks,citecolor=blue,urlcolor=blue]{hyperref}
\RequirePackage{graphicx}

\startlocaldefs
\def\bSig\mathbf{\Sigma}
\newcommand{\N}{\mathcal{N}}
\newtheorem{theorem}{Proposition}
\newtheorem{assumption}{Assumption}

\usepackage[figuresright]{rotating}
\usepackage{amsmath}
\usepackage{caption, subcaption}
\usepackage{multicol}
\usepackage{multirow}
\usepackage{color, colortbl}
\usepackage{graphicx}
\usepackage{booktabs}
\usepackage{float}
\usepackage{comment}
\usepackage{tikz, tikz-cd}
\usepackage{comment}
\usetikzlibrary{arrows, positioning, shapes.geometric}
\tikzstyle{startstop} = [rectangle, rounded corners, minimum width=1.5cm, minimum height=0.5cm,text centered, draw=black]
\tikzstyle{arrow} = [thick,->,>=stealth]

\endlocaldefs

\begin{document}

\begin{frontmatter}
\title{Estimating Causal Effects of HIV Prevention Interventions with Interference in Network-based Studies among People Who Inject Drugs}
%\title{A sample article title with some additional note\thanksref{t1}}
\runtitle{Causal Effects in Network-based Studies}
%\thankstext{T1}{A sample additional note to the title.}

\begin{aug}
%%%%%%%%%%%%%%%%%%%%%%%%%%%%%%%%%%%%%%%%%%%%%%
%%Only one address is permitted per author. %%
%%Only division, organization and e-mail is %%
%%included in the address.                  %%
%%Additional information can be included in %%
%%the Acknowledgments section if necessary. %%
%%%%%%%%%%%%%%%%%%%%%%%%%%%%%%%%%%%%%%%%%%%%%%
\author[A]{\fnms{TingFang} \snm{Lee}\ead[label=e1] {tingfanglee@uri.edu}},
\author[B]{\fnms{Ashley L.} \snm{Buchanan}\ead[label=e2] {buchanan@uri.edu}}, 
\author[C]{\fnms{Natallia V.} \snm{Katenka}\ead[label=e3]{nkatenka@uri.edu}},
\author[D]{\fnms{Laura} \snm{Forastiere} \ead[label=e4]{laura.forastiere@yale.edu }},
\author[E]{\fnms{M.  Elizabeth} \snm{Halloran}\ead[label=e5]{betz@fredhutch.org}},
\author[F]{\fnms{Samuel R.} \snm{Friedman} \ead[label=e6] {Samuel.Friedman@nyulangone.org}},
\\ \and 
\author[G]{\fnms{Georgios} \snm{Nikolopoulos}\ead[label=e7]{nikolopoulos.georgios@ucy.ac.cy}}

%%%%%%%%%%%%%%%%%%%%%%%%%%%%%%%%%%%%%%%%%%%%%%
%% Addresses                                %%
%%%%%%%%%%%%%%%%%%%%%%%%%%%%%%%%%%%%%%%%%%%%%%
\address[A]{Department of Pharmacy Practice, University of Rhode Island,
\printead{e1}}
\address[B]{Department of Pharmacy Practice, University of Rhode Island,
\printead{e2}}
\address[C]{Department of Computer Science and Statistics, University of Rhode Island, \printead{e3}}
\address[D]{School of Public Health, Yale University, \printead{e4}}
\address[E]{Biostatistics, Bioinformatics, and Epidemiology Program, Vaccine and Infectious Disease Division, Fred Hutchinson Cancer Center, and Department of Biostatistics, University of Washington, \printead{e5}}
\address[F]{Department of Population Health, NYU Grossman School of Medicine, \printead{e6}}
\address[G]{Medical School, University of Cyprus, \printead{e7}}

\end{aug}

\begin{abstract}
Evaluating causal effects in the presence of interference is challenging in network-based studies of hard-to-reach populations. Like many such populations, people who inject drugs (PWID) are embedded in social networks and often exert influence on others in their network. In our setting, the study design is observational with a non-randomized network-based HIV prevention intervention. Information is available on each participant and their connections that confer possible HIV risk through injection and sexual behaviors. We considered two inverse probability weighted (IPW) estimators to quantify the population-level spillover effects of non-randomized interventions on subsequent health outcomes. We demonstrated that these two IPW estimators are consistent, asymptotically normal, and derived a closed-form estimator for the asymptotic variance, while allowing for overlapping interference sets (groups of individuals in which the interference is assumed possible). A simulation study was conducted to evaluate the finite-sample performance of the estimators. We analyzed data from the Transmission Reduction Intervention Project, which ascertained a network of PWID and their contacts in Athens, Greece, from 2013 to 2015. We evaluated the effects of community alerts on subsequent HIV risk behavior in this observed network, where the connections or links between participants were defined by using substances or having unprotected sex together. In the study, community alerts were distributed to inform people of recent HIV infections among individuals in close proximity in the observed network. The estimates of the risk differences for spillover using either IPW estimator demonstrated a protective effect. The results suggest that HIV risk behavior could be mitigated by exposure to a community alert when an increased risk of HIV is detected in the network.
\end{abstract}

%
%  Please place your key words in alphabetical order, separated
%  by semicolons, with the first letter of the first word capitalized,
%  and a period at the end of the list.
%

\begin{keyword}
\kwd{Causal Interference}
\kwd{Interference/dissemination}
\kwd{Network Studies} 
\kwd{People who Use Drugs}
\kwd{HIV/AIDS} 
\kwd{Inverse Probability Weights}
\end{keyword}

\end{frontmatter}

\section{Introduction}
\label{s:intro}

The objective of this work is to evaluate causal effects in the presence of interference (also known as dissemination or spillover) where usual assumptions, such as partial or clustered interference \cite{hudgens2008toward, tchetgen2012causal}, may no longer hold. This proves to be a challenging problem in network-based studies of hidden or hard-to-reach populations, such as people who inject drugs (PWID), where participants are frequently recruited through contact tracing. Worldwide in 2011, an estimated 10$\%$ of new HIV infections occur because of injection drug use, and this proportion was 30$\%$ outside Africa \citep{prejean2011estimated, lansky2014estimating, mathers2008global}. 
In Greece through 2010, there were only a few sporadic cases of HIV transmission among PWID and the HIV epidemic was traditionally concentrated among men having sex with men. From 2002 to 2010, less than 20 HIV cases were reported annually among PWID, representing 2\% to 4\% of newly diagnosed HIV infections per year. In 2011, the number of reported HIV cases among PWID increased 16-fold from the number reported in 2010 to a total of 260 cases. The emergence of the HIV outbreak among PWID in Athens coincided with an economic recession, highlighting its possible role in the outbreak due to the temporal ordering \citep{econrecession2013}. Investigation of the outbreak demonstrated that clustered HIV transmission among PWID was rare until 2009. Starting in 2010, a large proportion of HIV sequences from newly diagnosed PWID could be grouped into PWID-specific phylogenetic clusters, indicating that parenteral transmission with contaminated syringes or other injecting equipment was now occurring in this population. Prior to 2011, prevention and harm reduction services, including medication for opioid use disorder and syringe exchange distribution programs, were available; however, access to these services remained low among PWID. Most of the newly diagnosed PWID (about 70\%) in 2011 were residents of Athens, suggesting that the outbreak was also geographically localized. \citep{aristotle, nikolopoulous2015bigevent}

Effective interventions were urgently needed to prevent further transmission in Athens. The Transmission Reduction Intervention Project (TRIP) was a successful attempt to recruit and intervene in this population by contact tracing the injection and sexual networks of recently-infected PWID. The program then referred  people found to be recently infected to engage in HIV treatment and care both to protect their own health and to reduce onward transmission of HIV to others, particularly during the early stage of HIV when there is a known increased risk of HIV transmission \cite{nikolopoulous2015bigevent}. Interestingly, this network study design can be used to investigate the connections or ties among people who are infected and uninfected, and thus can address questions about why certain groups of people who are uninfected remain that way despite having risk network ties to people who both have high viral loads and engage in risky behavior \citep{williams2018pockets}. The TRIP recruitment strategy successfully found more recently infected PWID than other strategies, such as a respondent driven sample or venue-based recruitment. These findings suggest that using strategically network-based approaches can accelerate seeking, testing, and treating recently-infected PWID. Moreover, reducing viral loads as early as possible is likely to decrease the expected number of transmissions in a community \citep{nikolopoulos2016network}.

Public health interventions often have disseminated effects, also known as indirect or spillover effects \citep{spillover2017, diffusion2020}. There can be disseminated effects of HIV behavioral interventions, suggesting that intervening among highly-connected individuals may maximize benefits to others \citep{Rewleye033759}. Akin to other populations, PWID are embedded in social networks and communities (e.g., injection drug, non-injection drug, sexual risk network) in which they possibly exercise an influence upon other members \citep{hayes2000design,ghosh2017social}. This influence can be measured as a \textit{disseminated} effect of specific interventions among individuals who were not exposed themselves but possibly receive intervention benefits from their connections to those exposed to the intervention. In PWID networks, interventions (e.g., educational training about HIV risk reduction, medical interventions such as pre-exposure prophylaxis, or treatment as prevention) may have disseminated effects, and intervention effects frequently depend on the network structure and intervention coverage levels. Disseminated effects may be larger in magnitude than direct effects (i.e., effect of receiving the intervention while holding the exposure of other individuals fixed), suggesting that an intervention has substantial effects in the network beyond those exposed themselves \citep{buchanan2018assessing}. 

The current methodologies used to evaluate direct and disseminated effects among members of hidden or hard-to-reach populations in networks remain limited. In particular, relatively few methodological approaches for observational network-based studies have been developed, and methods that incorporate the observed connections (links, ties, or edges) in the underlying network structure are needed to understand the spillover mechanisms. In our setting, connections in the network refer to sexual and/or drug use partnerships. Recent methodological developments relaxed the no interference assumption and allowed for interference within clusters, known as {\it partial interference} \citep{sobel2006randomized, hong2006evaluating, hudgens2008toward, tchetgen2012causal, liu2014large}. In partial interference approaches, a clustering of observations is used to define the interference set (e.g., study clusters, provider practices, or geographic location) that allow for interference within but not across clusters; however, the information on connections within a particular cluster is typically not measured or utilized in the analytical approach \citep{aronow2017samii}. Another approach defines interference by spatial proximity or network ties \citep{liu2016inverse, forastiere2016identification},  allowing for overlapping interference sets (i.e., groups of individuals in which interference is assumed to be possible). In \cite{liu2016inverse}, an IPW estimator was proposed for a generalized interference set that allowed for overlap between interference sets; however, the asymptotic variance was estimated under the assumption of partial interference defined by larger groupings or clusters of participants in the study. In a separate paper, the subclassification estimator and generalized propensity score were used to quantify effects, and a bootstrapping procedure with resampling at the individual-level or the cluster-level was used to quantify the variance \citep{forastiere2016identification}. However, these approaches either rely on partial interference defined by larger clusters or resort to bootstrapping to derive estimators of the variance. In practice, ignoring the overlapping interference sets while estimating the variance can lead to inaccurate inference and resampling approaches, particularly in a network setting, can also be computationally intensive. 

While previous work allows for overlapping interference sets for point estimation, the asymptotic variances were estimated under the assumption of partial interference or used bootstrapping techniques \citep{liu2016inverse, forastiere2016identification}. Our paper addresses an important gap by developing inverse probability weight estimators and deriving a closed-form variance estimator that allows for overlapping interference sets, possibly leading to a more statistically efficient estimator in network-based studies due to the use of additional information on connections between individuals. In our paper, we propose two inverse probability weighted (IPW) estimators where the interference set is defined as the set of the individual's nearest neighbors within a sociometric network; that is, a network in which all or some of participants' direct and indirect contacts are ascertained \citep{hadjikou2021drug}. The first IPW estimator is a novel application of the estimator of the approach in \citet{liu2016inverse} to a sociometric network-based study setting. Originally, the asymptotic variance estimators were developed for clustered observational studies without explicit consideration of the connections in the study. We relax the partial interference assumption for variance estimation such that interference sets are uniquely defined by nearest neighbors for each individual. The second IPW estimator uses a generalized propensity score developed by \citet{forastiere2016identification}; however, we propose a weighted estimator instead of a stratified estimator for comparison to the first IPW estimator in this paper. For both estimators, we assume that the nearest neighbors comprise the interference sets and use this structure to calculate a novel closed-form variance estimator by applying M-estimation. We focus on comparing these two alternative IPW estimators in a network study with a non-randomized intervention and statistical inference approaches using M-estimation.

The rest of the paper is structured as follows. In Sections 2, we introduce the TRIP study design and setting.  In Section 3 and 4, we define the notations and assumptions for nearest neighbors settings, then the estimands of interest for this setting. We provide definitions of the two IPW estimators with specific assumptions for each, and demonstrate that the estimator is consistent and asymptotically normal, and obtain a closed-form estimator of the corresponding variances in Section 5. In Section 6, a simulation study was conducted to demonstrate the finite-sample performance of both estimators and the results are summarized. The methods were then utilized to assess the direct and disseminated effects of community alerts on HIV risk behavior in the sociometric network study of PWID and their contacts, Transmission Reduction Intervention Project (TRIP) from 2013 to 2015 in Athens, Greece in Section 7. We discuss limitations of this approach and next steps for methodological work to quantify causal effects in network-based studies in Section 8. 

\section{TRIP Study Design}
\begin{figure}
\centering
\begin{tikzpicture}
\footnotesize
\node (start) [startstop, text width=4cm] {Allied prevention project and collaborating testing sites};
\node (enroll) [startstop, below =0.5cm of start, text width=4cm] {
Enrolled in TRIP};
\node (tested) [startstop, below = 0.5cm of enroll, text width=4cm] {Tested for HIV};
\draw[arrow](start) -- (enroll);
\draw[arrow] (enroll)--(tested);
\node (controlseed)[startstop, below = 1.5cm of tested, text width=2cm]{Control seeds with long-term infection};
\draw[arrow] (tested)--(controlseed);
\node(nrecentseed)[startstop, left = 0.2cm of controlseed,  text width=2cm]{Network of recent seeds};
\node(recentseed)[startstop, left = 0.9cm of nrecentseed, text width=2cm]{Recent seeds};
\node(ncontrolseed)[startstop, right = 0.9cm of controlseed,  text width=2cm]{Network of control seeds with long-term infection};
\node(negativecontrol)[startstop, right=0.2cm of ncontrolseed, text width=2cm]{Negative controls};

\node (HIVlonginf) [below =0.5cm of tested, xshift=-0.9cm, text width=2cm] {\footnotesize{HIV infection\\$\geq$ 6 months}};
\node (HIVnewinf) [below =0.5cm of tested, xshift=-6.4cm, text width=2cm]{\footnotesize{HIV infection $<$6 months}};
\node (HIVnoinf) [below = 0.5 of tested, xshift=4.4cm, text width=2.5cm]{\footnotesize{no HIV infection}};
\draw[arrow] (tested) -| (recentseed);
\draw[arrow] (tested) -| (negativecontrol);

\node(enrollment)[startstop, fill=lightgray, left= 0.3cm of recentseed]{Enrollment};
\node (contrace1)[left of = nrecentseed, xshift=-0.4cm, text width =1cm]{\tiny{Contact tracing}};
\node (contrace2)[right of = controlseed, xshift=0.7cm, text width =1cm]{\tiny{Contact tracing}};
\draw[-] (recentseed)--(nrecentseed);
\draw[-] (controlseed)--(ncontrolseed);

\node(ncontrolseed2)[startstop, below = 0.5cm of ncontrolseed,  text width=2cm]{Network of control seeds with long-term infection};
\node (controlseed2)[startstop, left = 0.9cm of ncontrolseed2, text width=2cm]{Control seeds with long-term infection};
\node(nrecentseed2)[startstop, left = 0.2cm of controlseed2,  text width=2cm]{Network of recent seeds};
\node(recentseed2)[startstop, left = 0.9cm of nrecentseed2, text width=2cm]{Recent seeds};
\node(negativecontrol2)[startstop, right=0.2cm of ncontrolseed2, text width=2cm]{Negative controls};
\node(followup) [startstop, fill=lightgray, left=0.3cm of recentseed2]{Follow-up};
\draw[arrow](controlseed)--(controlseed2);
\draw[arrow](nrecentseed)--(nrecentseed2);
\draw[arrow](recentseed)--(recentseed2);
\draw[arrow](ncontrolseed)--(ncontrolseed2);
\draw[arrow](negativecontrol)--(negativecontrol2);

\end{tikzpicture}
\caption{Flowchart of participant selection in TRIP \citep{pampaka2021mental}}\label{fig:flowTRIP}
\end{figure}
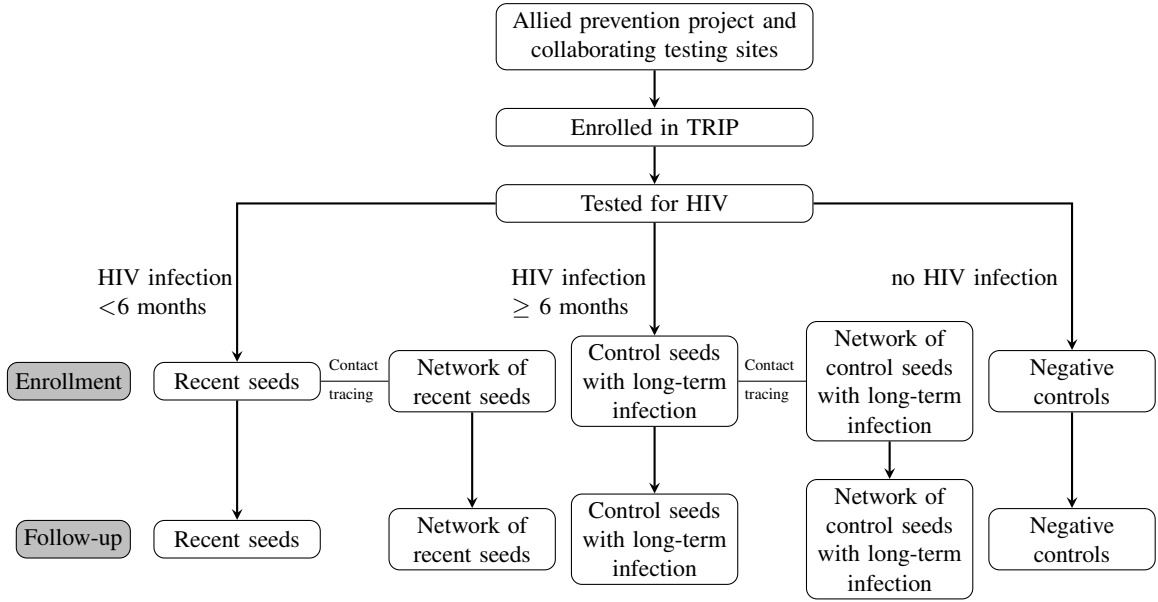

The Transmission Reduction Intervention Project (TRIP) included PWID and their HIV risk networks and initially found individuals who were recently diagnosed with HIV (known as seeds) and their possible HIV risk partners through sexual and injection routes of transmission \citep{nikolopoulos2016network,psichogiou2019identifying,  giallouros2021drug, hadjikou2021drug, pampaka2021mental}. TRIP also recruited seeds with long-term HIV infection. TRIP used contact network tracing (i.e., nomination and coupon referrals) and venue recruitment methods to locate those who were at risk for HIV infection based on their proximity in the network to other recently-infected individuals. PWID who were participants in the ARISTOTLE study at HIV testing centers in Athens were initially recruited into the TRIP study if they were found to be recently infected or long-term infected with HIV. ARISTOTLE was a seek, test, treat, and retain intervention that used respondent-driven sampling to target PWID residing in Athens and aimed to contribute to the control of HIV transmission among PWID in Greece \citep{aristotle}.  Each recently-diagnosed and long-term infected individual was asked to identify their recent sexual and drug use partners and their partners' partners in the six months prior to the interview. For the recently-diagnosed seeds, these direct contacts and their contacts' contacts were then recruited and asked to identify their sexual and drug use partners, who were also recruited and linked back to other individuals recruited in the study. For seeds with long-term HIV infection, their contacts were recruited (i.e., one wave of contact tracing) and these individuals were recruited and their connections to other participants were ascertained. If any of these contacts  were identified as recently infected with HIV, then their contacts and the contacts of their contacts (i.e., two waves of contact tracing) would be recruited as well and connections to other participants in the network were ascertained (Figure \ref{fig:flowTRIP}); otherwise, one wave of contact tracing was performed.  The study also recruited  HIV-negative individuals from allied projects who served as controls. The HIV-negative individuals were isolates (i.e., no connections to others in the network) unless reported as a contact by another participant. This resulted in a network consisting of individuals recently diagnosed with HIV and their possible HIV risk partners and the connections in the network were defined by sex or injection drug use partnerships. This information was used to create a final observed network in which each recruited individual is linked to all other individuals who named them as a contact or was named as a contact by them, regardless of recruitment order. 

Participants were interviewed at a baseline visit and 6-months after the baseline visit using a questionnaire to ascertain demographics, sexual and injection behaviors and partners in the prior 6 months, drug treatment, and antiretroviral treatment.   In addition to HIV testing, the study provided access to treatment as prevention (TasP) for those with HIV, referrals for medical care, and distributed community alerts to inform members of the community about temporary increases in the risk for HIV acquisition. These alerts were paper flyers provided to participants and posted in locations frequented by members of the local PWID community. Participants were followed to ascertain demographics, risk behaviors, and substance use through interviews, HIV serostatus, timing of HIV infection, and HIV disease markers, including HIV viral load, through phylogenetic techniques approximately 6 months later. Complete details on the study design and recruitment can be found in \citet{nikolopoulos2016network,psichogiou2019identifying,  giallouros2021drug, hadjikou2021drug, pampaka2021mental}.

\begin{figure}
    \centering
    \includegraphics[scale=0.3]{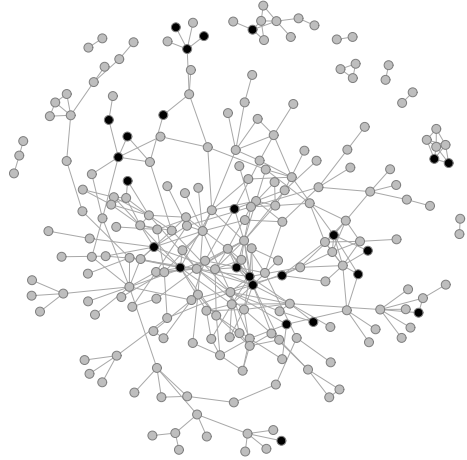}
    \caption{The TRIP network consisted of 10 connected components. The size of each component was $\{185, 9, 6, 3, 3, 2, 2, 2, 2, 2\}$. The size of each component after using community detection to further divide the network into 20 components is $\{28, 26, 23, 19, 19, 18, 15, 12, 10, 9, 8, 7, 6, 3, 3, 2, 2, 2, 2, 2\}$. Dark shaded nodes represent the participants who were exposed to community alert and gray shaded nodes represent participants who were not exposed.}
    \label{fig:trip_network}
\end{figure}

For this study, we used data from the Athens, Greece site which was collected from 2013 to 2015 during the HIV outbreak that began following the economic recession in 2008 \citep{nikolopoulous2015bigevent,williams2018pockets}. The network structure in TRIP included 356 participants and 542 shared connections. One of the participant was recruited twice as a network member of a recent seed and as a network member of a control seeds with long-term HIV infection. In our analysis, we only used the information for this participant corresponding to their records as a network member of a recent seed. In the network, 79 participants were isolates (i.e. not sharing connection with other network members) and removed for our analysis as spillover is not possible for isolates. In addition, 2 participants were removed due to missing values on HIV risk behavior in the past 6 months reported at baseline and 59 participants were removed from the network due to loss to follow-up that resulted in missing information at the 6-month visit, including HIV risk behavior, which was the the outcome of interest. Figure \ref{fig:trip_network} represents the TRIP network with 216 participants after excluding isolates who were participants not connected with any other participants in the network and 25 participants (11.6\%) of the 216 participants were exposed to the community alerts. The network characteristics and distribution of participant attributes are summarized in Table \ref{tab:descripitve_stat}. %The TRIP network has average degree $3.35$ (SD=2.75) and density $0.0155$. The transitivity is $0.25$, where the transitivity measures the tendency of the nodes to cluster together. High transitivity means that the network contains communities or groups of nodes that are densely connected internally. The assortativity is $0.24$ which quantifies the extent to which connected nodes share similar properties.

\begin{table}
\centering
\caption{TRIP  network characteristics and participant attribute  variables after excluding isolates and 59 participants (21\%) who were lost to follow up before their six-month visit$^1$.}
    \begin{tabular}{lrr}
    \toprule
    \multirow{6}{*}{Network Characteristics} & Nodes & 216 \\
         & Edges & 362 \\
         & Components & 10\\
         & Average Degree (SD) & 3.35 (2.75) \\
         & Density & 0.016 \\
         & Transitivity & 0.25 \\
         & Assortativity & 0.25 \\
    \arrayrulecolor[rgb]{ .8,  .8,  .8}\midrule
      \multirow{2}{*}{{\it Baseline Visit}} & & \\
      & & \\
        \midrule 
    \multirow{2}{*}{Community alert} & Exposed & 25 (11.6\%) \\
         & Not Exposed & 191 (88.4\%) \\
    \midrule
    \multirow{2}{*}{HIV Status} & Positive & 113 (52.3\%) \\
         & Negative & 103 (47.7\%) \\
    \midrule
    \multirow{2}{*}{Date of first interview} & Before ARISTOTLE ended & 110 (50.9\%) \\
         & After ARISTOTLE ended & 106 (49.1\%) \\
    \midrule
    \multirow{4}{*}{Education} & Primary School or less & 64 (29.6\%) \\
         & High School (first 3 years) & 68 (31.5\%) \\
         & High School (last 3 years) & 52 (24.1\%) \\
         & Post  High School & 32 (14.8\%) \\
    \midrule
    \multirow{4}{*}{Employment status} & Employed & 33 (15.3\%) \\
         & Unemployed; looking for work & 54 (25.0\%) \\
         & Can't work; health reason & 101 (46.8\%) \\
         & Other & 28 (12.9\%)\\
    \midrule
    Shared injection equipment & Yes  & 159 (73.6\%)\\
    in last 6 months & No   & 57 (26.4\%) \\
    \midrule
    \multirow{2}{*}{{\it Six-month Visit}} & & \\
      & & \\
        \midrule 
    Outcome: sharing injection  & Yes  & 92 (42.6\%) \\
    equipment at the 6-month visit  & No   & 124 (57.4\%)\\
    \arrayrulecolor[rgb]{ 0,  0,  0}\bottomrule
    \end{tabular}%
    
    \footnotesize{$^1$ The transitivity measures the density of triads in a network. The assortativity quantifies the extent to which connected nodes share similar properties. }
\label{tab:descripitve_stat}
\end{table}

\section{Notation}
We employ a potential outcomes framework for causal inference and assume the sufficient conditions for valid estimation of causal effects, which have been well-described \citep{ogburn2014causal, liu2016inverse, forastiere2016identification}. However, we relax the no dissemination or interference assumption \citep{rubin1980}. In our setting, we evaluate the effect of a non-randomized intervention on a subsequent outcome in an observed network, where information is available on the nodes (i.e., each participant) and their links (i.e., HIV risk connections through sexual or injection behavior). We evaluate the effect of being exposed to community alerts on HIV risk behaviors (i.e., sharing injection equipment) reported at the 6-month follow up. According to the network-based study design of TRIP that recruited at least one wave of contact tracing for each participant of an HIV-infected seed, we anticipate that there could be dissemination or spillover between two individuals connected by an link (i.e., possible influence of their neighbors' intervention exposure on an individual's outcome).  Based on reported connections, we assume that smaller groupings or neighbors for each individual can be identified in the data. Following \cite{forastiere2016identification} and \cite{liu2016inverse}, we make the nearest neighbors interference assumption (NIA). The NIA is a network analog to the partial interference assumption used for clusters \citep{sobel2006randomized,hudgens2008toward}; however, partial interference does not assume a unique interference set for each individual, but instead the set is the same for all individuals in a cluster.  The NIA assumption applies to the nearest neighbors uniquely defined for each participant in the study, so the connections between individuals and their neighbors can now be explicitly considered in the estimands and estimation. This implies that the potential outcomes of a participant depend only on their own exposure and that of their nearest neighbors and not on the exposures of others in the network beyond the nearest neighbors, positing that an individual only has spillover from their first degree contacts. In other words, if the exposures of an individual and their neighbors are held fixed, then changing the exposures of others outside the nearest neighbors and the individual does not change the outcome for the individual. %We assume that the nearest neighbors sets are fixed and known and that the observed study network is complete and the baseline covariates are independent between individuals.

Consider a finite population of $n$ individuals and each individual self-selects their exposure to a study intervention. Let  $i=1, ... , n$ denote each participant in the study and let $A_i$ be the binary exposure of participant $i$ with $A_i=1$ if exposed to an intervention and $0$, otherwise. Let $Z_i$ denote the vector of pre-exposure covariates for participant $i$. These participants are connected through an observed network $\mathcal{C}$ that can be represented by a binary adjacency matrix $E(\mathcal{C}) = [e_{ij}]_{i,j=1}^n \in \lbrace 0,1 \rbrace^{n \times n}$, with $e_{ij}=1$ if participants $i$ and $j$ share an edge or connection, and $e_{ij}=0$, otherwise. We assume $e_{ii}=0$. Each participant is represented as a node in the network. The set of nodes in network $\mathcal{C}$ is denoted by $V(\mathcal{C})$. Given an observed network, a component is a connected subnetwork that is not part of any larger connected subnetwork. A network that is itself connected has exactly one component. If $\mathcal{C}$ has $m$ components, we denote the components $\{C_\nu|\nu=1, \hdots, m\}$. Denote the nearest neighbors of participant $i$ by $\N_i=\lbrace j: e_{ij}=1 \rbrace$ and  ${\N_i^*}=\N_i\cup\{i\}$ denote the nearest neighbors and participant $i$.  The degree of individual $i$ (or number of nearest neighbors) is denoted as $d_i = \sum_{j=1}^n e_{ij} =|\mathcal{N}_i|$. We denote the vector of intervention exposures for the nearest neighbors for participant $i$ as ${A}_{\mathcal{N}_i}=  [A_{j}]_{j:e_{ij}=1}$. In this setting, the outcome of participant $i$ depends not only on their own exposure, but also on the vector of their neighbors' exposures ${A}_{\mathcal{N}_i}$ (NIA). In other words, we let $\N_i$ be the interference set of individual $i$ in which the neighbors' exposures may affect the outcome of individual $i$. We also denote the vector of pre-exposure covariates for the nearest neighbors for participant $i$ as  ${Z}_{\mathcal{N}_i}= [Z_{j}]_{j:e_{ij}=1}$. Denote realizations of exposures $A_i$ by $a_i$ and ${A}_{\mathcal{N}_i}$ by $a_{\mathcal{N}_i}$. Similarly, denote realizations of covariates  $Z_i$ by $z_i$ and ${Z}_{\mathcal{N}_i}$ by $z_{\mathcal{N}_i}$. 

Let $y_i(a_i, {a}_{\mathcal{N}_i})$ denote the potential outcome of individual $i$ if they received intervention $a_i$ and their nearest neighbors received the vector of interventions denoted by ${a}_{\mathcal{N}_i}$. Let $Y_i = y_i(A_i, {A}_{\mathcal{N}_i})$ denote the observed outcome, which holds by causal consistency. Therefore, the potential outcomes are assumed to be deterministic functions and the observed outcomes are assumed to be random variables. In our study setting, $A_i$ represents an indicator for whether participant $i$ is exposed to community alerts and the pre-exposure covariates include HIV status ascertained in the TRIP study, date of first interview, education status, employment status, and report of shared drug use equipment (e.g. syringe) in last 6 months prior  to baseline. The observed outcome $Y_i$ is the status of sharing injection equipment in the last 6 months prior to the 6-month follow-up visit. 

In this paper, we define average potential outcomes using a Bernoulli allocation strategy \citep{tchetgen2012causal}, where $\alpha$ represents the counterfactual scenario in which individuals in $\mathcal{N}_i$ receive the exposure with probability $\alpha$ and we refer to this parameter as the \textit{intervention} coverage for the nearest neighbors. This is essentially like  standardizing the observed exposure vectors to study population in which the exposure assignment mechanism follows a Bernoulli distribution with probability $\alpha$. This allows stochasticity in the intervention assignment for individuals who are possibly members of more than one nearest neighbors. In the observational study, we are not assuming that $A_1,\cdots, A_n$ are independent Bernoulli random variables; however, this distribution of exposure is used to define the counterfactuals. We use information collected in a sociometric network with a non-randomized intervention for estimation. Let $\pi(a_{\N_i};\alpha)=\alpha^{\sum_{j\in \N_i}a_j}(1-\alpha)^{d_i-\sum_{j \in \N_i}a_j}$ denote the probability of  the nearest neighbors of individual $i$ receiving intervention exposure $a_{\N_i}$  under allocation strategy $\alpha$. The allocation strategy $\alpha$ can also be considered as the intervention coverage level for the nearest neighbors. Let $\pi(a_i;\alpha)=\alpha^{a_i}(1-\alpha)^{1-a_i}$ denote the probability of individual $i$ receiving exposure $a_i$ and $\pi(a_i, a_{\N_i};\alpha)=\pi(a_{\N_i};\alpha)\pi(a_i;\alpha)$ denote the probability of individual $i$ together with their nearest neighbors receiving the set of exposures $(a_i,a_{\N_i})$.

\section{Estimands}
We follow notations from \cite{liu2016inverse} to define the estimands. Define $\bar{y}_i(a,\alpha) =\sum_{a_{\mathcal{N}_{i}}} y_i(a_i=a,{a}_{\mathcal{N}_i})\pi({a}_{\mathcal{N}_i};\alpha)$ to be the average potential outcome for individual $i$ under allocation strategy $\alpha$ and exposure $a_i=a$ where the summation is over all $2^{d_i}$ possible values of ${a}_{\mathcal{N}_i}$. Averaging over all individuals, we define the population average potential outcome as $\bar{y}(a,\alpha)=\sum_{i=1}^n \bar{y}_i(a,\alpha)/n$. We also define the marginal average potential outcome for individual $i$ under allocation strategy $\alpha$ by $\bar{y}_i(\alpha)=\sum_{a_i,a_{\mathcal{N}_{i}}} y_i(a_i,a_{\mathcal{N}_i})\pi(a_i,{a}_{\mathcal{N}_i};\alpha)$ and define the marginal population average potential outcome as $\bar{y}(\alpha)=\sum_{i=1}^n \bar{y}_i(\alpha)/n$. 

We consider different contrasts of these average causal effects often of interest in network-based studies. We define these on the risk difference scale and analogous effects can be defined on the ratio scale. The direct effect is defined as $\overline{DE}(\alpha)=\bar{y}(1,\alpha)-\bar{y}(0,\alpha)$, which compares the average potential outcomes when a participant is exposed to the intervention compared to when a participant is not exposed under allocation strategy $\alpha$. For example, in TRIP study, the direct effect is a difference in the risk of reporting HIV risk behaviors  when a participant is exposed to community alerts versus when a participant is not exposed with  $100\cdot\alpha$\% of their nearest neighbors exposed to alerts. The disseminated (i.e., indirect or spillover) effect is $\overline{IE}(\alpha_1, \alpha_0)=\bar{y}(0,\alpha_1)-\bar{y}(0,\alpha_0)$, which compares the average potential outcomes of unexposed individuals  under two different allocation strategies $\alpha_1$ and $\alpha_0$. The composite or total effect is defined as $\overline{TE}(\alpha_1, \alpha_0)=\bar{y}(1,\alpha_1)-\bar{y}(0,\alpha_0)$, which is a function of both the direct and disseminated effects and is a measure of the maximal intervention effect (assuming that $\alpha_1>\alpha_0$), comparing average potential outcomes for exposed participants under allocation strategy $\alpha_1$ to unexposed participants under allocation strategy $\alpha_0$. Lastly, the overall effect is  $\overline{OE}(\alpha_1, \alpha_0)=\bar{y}(\alpha_1)-\bar{y}(\alpha_0)$, which is the difference in average potential outcomes under two different allocation strategies. 

\section{IPW Identification Assumptions and Estimators}

\begin{figure}
    \centering
    \includegraphics[scale=0.3]{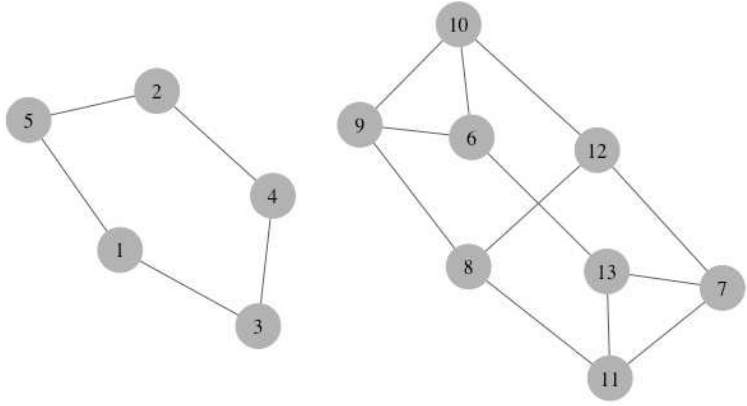}
    \caption{A sample network with two components. $C_1= \{1, 2, 3, 4, 5\}$ and $C_2=\{6, 7, 8, 9, 10, 11, 12, 13\}$. The nearest neighbors of node $2$ are $\N_2=\{4, 5\}$, of node 3 are $N_3=\{1, 4\}$, and of node $6$ are $\N_6=\{9, 10, 13\}$.}
    \label{fig:sample_network}
\end{figure}

In an observational study of a network, interventions are typically not randomized at either the network or individual-level, but rather individuals and their nearest neighbors typically self-select their own exposures.  Therefore, identification of causal effects does not benefit from exchangeability achieved by randomization and adjustment for a sufficient set of pre-exposure covariates at both the individual- and network-level is needed to quantify causal effects. In this section, we apply two different IPW estimators \citep{liu2016inverse, forastiere2016identification} to a setting with the interference set defined by nearest neighbors in observed networks with components. We assume that the observed network can be expressed as the union of $m$ components denoted by $C_1, C_2, \hdots, C_m$ for $\nu = 1, \hdots, m$ (Figure \ref{fig:sample_network}). We quantify the variance accounting for correlation within components of the full observed network. Importantly, we now incorporate the nearest neighbor structure in the estimating equations used to calculate the closed-form variances because this better reflects the underlying structure through which dissemination operates in the observed network. Individuals who share a connection or link are more likely to influence each other, as opposed to individuals who are clustered together, possibly in a large grouping like a component. In \cite{liu2016inverse}, information on their connections or distance in the network between individuals is either not available or not used for statistical inference and the assumption is nonetheless made that these individuals could all possibly influence each other within the set (i.e., a generalized interference set). 

\subsection{Assumptions}
\begin{assumption} (Exchangeability) \label{assump:exchange}
Assume that conditional on pre-exposure covariate vector $Z_i$ and the covariates of their nearest neighbors ${Z}_{\mathcal{N}_i}$, the intervention allocation for individual $i$ and their nearest neighbors $\mathcal{N}_i$ is independent of all potential outcomes 
\begin{align} 
&\Pr(A_i=a_i, {A}_{\mathcal{N}_i}={a}_{\mathcal{N}_i}|Z_i=z_i,{Z}_{\mathcal{N}_i}={z}_{\mathcal{N}_i})\notag\\
=&\Pr(A_i=a_i, {A}_{\mathcal{N}_i}={a}_{\mathcal{N}_i} |Z_i=z_i, {Z}_{\mathcal{N}_i}={z}_{\mathcal{N}_i}, y_1(\cdot),\ldots, y_n(\cdot)).\notag\end{align}
\end{assumption} 

\begin{assumption}\label{assump:positive}
(Positivity) Assume that 
$\Pr(A_i=a_i|Z_i=z_i)>0$ and $\Pr(A_i=a_i, A_{\N_i}=a_{\N_i}|Z_i=z_i, Z_{\N_i}=z_{\N_i})>0$ for all $a_i$, $a_{\N_i}$, $z_i$, and $z_{\N_i}$.
\end{assumption}

\begin{assumption}\label{assump:irrelavance}
(Treatment variation irrelavance) We assume that the treatment or intervention assignment mechanism does not affect the outcome. More precisely, if there are different versions of the intervention, we assume that those are irrelevant for the causal contrasts of interest and that we have one version of intervention and one version of no intervention. \citep{forastiere2016identification}. 
\end{assumption}

\begin{assumption}\label{assump:local_nn}
(Conditional exposure independence) Conditional on the exposure and covariates for individual $i$ and their neighbors $\N_i$ and the neighbor-level random effect $b_{\N_i^*}$, the exposure $A_i$ for individual $i$ and the exposure for the neighbors $A_{\N_i}$ are independent. That is, given the nearest neighbor-level random effect $b_{\N_i^*}$ and $b_{\N_i^*}$,  
$$A_i|A_{\N_i}, Z_{\N_i}, b_{\N_i^*} \perp A_j|A_{\N_j}, Z_{\N_j}, b_{\N_i^*}.$$
\end{assumption}
The nearest neighbor-level random effect $b_{\N_i^*}$ accounts for possible correlation of exposures among individual $i$ and their neighbors $\N_i$. This assumption is used to estimate the propensity score of IPW$_1$ (defined in Section 5.2).

\begin{assumption}\label{assump:nni}
(Nearest neighbors interference) The outcome for an individual depends their own exposure and the exposures of only other individuals who are their nearest neighbors \citep{forastiere2016identification}. By consistency, the following holds:
$$Y_i=Y_i(A_i, A_{\N_i}).$$
\end{assumption}
For example, in Figure \ref{fig:sample_network}, $Y_1=Y_1(a_1, (a_3, a_5))$, which is that the outcome for individual 1   is affected by their own exposure and the exposure of individual 3 and 5 only and no other individuals' exposures in either the component or network.

\begin{assumption}\label{assump:stratifi}
(Stratified interference) The outcome for an individual depends on their own exposure and on the total number of exposed nearest neighbors \citep{hudgens2008toward, sobel2006randomized}. 
\end{assumption}

\begin{assumption}\label{assump:reducible} (Reducible propensity score assumption)
The individual exposure $A_i$ does not depend on neighbors' covariates $Z_{\N_i}$ and neighbors' exposures $A_{\N_i}$ do not depend on individual covariates $Z_i$ \citep{forastiere2016identification}.
\begin{center}
    $P(A_i|Z_i, Z_{\N_i})=P(A_i|Z_i)$ and $P(A_{\N_i}|A_i, Z_i, Z_{\N_i})=P(A_{\N_i}|A_i, Z_{\N_i})$.
\end{center}
\end{assumption}

Exchangeability (Assumption \ref{assump:exchange}), positivity (Assumption \ref{assump:positive}), and treatment variation irrelevance (Assumption \ref{assump:irrelavance}) are necessary assumptions for causal inference under the potential outcomes framework \citep{rubin1980}. Due to the lack of randomization of the intervention, we require a conditional exchangeability assumption for both the individual and their neighbors, which allows for identification of causal contrasts related to both the individual's exposure and the allocation strategy for their neighbors. The positivity assumption ensures we have individuals and their neighbors exposed (and not exposed) at each level of the covariates. We also assume treatment variation irrelevance for the intervention, which ensures we have only one version of being exposed to the intervention and one version of not being exposed, which clarifies how we define the potential outcomes related to each intervention exposure. In this work, we assume that only the first degree neighbors' exposures can influence an individual's outcome, which allows us to focus locally in the network to evaluate spillover. Assumption \ref{assump:stratifi} and \ref{assump:reducible} apply to IPW$_2$ only and are discussed in Section 5.2; however, assumption \ref{assump:stratifi} may also be applied to IPW$_1$ (defined in Section 5.2) when there are concerns about positivity violations.

\subsection{Estimators}
Under the Assumptions \ref{assump:exchange}, \ref{assump:positive}, \ref{assump:irrelavance}, \ref{assump:local_nn} and \ref{assump:nni}, the first IPW estimator is an adaptation of the one proposed by \citet{liu2016inverse}, and we  define the interference sets by the nearest neighbors for each individual in the observed network, and then use this nearest neighbor structure within each component when deriving the closed-form variance estimator in Section 5.3. Define the IPW estimator for exposure $a$ with allocation strategy $\alpha$ as
\begin{equation}\widehat{Y}^{IPW_1}(a, \alpha)=\frac{1}{n}\sum_{i=1}^n \frac{y_i(A_i, A_{\N_i})I(A_i=a)\pi(A_{\N_i};\alpha)}{f_1(A_i, A_{\N_i}|Z_i, Z_{\N_i})},\label{eq:eq1}\end{equation} where  $f_1(A_i, A_{\N_i}|Z_i, Z_{\N_i})$ is the nearest neighbors-level exposure propensity score. We assume that conditional on the nearest neighbor-level random effect and the exposure and covariates for individual $i$ and the neighbors $\N_i$, the exposures of nearest neighbors $A_{\mathcal{N}_i}$ and the exposure of individual $A_i$ are independent. In other words, the dependency between the exposures for individual $i$ and their neighbors is captured by both the fixed exposures and covariates and nearest neighbor-level random effect. To model the propensity score, the probability of exposure following a Bernoulli distribution and conditional on observed baseline covariates is given by 
$$f_1(A_i, A_{\N_i}|Z_i, Z_{\N_i})=\int_{-\infty}^\infty\prod_{j \in \N_i^*}p_j^{A_j}(1-p_j)^{1-A_j}f(b_{\N_i^*}; 0,  \psi)db_{\N_i^*},$$
where $\N_i^*=\N_i\cup\{i\}$, $$p_j=\mbox{Pr}(A_j=1|Z_j, b_{\N_i^*})=\mbox{logit}^{-1}(Z_j\cdot\gamma+b_{\N_i^*}),$$ and $f(b_{\N_i^*}; 0, \psi)\sim N(0, \psi)$. Here, $b_{\N_i^*}$ is the nearest neighbors-level random effect accounting for possible correlation of exposures among individual $i$ and their neighbors $\N_i$.

\noindent The marginal population-level average potential outcome estimator is
\begin{equation}\widehat{Y}^{IPW_1}(\alpha)=\frac{1}{n}\sum_{i=1}^n \frac{y_i(A_i, A_{\N_i})\pi(A_i, A_{\N_i};\alpha)}{f_1(A_i, A_{\N_i}|Z_i, Z_{\N_i})}.\label{eq:eq1}\end{equation} 
  
Under the Assumptions \ref{assump:exchange}, \ref{assump:positive}, \ref{assump:irrelavance}, \ref{assump:nni}, \ref{assump:stratifi} and \ref{assump:reducible}, the second IPW estimator uses an individual and nearest neighbors propensity score as defined in \cite{forastiere2016identification}. The potential outcomes of individual $i$ depend on the total number of exposed neighbors, $s_i=\sum_{j \in \N_i} a_j$ (and let $S_i=\sum_{j \in \N_i} A_j$). In particular, $$y(a_i, a_{\N_i})=y(a_i, s_i).$$ The IPW estimator for exposure $a$ with coverage $\alpha$ is defined as
\begin{equation}\widehat{Y}^{IPW_2}(a;\alpha)=\frac{1}{n}\sum_{i=1}^n\frac{y_i(A_i, S_i)I(A_i=a)\pi(S_i;\alpha)}{f_2(A_i, S_i|Z_i, Z_{\N_i})},\label{eq:eq3}\end{equation} and the IPW marginal estimator as
\begin{equation}\widehat{Y}^{IPW_2}(\alpha)=\frac{1}{n}\sum_{i=1}^n\frac{y_i(A_i, S_i)\pi(A_i, S_i;\alpha)}{f_2(A_i, S_i|Z_i, Z_{\N_i})}.\label{eq:eq4}\end{equation} Let  $$\pi(S_i;\alpha)={d_i\choose S_i}\alpha^{S_i}(1-\alpha)^{d_i-S_i}$$ be the probability of individual $i$ has $S_i$ exposed neighbors and
$$\pi(A_i, S_i;\alpha)=\pi(S_i;\alpha) \pi(A_i;\alpha)$$ denote the probability of exposure for individual $i$ together with $S_i$ exposed neighbors.

The propensity score $f_2(A_i, S_i|Z_i, Z_{\N_i})$ is the joint probability distribution of individual exposure and nearest neighbors exposure given the covariates $Z_i$ and $Z_{\N_i}$. Here, we express this as a product of the individual propensity score,  $f_{22}(A_i|Z_i)$, and nearest neighbors propensity score, $f_{21}(S_i|A_i, Z_{\N_i})$. 

We assume that the individual exposure $A_i$ follows a Bernoulli distribution 
$$P(A_i=a_i|Z_i)=p_{2, i}^{A_i}(1-p_{2, i})^{1-A_i}$$
with probability $p_{2, i}$ defined as the individual propensity score, modeled as a function of a covariate vector using a logit link
$$p_{2, i}=\mbox{Pr}(A_i=1|Z_i)=\mbox{logit}^{-1}(Z_i\cdot\gamma).$$
Furthermore, we assume that the total number of exposed neighbors $\sum A_{\mathcal{N}_i}$ follows a binomial distribution
 $$P(S_i=s_i| A_i, Z_{\N_i})={d_i \choose S_i}p_{1, i}^{S_i}(1-p_{1, i})^{d_i-S_i}$$
with probability $p_{1, i}$  modeled as a function of the nearest neighbors covariate vector using a logit link
$$p_{1, i}=\mbox{Pr}(S_i=s_i|A_i, Z_{\N_i})=\mbox{logit}^{-1}(A_i\beta + h(Z_{\N_i})\cdot\delta'),$$ where $h(Z_{\N_i})$ is an aggregate function of the vector $Z_{\N_i}$. For instance, the proportion of females or males in the nearest neighbors or average age of an individual's nearest neighbors.

We assume that conditional on the nearest neighbors covariates and the exposure for individual $i$, the exposures of nearest neighbors $A_{\mathcal{N}_i}$ are independent and identically distributed. In other words, the dependency between neighbors' exposure is captured by the correlation with the exposure for individual $i$ and the covariates of the nearest neighbors.\footnote{In principle, we could compute the nearest neighbors propensity score $f_{21}(S_i|A_i, Z_{\N_i})$ as a product of the individual propensity scores for all neighbors for all exposure combinations $a_{\N_i}$ such that $S_i=s_i$ under the assumption of independence of $A_i$ given a nearest neighbor-level random effects and individual exposure and covariates. This would be one correct way of computing the nearest neighbors propensity score. Instead in this estimator, we use an alternative solution where the nearest neighbors propensity score is estimated assuming a binomial model conditional on a summary statistics of the nearest neighbors covariates. This approach, while approximate, is more straightforward and works when the dependency among neighbors' exposures cannot be attributed to a latent factor shared by all units belonging to the same nearest neighbor set in the network.} Therefore, 
the propensity score $f_2(A_i, S_i|Z_i, Z_{\N_i})$ can be factor into two marginal distributions $f_{21}$ and $f_{22}$ as follows:
\begin{align}
f_2(A_i, S_i|Z_i, Z_{\N_i})&=f_{21}(S_i|A_i, Z_{\N_i})f_{22}(A_i|Z_i)\notag\\
&={d_i \choose S_i} p_{1, i}^{S_i}(1-p_{1, i})^{d_i-S_i}\cdot p_{2, i}^{A_i}(1-p_{2, i})^{1-A_i}\notag
\end{align}

Under allocation strategy $\alpha, \alpha_0$, and $\alpha_1$, we consider the following risk difference estimators of the direct, disseminated (indirect), composite (total), and overall effects:
\begin{align}
    & \widehat{DE}_r(\alpha)=\widehat{Y}^{IPW_r}(1, \alpha)-\widehat{Y}^{IPW_r}(0, \alpha),\notag\\
    &\widehat{IE}_r(\alpha_1, \alpha_0)=\widehat{Y}^{IPW_r}(0, \alpha_1)-\widehat{Y}^{IPW_r}(0, \alpha_0),\notag\\
    &\widehat{TE}_r(\alpha_1, \alpha_0)=\widehat{Y}^{IPW_r}(1, \alpha_1)-\widehat{Y}^{IPW_r}(0, \alpha_0),\notag\\
    &\widehat{OE}_r(\alpha_1, \alpha_0)=\widehat{Y}^{IPW_r}(\alpha_1)-\widehat{Y}^{IPW_r}(\alpha_0),\notag
\end{align}
where $r=1, 2$ corresponds to the two IPW estimators that we defined above.

\begin{theorem}
If the propensity scores $f_1(A_i, A_{\N_i}|Z_i, Z_{\N_i})$ and $f_2(A_i, S_i|Z_i, Z_{\N_i})$ are known, then $E[\widehat{Y}^{IPW_r}(a, \alpha)]=\bar{y}(a, \alpha)$ and $E[\widehat{Y}^{IPW_r}(\alpha)]=\bar{y}(\alpha)$.
\end{theorem}
Proof of Proposition 1 is shown in Appendix A. Using these unbiased estimators when the propensity score is known, the estimation of the causal effects will also be unbiased because the causal effects are contrasts of these marginal quantities. 

\subsection{Large sample properties of the inverse probability of sampling weighted estimator}
The large sample variance estimators can be derived using M-estimation theory \citep{mestimator2013}.  We assume that the observed network can be expressed as the union of components; that is, non-overlapping groups of individuals \citep{liu2016inverse}. Consider a social network with $n$ individuals and $m$ components denoted by $\{C_1, C_2, \cdots, C_m\}$ with $\nu = 1,\hdots, m$. Let $Y_{\nu i}, A_{\nu i}, Z_{\nu i}$ denote the outcome, exposure, and covariates for individual $i$ in component $\nu$, respectively. Let $V(C_\nu)$ be the set of nodes in $C_\nu$, and $Y_\nu=\{Y_{\nu i}|i \in V(C_\nu)\}$, $A_\nu=\{A_{\nu i}|i \in V(C_\nu)\}$, $Z_\nu=\{Z_{\nu i}|i \in V(C_\nu)\}$. The observable random variables $(Y_\nu, A_\nu, Z_\nu)$ for $\nu=1, \hdots, m$ are assumed to be independent but not necessarily identically distributed with distribution $F_\nu$. 
We assume that the $m$ components are a random sample from the infinite super-population of groups and the size of each component is bounded \citep{mestimator2013}. 

Recall, for IPW$_1$, the parameters of the exposure propensity score model include coefficients for the fixed effects and the random effect, while for IPW$_2$, the parameters include coefficients for the fixed effects from two logistic models (see Section 5.2). Let $\Theta=\{\gamma, \psi\}$ the set of coefficients of fixed effects and the random effects in the propensity score $f_1$ when using IPW$_1$, and $\Theta=\{\gamma, \beta, \delta'\}$ be the set of coefficients in the propensity score $f_2$ when using IPW$_2$. To generalize notation, we set the dimension of $\Theta$ to be $p$ and refer to these parameters as $\eta$ in the estimating equations below. Let $Y_{C_\nu}=(Y_{ C_\nu0}, Y_{C_\nu1}, Y_{C_\nu2})$ be the component-level average potential outcomes defined as
\begin{align}
    Y_{C_\nu0} &=\sum_{i \in V(C_\nu), a_{\N_i}}y_i(a_i=0, a_{\N_i})\pi(a_{\N_i};\alpha),\notag\\
    Y_{C_\nu1} &= \sum_{i \in V(C_\nu), a_{\N_i}}y_i(a_i=1, a_{\N_i})\pi(a_{\N_i};\alpha),\notag\\
    Y_{C_\nu2} &=\sum_{i \in V(C_\nu), a_i, a_{\N_i}}y_i(a_i, a_{\N_i})\pi(a_i, a_{\N_i};\alpha).\notag
\end{align}
 To conduct inference, we use $m$ independent components, while preserving the underlying connections of an individual's nearest neighbors comprising the network structure of each component. That is, by extending \cite{liu2016inverse}, every individual is now assigned their own propensity score based on the observed network structure defined by their nearest neighbors (see Section 5.2). Whereas in \cite{liu2016inverse}, statistical inference was conducted by assuming partial interference in which the study population was partitioned into non-overlapping groups and all individuals in a group were assigned one group-level propensity score. To simplify the notation in this section, we write the propensity score $f_2(A_{\nu i}, S_{\nu i}|Z_{\nu i}, Z_{\N_{\nu i}})$ as $f_2(A_{\nu i}, A_{\N_{\nu i}}|Z_{\nu i}, Z_{\N_{\nu i}})$  and let the observed outcome for individual $i$ in component $\nu$ be denoted by $Y_{\nu i}=Y_{\nu i}(A_{\nu i}, A_{\N_{\nu i}})$. Note that the potential outcomes are random due to the random sampling of the $m$ components. With this partition of the network, the inverse probability weighted estimator for exposure $a$ and strategy $\alpha$ presented in Section 5.2 equals
\begin{equation}
    \widehat{Y}^{IPW_r}(a, \alpha)=\frac{1}{n}\sum_{\nu=1}^m\sum_{i\in V(C_{\nu})} \frac{Y_{\nu i}I(A_{\nu i}=a)\pi(A_{\N_{\nu i}};\alpha)}{f_r(A_{\nu i}, A_{\N_{\nu i}}|Z_{\nu i}, Z_{\N_{\nu i}})}.\label{eq:eq5}\end{equation} 

 Let $\theta=(\Theta, \theta_{0\alpha}, \theta_{1\alpha}, \theta_\alpha)$, where $\theta_{0\alpha}=\bar{y}(0, \alpha)=1/n\sum _{\nu=1}^m Y_{C_\nu0}$, $\theta_{1\alpha}=\bar{y}(1, \alpha)=1/n\sum _{\nu=1}^m Y_{C_\nu1}$, and $\theta_{\alpha}=\bar{y}(\alpha)=1/n\sum _{\nu=1}^m Y_{ C_\nu2}$. Let $\hat{\theta}=(\hat{\Theta}, \hat{\theta}_{0\alpha}, \hat{\theta}_{1\alpha}, \hat{\theta}_{\alpha})$. Similar to the approach in \cite{liu2016inverse}, let the average component size in the study population be defined as $k=E[|V(C_\nu)|]$, which is the mean component size in the population. We use this to redefine the inverse probability weighted estimators in equation \ref{eq:eq5} because equally weighting individuals ignoring components may result in biased estimators (\cite{basse2018analyzing}). With the average component size, the inverse probability weighted estimator for exposure $a$ and strategy $\alpha$ presented in Section 5.2 equals
\begin{equation}
    \widehat{Y}^{IPW_r}(a, \alpha)=\frac{1}{m}\sum_{\nu=1}^m \frac{1}{k}\sum_{i\in V(C_{\nu})} \frac{Y_{\nu i}I(A_{\nu i}=a)\pi(A_{\N_{\nu i}};\alpha)}{f_r(A_{\nu i}, A_{\N_{\nu i}}|Z_{\nu i}, Z_{\N_{\nu i}})}.\label{eq:eq6}\end{equation} 
 
 The estimating equations corresponding to the estimator in equation \eqref{eq:eq5} are defined as follows 
$$\psi_{\eta}(Y_\nu, A_\nu, Z_\nu; \theta)=\frac{1}{k}\sum_{i \in V(C_\nu)}\frac{\partial \log f_r(A_{\nu i}, A_{\N_{\nu i}}|Z_{\nu i}, Z_{\N_{\nu i}})}{\partial \eta}, \eta\in\Theta,$$
$$\psi_0(Y_\nu, A_\nu, Z_\nu; \theta; \alpha)=\frac{1}{k}\sum_{i \in V(C_\nu)}\biggl\{\frac{Y_{\nu i}I(A_{\nu i}=0)\pi(A_{\N_{\nu i}};\alpha)}{f_r(A_{\nu i}, A_{\N_{\nu i}}|Z_{\nu i}, Z_{\N_{\nu i}})}\biggl\}-\theta_{0, \alpha},$$
$$\psi_1(Y_\nu, A_\nu, Z_\nu; \theta; \alpha)=\frac{1}{k}\sum_{i \in V(C_\nu)}\biggl\{\frac{Y_{\nu i}I(A_{\nu i}=1)\pi(A_{\N_{\nu i}};\alpha)}{f_r(A_{\nu i}, A_{\N_{\nu i}}|Z_{\nu i}, Z_{\N_{\nu i}})}\biggl\}-\theta_{1, \alpha},$$ and $$\psi_2(Y_\nu, A_\nu, Z_\nu; \theta; \alpha)=\frac{1}{k}\sum_{i \in V(C_\nu)}\biggl\{\frac{Y_{\nu i}\pi(A_{\nu i}, A_{\N_{\nu i}};\alpha)}{f_r(A_{\nu i}, A_{\N_{\nu i}}|Z_i, Z_{\N_{\nu i}})}\biggl\}-\theta_{\alpha}.$$ 

Let 
\begin{align*}
\psi_{\nu}(Y_\nu, A_\nu, Z_\nu; \theta)=\begin{pmatrix}\psi_{\eta}(Y_\nu, A_\nu, Z_\nu; \theta) \\ \psi_0(Y_\nu, A_\nu, Z_\nu; \theta; \alpha) \\ \psi_1(Y_\nu, A_\nu, Z_\nu; \theta; \alpha) \\ \psi_2(Y_\nu, A_\nu, Z_\nu; \theta; \alpha)\end{pmatrix}_{\eta\in\Theta},\end{align*}
such that $\displaystyle \sum_{\nu=1}^m\psi_{\nu}(Y_\nu, A_\nu, Z_\nu; \hat{\theta})=0$. Note that $\hat{\theta}$ is the solution for $\theta$ for this vector of estimating equations. In addition,  $E[\psi_{\nu}(Y_\nu, A_\nu, Z_\nu; \theta)]=0$ \citep{mestimator2013}. 
Let  $A(\theta)=E[-\dot{\psi}_{\nu}(Y_\nu, A_\nu, Z_\nu; \theta)]$ and $B(\theta)=E[\psi_{\nu}(Y_\nu, A_\nu, Z_\nu; \theta)\psi_{\nu}(Y_\nu, A_\nu, Z_\nu; \theta)^T]$ with the expectation take across all $m$ components in the population.
\begin{theorem} Under suitable regularity conditions and due to the unbiased estimating equations, as $m\rightarrow \infty$, $\hat{\theta}$ converges in probability to $\theta$ and
$\sqrt{m}(\hat{\theta}-\theta)$ converges in distribution to $N(0, \Sigma)$, where the covariance matrix is given by $$\Sigma=\frac{1}{m}A(\theta)^{-1}B(\theta)A(\theta)^{-T}.$$
\end{theorem}
Additional details for Proposition 2 are shown in Appendix B. A consistent sandwich estimator of $\Sigma$ is given in Appendix B. We demonstrate how to obtain the variance for the estimator of the disseminated effect $\widehat{IE}_r(\alpha_1, \alpha_0)$. An analogous procedure can be performed to obtain the variance for the estimators of the direct, overall and total effects. Let 
\begin{align*}
\psi_{\nu}(Y_\nu, A_\nu, Z_\nu; \theta)=\begin{pmatrix}\psi_{\eta}(Y_\nu, A_\nu, Z_\nu; \theta) \\ \psi_0(Y_\nu, A_\nu, Z_\nu; \theta; \alpha_1) \\ \psi_0(Y_\nu, A_\nu, Z_\nu; \theta; \alpha_0) \\ \psi_1(Y_\nu, A_\nu, Z_\nu; \theta; \alpha_1) \\ \psi_1(Y_\nu, A_\nu, Z_\nu; \theta; \alpha_0) \\ \psi_2(Y_\nu, A_\nu, Z_\nu; \theta; \alpha_1)\\\psi_2(Y_\nu, A_\nu, Z_\nu; \theta; \alpha_0)\end{pmatrix}_{\eta\in\Theta}.\end{align*} Followed with Slutsky’s Theorem and an application of the Delta method as $m\rightarrow \infty$, $\widehat{IE}_r(\alpha_1, \alpha_0)$ is a consistent estimator of $\overline{IE}(\alpha_1, \alpha_0)$ and $\sqrt{m}(\widehat{IE}_r(\alpha_1, \alpha_0)-\overline{IE}(\alpha_1, \alpha_0))$ converges in distribution to $N(0, \Sigma_{IE})$, where $\Sigma_{IE}=\lambda^T\Sigma \lambda$ and $\lambda = (0_{1 \times p}, 1, -1, 0, 0, 0, 0)^T$. A consistent sandwich estimator of the variance of $\overline{IE}(\alpha_1, \alpha_0)$ is given in Appendix B. This variance estimator can be used to construct Wald-type confidence intervals (CIs) for the disseminated effects. 

\section{Simulation}
A simulation study was conducted to evaluate the performance of the two IPW estimators and their corresponding closed-form variance estimators. We focused on the evaluation of the finite sample bias and coverage of the corresponding 95\% Wald-type confidence intervals. The network characteristics (e.g., number of components, number of nodes in each component) and  parameters of potential outcome models were motivated using empirical estimates from the TRIP data. In this simulation study, we considered regular network where each node has the same number of neighbors. We first generate $m$ network components as regular networks of degree four for each node. The number of nodes in each component is sampled from a Poisson distribution with average 10. We conducted several simulations where the numbers of components $m$ is from the set $\{10, 50, 100, 150, 200\}$. Given a generated network, a total of 1,000 data sets were simulated in the following steps. 

\begin{itemize}
\item[Step 1.] A baseline covariate was randomly generated as  $Z_{i}\sim \text{Bernoulli}(0.5)$. We then generated all possible potential outcomes $$y_i(a_i, a_{\N_i})=\text{Bernoulli}(p={\rm logit}^{-1}(-1.75 + 0.5\cdot a_i + \dfrac{s_i}{d_i}  -1.5\cdot a_i\dfrac{s_i}{d_i}  +0.5\cdot Z_i)).$$ 
\item[Step 2.] Assign the random effect to each component in the network $b_\nu\sim N(0, 0.5^2)$ to allow for correlation between the outcomes within components. The exposure was generated as 
$$A_i = \text{Bernoulli}(p={\rm logit}^{-1}\lbrace 0.7-1.4\cdot Z_i +b_\nu\rbrace).$$
\item[Step 3.] We then obtain the corresponding observed outcomes from the potential outcomes that we generated in Step 1. The true parameters were calculated by averaging the potential outcomes as described in Section 4 that we generated in Step 1. 
\end{itemize} 

For each simulated data set, the $\widehat{Y}^{IPW_1}(a, \alpha)$, $\widehat{Y}^{IPW_1}(\alpha)$, $\widehat{Y}^{IPW_2}(a, \alpha)$, and $\widehat{Y}^{IPW_2}(\alpha)$ were computed for $a=0, 1$ and $\alpha=0.25, 0.5, 0.75$. The estimated  standard errors were derived using the appropriate entries from the variance matrix in Appendix B, then averaged across simulations to obtain the average standard error (ASE). Empirical standard error (ESE) was the standard deviation of estimated means across all simulated data sets. Empirical coverage probability (ECP) is the proportion of the instances that the true parameters were contained in the Wald-type 95\% confidence intervals based on the estimated standard errors among the 1000 simulations with a margin of error equal to 0.014. 
In our main scenario, we simulated networks with component size in average 10 and increased the number of components to evaluate the performance of IPW$_1$ and IPW$_2$ for estimation of the average potential outcomes (Tables A1 to A5). The complete simulation results are summarized in Appendix C.  

\begin{figure}[htbp]
\caption{Absolute bias (left) estimator and corresponding Wald 95\% confidence intervals empirical coverage probability (ECP) (right) of IPW$_1$ (top) and IPW$_2$ (bottom) for different number of components in the network}
 \includegraphics[scale=0.53]{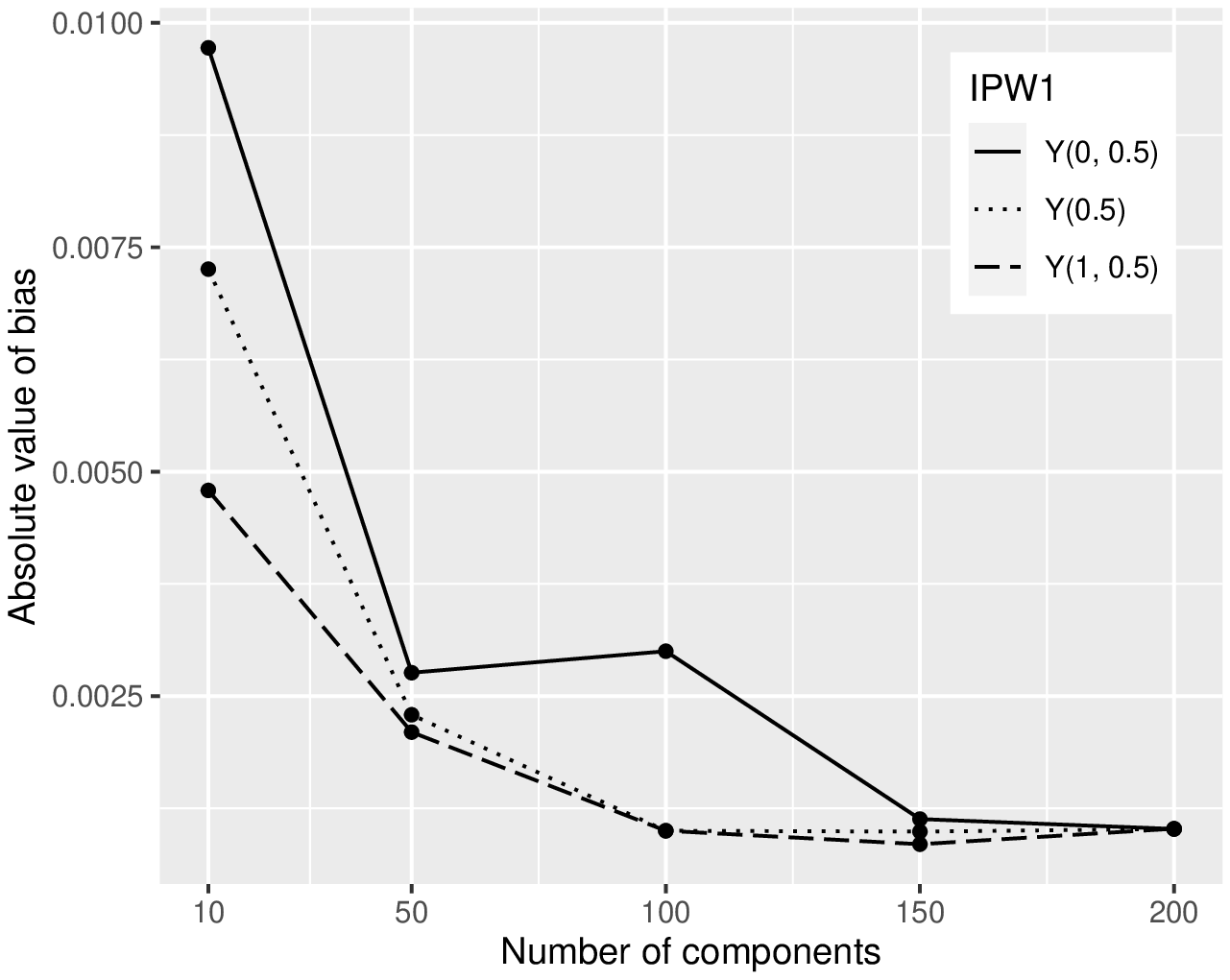}\includegraphics[scale=0.53]{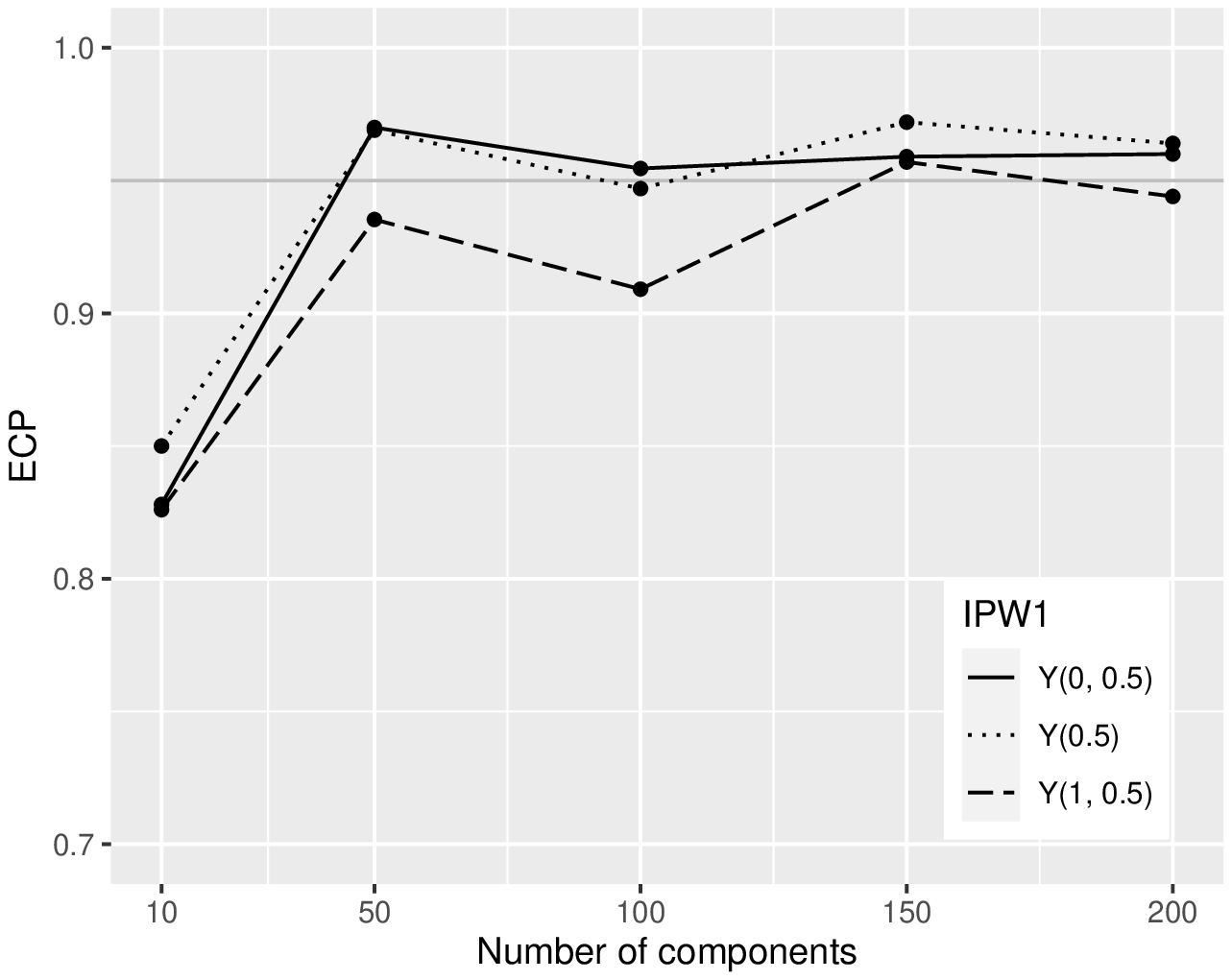}
 \includegraphics[scale=0.53]{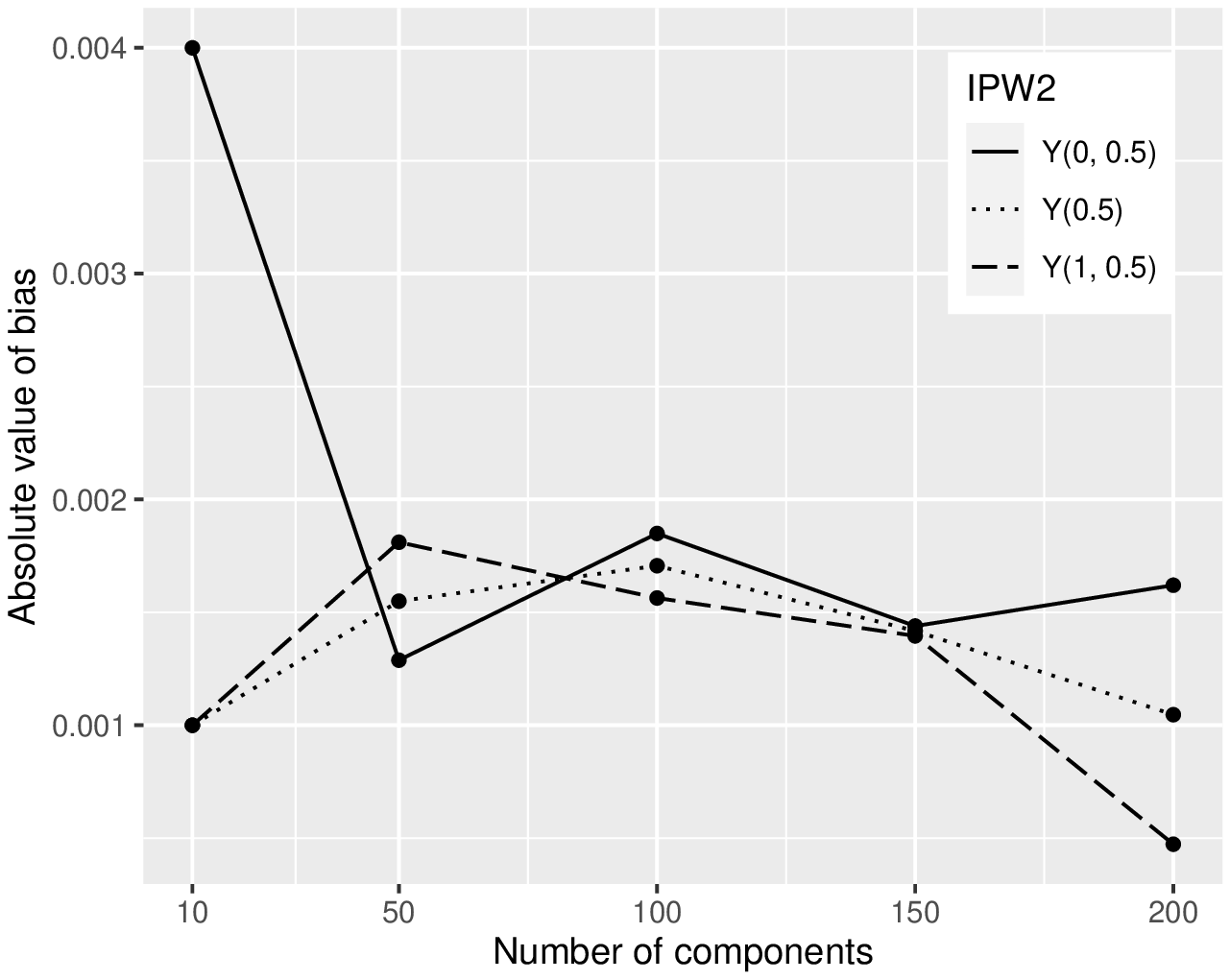}\includegraphics[scale=0.53]{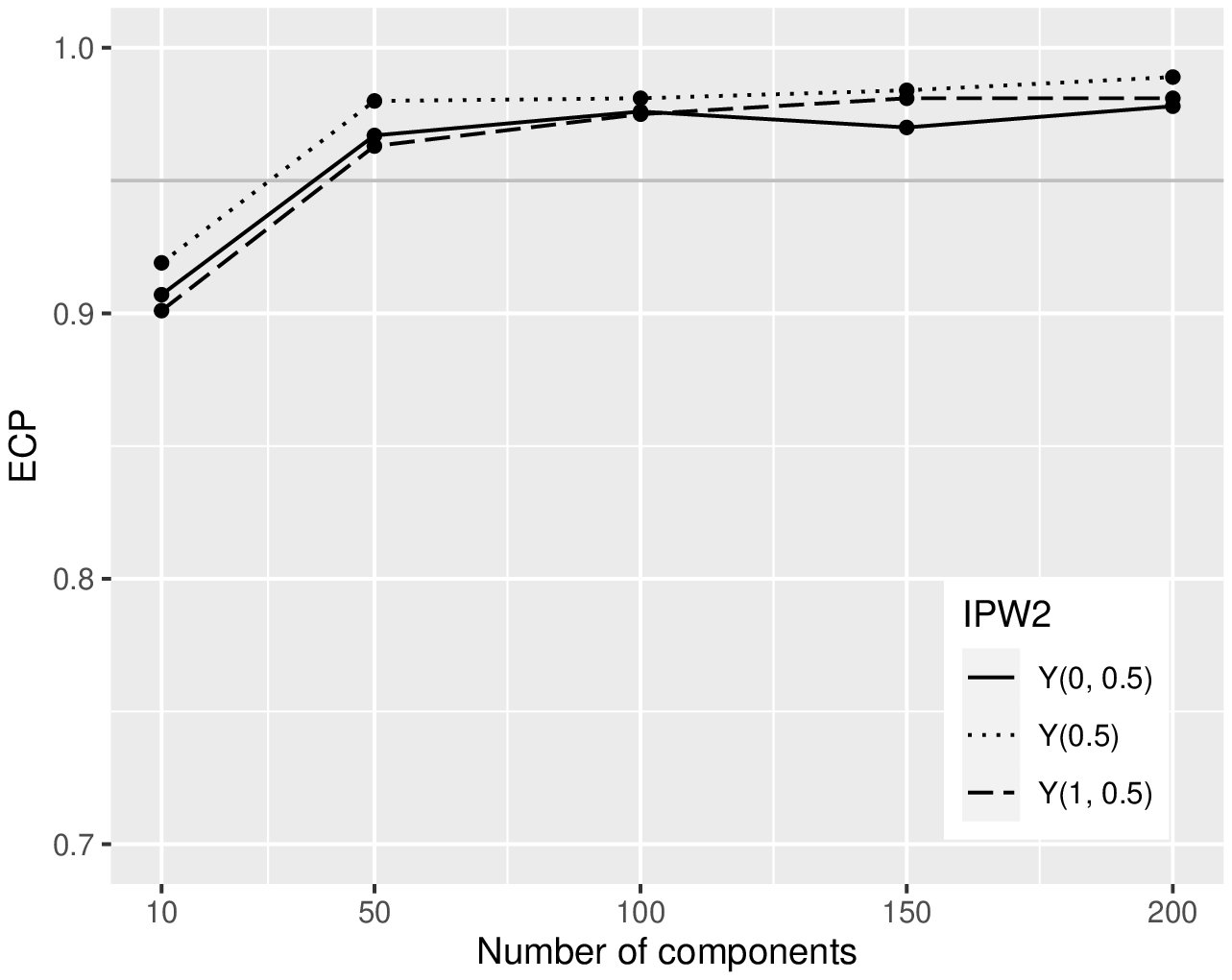}
\label{fig:biasecp}
\end{figure}

Figure \ref{fig:biasecp} shows that the finite sample bias approaches zero and ECPs approach the nominal 0.95 level when the number of components increases from 10 to 200. In Table \ref{tab:200comp}, the ECPs of the estimator IPW$_1$ under all allocation strategies were close to the nominal level and ECPs of IPW$_2$ approach the nominal level when the allocation strategies had a coverage level around 50$\%$ in the observed data. To compare the performance of our variance estimator to an estimator for the asymptotic variance that assumes partial interference \citep{liu2016inverse}, we used observed components in the network as groups to define partial interference sets. The partial interference assumption for variance estimation resulted in higher ASE and ECP, as compared to the asymptotic variance defined in Appendix B, which was closer to the ESE (Figure \ref{fig:liuasymp}).

\begin{figure}[htbp]
\caption{Given a network with 100 components, comparison of the average empirical standard error (ESE), the average standard error (ASE) based on variance estimator in Appendix B, and average standard error (Liu ASE) based on variance estimator in \citet{liu2016inverse}  of the average potential outcomes under allocation strategies 25\%, 50\%, and 75\%.}
\includegraphics[scale=0.99]{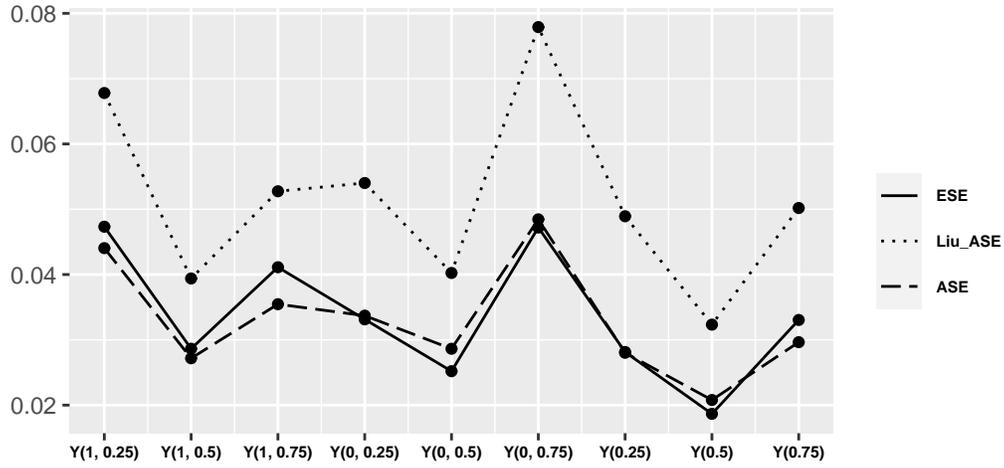}
\label{fig:liuasymp}
\end{figure}

\begin{figure}[htbp]
 \caption{The frequency of the proportion of exposed neighbors in one simulated data when using the exposure generating models $A_i={\rm Bern}({\rm expit}(0.7-1.4\cdot Z_i+b_\nu))$ (left) and $A_i={\rm Bern}({\rm expit}(-0.5-1.5\cdot Z_i+b_\nu))$ (right)}
 \includegraphics[scale=0.53]{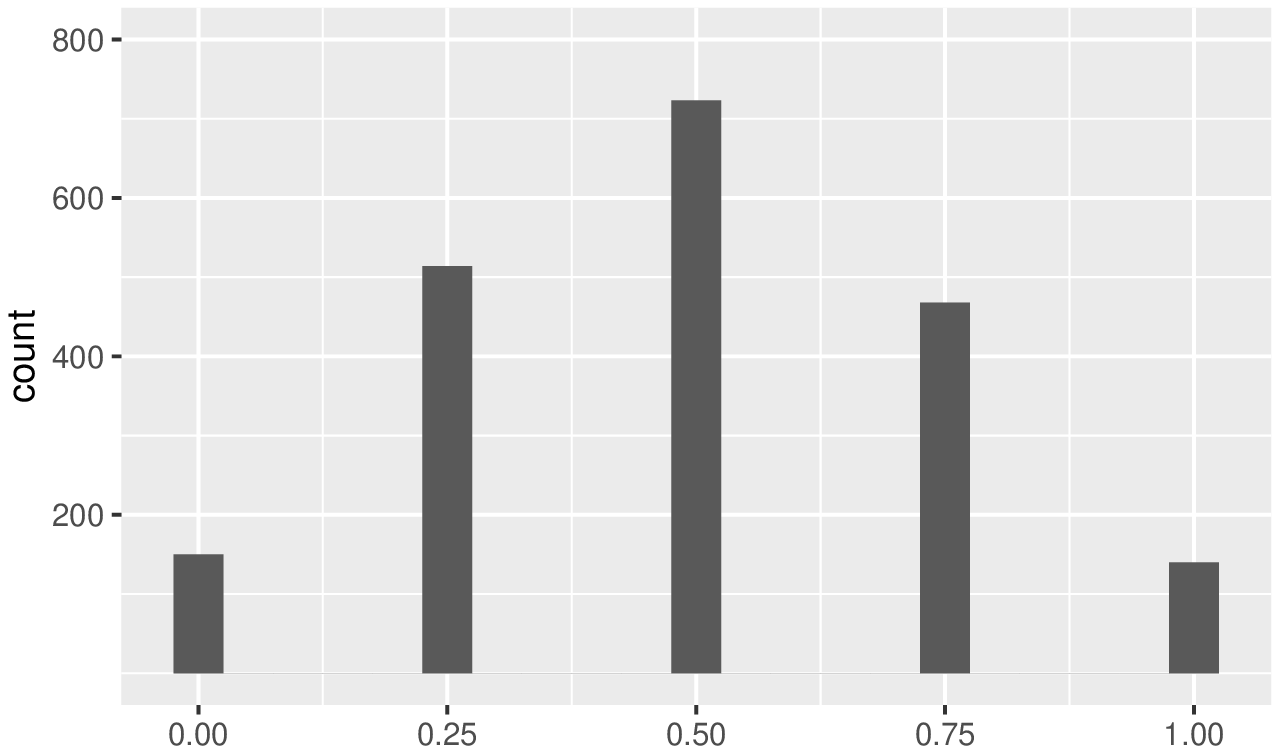}\includegraphics[scale=0.53]{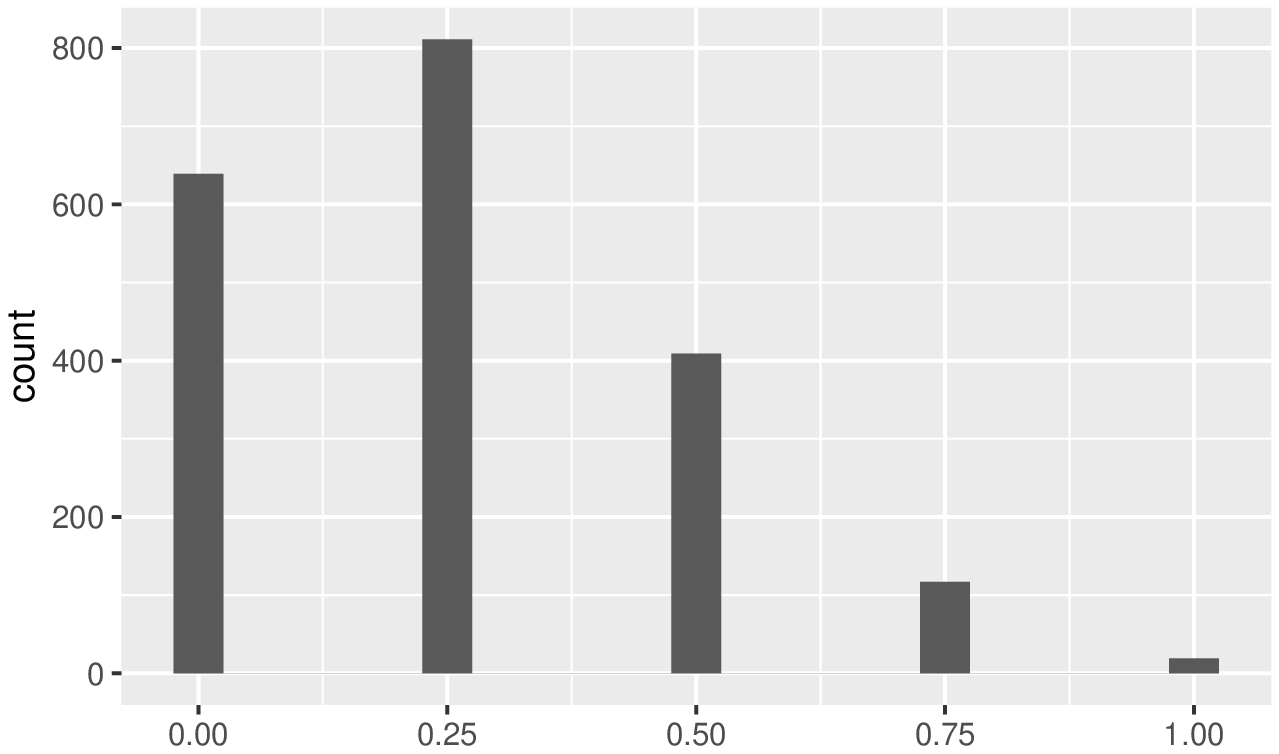}
 \label{fig:distribution}
 \end{figure}

In addition to the main simulation scenarios that vary the number of components, we also used a regular network of degree 4 with 100 components to compare scenarios with a different exposure generating mechanism without random effects, and a scenario in which the stratified interference assumption is violated. In addition to this simulated regular network, we used the TRIP network structure to investigate the performance when community detection was used to further divide the network to larger number of component in the network. We also considered a scenario where we regenerated the network in each simulated dataset. Specifically, we considered the following additional five scenarios:
\begin{itemize}
\item[1.] We used the exposure generating mechanism without random effects as one way to misspecify the propensity score $$A_i={\rm Bern}(p={\rm logit}^{-1}(0.7-1.4\cdot Z_i)).$$
In Table \ref{tab:noranef}, the ECPs of IPW$_2$ were below the nominal level when the exposure mechanism was misspecified, while finite sample performance of IPW$_1$ remained largely similar to settings with a correctly specified exposure mechanism.    

\item[2.] We used a different exposure generating model given by 
$$A_i={\rm Bern}(p={\rm logit}^{-1}(-0.5-1.5\cdot Z_i+b_\nu)).$$ Unlike the previous exposure generating model, this model results in more individuals who have none or 25\% of their neighbors  exposed in the simulated data (Figure \ref{fig:distribution}). In Table \ref{tab:trtdist}, both IPW estimators have higher ECPs for allocation strategy 25\% (IPW$_1$: 94\%, IPW$_2$: 97\%) and lower at allocation strategy 75\% (IPW$_1$: 68\%, IPW$_2$: 71\%) in this scenario, suggesting that the finite sample performance of both estimators for the point estimates and ASEs were better under allocation strategies $\alpha$ for which there were more individuals with $100\cdot\alpha\%$ of their neighbors exposed in the simulated data. 

\item[3.] We considered an outcome model where the stratified interference assumption was violated while the exposure generating model was defined as in Step 2 $$A_i=\mbox{Bernoulli}(p=\mbox{logit}^{-1}\{0.7-1.4\cdot Z_i+b_\nu\}).$$ We used the potential outcome model $y_i(a_i, a_{\N_i})$ given by
$$\text{Bern}(p={\rm logit}^{-1}(-1.75 + 0.5\cdot a_i - 2\cdot \sum_{j \in \N_i} \dfrac{{\scriptstyle I(Z_i=Z_j)}\cdot a_j}{d_i}  +5\cdot \sum_{j\in \N_i}\dfrac{{\scriptstyle I(Z_i\neq Z_j)}\cdot a_j}{d_i}  +0.5\cdot Z_i)).$$
The simulation results in Table \ref{tab:wrongpotentialmodel} showed that both estimators did not perform well with respect to the point estimates, as the magnitude of absolute bias was larger. The ECPs of IPW$_1$ were all greater than 95\% which suggested over-coverage. The ECPs of IPW$_2$ had coverage above the nominal level  or  slightly below the nominal level of 95\%.

\item[4.] We considered the network structure similar to our motivating study TRIP. Based on previous simulation results, a small number of components may result in poor finite-sample performance of variance estimators. To increase the number of components for estimation of the asymptotic variance of the estimated causal effects, we employed an efficient modularity-based, fast greedy approach to detect communities to further divide  large connected components of the TRIP network into a total of 20 smaller components. Modularity takes large values when there are more substantial connections among some individuals than expected if the connections were randomly assigned \citep{PhysRevE.70.066111}. More precisely, each node initially belongs to a separate component, and components are merged iteratively such that each merge yields the largest increase in the current value of modularity. The algorithm stops when it is not possible to increase the modularity any further. As a result, components each comprise a unique set of participants and there are more links between the participants within components than across components in the TRIP network. By ignoring links across components, we treat the obtained communities as independent units to possibly improve the  estimation of the variance. Importantly, we still define the interference sets using the nearest neighbors for point estimation of the causal effects. In this scenario, we use potential outcome model in Step 1 and exposure generating model in Step 2. The simulation results on the TRIP network with and without community detection (Table \ref{tab:trip_sim}) demonstrated that the ECPs of both IPW$_1$ and IPW$_2$ had coverage above the nominal level (97\%-100\%) when using the TRIP network with only 10 components. After further divide the network using community detection, the ECPs have coverage slightly below the nominal level in some cases. To simulate more realistic covariates, we considered a scenario with additional baseline covariates in a TRIP network with 20 components. Two binary variables, $Z_{1, i}\sim \mbox{Bernoulli}(0.5)$ and $Z_{2,i}\sim\mbox{Bernoulli}(0.5)$, and two continuous variables, $Z_{3, i}\sim N(1, 0.5^2)$ and $Z_{4, i}\sim N(0, 1)$, were added into the exposure generating model
$$A_i=\mbox{Bern}(\mbox{logit}^{-1}(-1.4\cdot Z_{1,i}+2\cdot Z_{2,i}-1.5\cdot Z_{3,i}+1.2\cdot Z_{4,i})).$$ The results are summarized in Table \ref{tab:TRIP_sim_4cov}. The estimators had coverage below the nominal level using IPW$_1$; however, IPW$_2$ performed slightly better in terms of ECP.

\item[5.] In the previous scenarios, we considered a network that is generated once (or fixed) and simulated 1000 datasets based on the one network. To evaluate the impact of uncertainty in the network structure, we also considered a scenario where we regenerated the network in each simulated dataset. We first generate a degree four regular network with 100 components, then Step 1-3, repeated 1000 times for each dataset (Table \ref{tab:multinetwork}). The results are mostly comparable to the results that  generated one simulated network and simulated 1000 datasets on a fixed network (Table \ref{tab:100comp}). 
\end{itemize} 

% Table generated by Excel2LaTeX from sheet 'Sheet1'
\begin{table}[htbp]
  \centering
  \caption{Results from 1000 simulated datasets on a network with 100 components for IPW$_1$ (left) and IPW$_2$ (right) for exposed ($a=1$), not exposed ($a=0$), and marginal estimators under allocation strategies 25\%, 50\%, and 75\% using exposure generating model $A_i={\rm Bern}({\rm expit}(0.7-1.4\cdot Z_i)).$}
    \begin{tabular}{l|rrrr|rrrr}
    \toprule
    \rowcolor[rgb]{ .929,  .929,  .929}      & \multicolumn{4}{c|}{IPW1} & \multicolumn{4}{c}{IPW2} \\
    \rowcolor[rgb]{ .929,  .929,  .929}      & \multicolumn{1}{c}{Bias} & \multicolumn{1}{c}{ESE} & \multicolumn{1}{c}{ASE} & \multicolumn{1}{c|}{ECP} & \multicolumn{1}{c}{Bias} & \multicolumn{1}{c}{ESE} & \multicolumn{1}{c}{ASE} & \multicolumn{1}{c}{ECP} \\
    \midrule
    \midrule
    $\widehat{Y}(1, 0.25)$ & 0.0013 & 0.048 & 0.046 & 0.91 & 0.0149 & 0.039 & 0.036 & 0.86 \\
    $\widehat{Y}(1, 0.5)$ & -0.0003 & 0.027 & 0.028 & 0.96 & 0.0016 & 0.022 & 0.024 & 0.96 \\
    $\widehat{Y}(1, 0.75)$ & -0.0032 & 0.041 & 0.041 & 0.93 & 0.0093 & 0.035 & 0.033 & 0.88 \\
    $\widehat{Y}(0, 0.25)$ & -0.0051 & 0.039 & 0.040 & 0.94 & 0.0093 & 0.033 & 0.031 & 0.90 \\
    $\widehat{Y}(0, 0.5)$ & -0.0018 & 0.028 & 0.029 & 0.95 & 0.0018 & 0.023 & 0.025 & 0.97 \\
    $\widehat{Y}(0, 0.75)$ & 0.0004 & 0.053 & 0.051 & 0.92 & 0.0195 & 0.045 & 0.042 & 0.83 \\
    $\widehat{Y}(0.25)$ & -0.0035 & 0.032 & 0.032 & 0.94 & 0.0107 & 0.027 & 0.026 & 0.88 \\
    $\widehat{Y}(0.5)$ & -0.0010 & 0.021 & 0.021 & 0.97 & 0.0017 & 0.016 & 0.018 & 0.97 \\
    $\widehat{Y}(0.75)$ & -0.0023 & 0.033 & 0.030 & 0.93 & 0.0118 & 0.029 & 0.028 & 0.86 \\
    \bottomrule
    \end{tabular}%
  \label{tab:noranef}%
\end{table}%

\begin{table}[htbp]
  \centering
  \caption{Results from 1000 simulated datasets on a network with 100 components for IPW$_1$ (left) and IPW$_2$ (right) for exposed ($a=1$), not exposed ($a=0$), and marginal estimators under allocation strategies 25\%, 50\%, and 75\% using an outcome model where the stratified interference assumption is violated.}
    \begin{tabular}{lr|rrrr|rrrr}
    \toprule
    \rowcolor[rgb]{ .949,  .949,  .949}       &       & \multicolumn{4}{c|}{IPW$_1$}   & \multicolumn{4}{c}{IPW$_2$} \\
    \rowcolor[rgb]{ .949,  .949,  .949}       & \multicolumn{1}{c|}{True} & \multicolumn{1}{c}{Bias} & \multicolumn{1}{c}{ESE} & \multicolumn{1}{c}{ASE} & \multicolumn{1}{c|}{ECP} & \multicolumn{1}{c}{Bias} & \multicolumn{1}{c}{ESE} & \multicolumn{1}{c}{ASE} & \multicolumn{1}{c}{ECP} \\
    \midrule
    \midrule
    $\widehat{Y}(1, 0.25)$ & 0.9965 & 0.0039 & 0.068 & 0.074 & 0.96  & 0.0345 & 0.090  & 0.074 & 0.84 \\
    $\widehat{Y}(1, 0.5)$ & 0.9885 & 0.0041 & 0.033 & 0.047 & 0.99  & 0.0051 & 0.006 & 0.048 & 1.00 \\
    $\widehat{Y}(1, 0.75)$ & 0.9724 & -0.0134 & 0.067 & 0.077 & 0.96  & -0.1540 & 0.115 & 0.148 & 0.93 \\
     $\widehat{Y}(0, 0.25)$ & 0.9943 & -0.0179 & 0.066 & 0.078 & 0.97  & -0.1682 & 0.121 & 0.154 & 0.92 \\
     $\widehat{Y}(0, 0.5)$ & 0.9821 & 0.0015 & 0.034 & 0.048 & 0.99  & 0.0032 & 0.008 & 0.049 & 1.00 \\
     $\widehat{Y}(0, 0.75)$ & 0.9583 & 0.0060 & 0.069 & 0.074 & 0.95  & 0.0398 & 0.085 & 0.070  & 0.81 \\
    $\widehat{Y}(0.25)$ & 0.9949 & -0.0125 & 0.050  & 0.063 & 0.98  & -0.1175 & 0.100   & 0.122 & 0.92 \\
    $\widehat{Y}(0.5)$ & 0.9853 & 0.0028 & 0.022 & 0.040  & 1.00     & 0.0042 & 0.005 & 0.035 & 1.00 \\
    $\widehat{Y}(0.75)$ & 0.9689 & -0.0086 & 0.050  & 0.062 & 0.97  & -0.1056 & 0.094 & 0.117 & 0.94 \\
    \hline
    \end{tabular}%
  \label{tab:wrongpotentialmodel}%
\end{table}%

% Table generated by Excel2LaTeX from sheet 'Sheet1'
\begin{table}[htbp]
  \centering
  \caption{Results from 1000 simulated datasets on TRIP network for 10 components (left) and using community detection to further divide the network to 20 components (right) for exposed ($a=1$), not exposed ($a=0$), and marginal estimators under allocation strategies 25\%, 50\%, and 75\%.}
    \begin{tabular}{lc|rcrc|rcrc}
    \toprule
    \rowcolor[rgb]{ .929,  .929,  .929}      &      & \multicolumn{4}{c|}{10 components} & \multicolumn{4}{c}{20 components} \\
    \rowcolor[rgb]{ .929,  .929,  .929}      &      & \multicolumn{2}{c}{IPW$_1$} & \multicolumn{2}{c|}{IPW$_2$} & \multicolumn{2}{c}{IPW$_1$} & \multicolumn{2}{c}{IPW$_2$} \\
    \rowcolor[rgb]{ .929,  .929,  .929}      & True & \multicolumn{1}{c}{Bias} & ECP  & \multicolumn{1}{c}{Bias} & ECP  & \multicolumn{1}{c}{Bias} & ECP  & \multicolumn{1}{c}{Bias} & ECP \\
    \midrule
    \midrule
    $\widehat{Y}(1, 0.25)$ & 0.2473 & 0.0098 & 0.986 & 0.0167 & 0.988 & 0.0098 & 0.849 & 0.0036 & 0.890 \\
    $\widehat{Y}(1, 0.5)$ & 0.2265 & 0.0021 & 0.998 & 0.0112 & 0.986 & 0.0021 & 0.946 & 0.0064 & 0.920 \\
    $\widehat{Y}(1, 0.75)$ & 0.2058 & $-$0.0020 & 0.987 & 0.0126 & 0.997 & $-$0.0020 & 0.894 & 0.0057 & 0.943 \\
    $\widehat{Y}(0, 0.25)$ & 0.2304 & $<$0.0001 & 0.996 & 0.0046 & 0.999 & $<$0.0001 & 0.920 & 0.0021 & 0.968 \\
    $\widehat{Y}(0, 0.5)$ & 0.2778 & 0.0010 & 1.000    & 0.0029 & 1.000    & 0.0010 & 0.954 & 0.0017 & 0.974 \\
    $\widehat{Y}(0, 0.75)$ & 0.3275 & 0.0073 & 0.996 & 0.0038 & 1.000    & 0.0073 & 0.896 & 0.0019 & 0.992 \\
    $\widehat{Y}(0.25)$ & 0.2346 & 0.0025 & 0.999 & 0.0133 & 0.971 & 0.0025 & 0.943 & 0.0061 & 0.915 \\
    $\widehat{Y}(0.5)$ & 0.2521 & 0.0015 & 1.000    & 0.0121 & 0.993 & 0.0015 & 0.982 & $<$0.0001 & 0.917 \\
    $\widehat{Y}(0.75)$ & 0.2362 & 0.0004 & 1.000    & 0.0130 & 0.998 & 0.0004 & 0.937 & 0.0046 & 0.940 \\
    \bottomrule
    \end{tabular}%
  \label{tab:trip_sim}%
\end{table}%

% Table generated by Excel2LaTeX from sheet 'Sheet1'
\begin{table}[htbp]
  \centering
  \caption{Results from 1000 simulated datasets on TRIP network further dividing the network into 20 components using $IPW_1$ (left) and $IPW_2$ (right) for exposed $(a=1)$, not exposed $(a=0)$, and marginal estimators under allocation strategies 25\%, 50\%, and 75\% with exposure generating model $A_i=\mbox{Bern}(\mbox{logit}^{-1}(-1.4\cdot Z_{1,i}+2\cdot Z_{2,i}-1.5\cdot Z_{3,i}+1.2\cdot Z_{4,i}))$.}
    \begin{tabular}{lr|rrrr|rrrr}
    \toprule
    \rowcolor[rgb]{ .949,  .949,  .949}      &      & \multicolumn{4}{c|}{IPW1} & \multicolumn{4}{c}{IPW2} \\
    \rowcolor[rgb]{ .949,  .949,  .949}      & \multicolumn{1}{c|}{True} & \multicolumn{1}{c}{Bias} & \multicolumn{1}{c}{ESE} & \multicolumn{1}{c}{ASE} & \multicolumn{1}{c|}{ECP} & \multicolumn{1}{c}{Bias} & \multicolumn{1}{c}{ESE} & \multicolumn{1}{c}{ASE} & \multicolumn{1}{c}{ECP} \\
    \midrule
    \midrule
    $\widehat{Y}(1, 0.25)$ & 0.2493 & 0.0445 & 0.161 & 0.095 & 0.61 & 0.0409 & 0.100 & 0.075 & 0.71 \\
    $\widehat{Y}(1, 0.5)$ & 0.2274 & 0.0295 & 0.179 & 0.094 & 0.64 & 0.0197 & 0.089 & 0.073 & 0.80 \\
    $\widehat{Y}(1, 0.75)$ & 0.2057 & 0.0194 & 0.290 & 0.111 & 0.57 & 0.0422 & 0.093 & 0.065 & 0.64 \\
    $\widehat{Y}(0, 0.25)$ & 0.2295 & 0.0160 & 0.141 & 0.087 & 0.72 & 0.0214 & 0.064 & 0.060 & 0.86 \\
    $\widehat{Y}(0, 0.5)$ & 0.2765 & 0.0306 & 0.149 & 0.092 & 0.69 & 0.0318 & 0.063 & 0.067 & 0.86 \\
    $\widehat{Y}(0, 0.75)$ & 0.3264 & 0.0589 & 0.189 & 0.113 & 0.59 & 0.0923 & 0.094 & 0.076 & 0.56 \\
    $\widehat{Y}(0.25)$ & 0.2345 & 0.0231 & 0.115 & 0.077 & 0.71 & 0.0263 & 0.054 & 0.055 & 0.84 \\
    $\widehat{Y}(0.5)$ & 0.2520 & 0.0300 & 0.118 & 0.079 & 0.70 & 0.0258 & 0.053 & 0.059 & 0.87 \\
    $\widehat{Y}(0.75)$ & 0.2358 & 0.0293 & 0.227 & 0.099 & 0.60 & 0.0547 & 0.072 & 0.058 & 0.62 \\
    \bottomrule
    \end{tabular}%
  \label{tab:TRIP_sim_4cov}%
\end{table}%

% Table generated by Excel2LaTeX from sheet 'Sheet1'
\begin{table}[htbp]
  \centering
  \caption{Results from 1000 simulated datasets with the network regenerated for each dataset with 100 components for IPW$_1$ (left) and IPW$_2$ (right) for exposed ($a=1$), not exposed ($a=0$), and marginal estimators under allocation strategies 25\%, 50\%, and 75\%.}
    \begin{tabular}{lc|rcrc|rcrc}
    \toprule
    \rowcolor[rgb]{ .929,  .929,  .929}     &      & \multicolumn{4}{c|}{IPW1} & \multicolumn{4}{c}{IPW2} \\
    \rowcolor[rgb]{ .929,  .929,  .929}     & True & \multicolumn{1}{c}{Bias} & \multicolumn{1}{c}{ESE} & \multicolumn{1}{c}{ASE} & \multicolumn{1}{c|}{ECP} & \multicolumn{1}{c}{Bias} & \multicolumn{1}{c}{ESE} & \multicolumn{1}{c}{ASE} & \multicolumn{1}{c}{ECP} \\
    \midrule
    \midrule
    $\widehat{Y}(1, 0.25)$ & 0.2489 & 0.0023 & 0.0487 & 0.0467 & 0.89 & -0.0099 & 0.0518 & 0.0457 & 0.93 \\
    $\widehat{Y}(1, 0.5)$ & 0.2270 & 0.0018 & 0.0271 & 0.0274 & 0.94 & -0.0037 & 0.0324 & 0.0317 & 0.96 \\
    $\widehat{Y}(1, 0.75)$ & 0.2058 & -0.0008 & 0.0425 & 0.0500 & 0.90  & -0.0052 & 0.0275 & 0.0274 & 0.96 \\
    $\widehat{Y}(0, 0.25)$ & 0.2281 & $<$0.0001 & 0.0366 & 0.0491 & 0.93 & -0.0013 & 0.0223 & 0.0237 & 0.97 \\
    $\widehat{Y}(0, 0.5)$ & 0.2745 & 0.0015 & 0.0257 & 0.0298 & 0.97 & -0.0023 & 0.0230 & 0.0246 & 0.96 \\
    $\widehat{Y}(0, 0.75)$ & 0.3249 & 0.0050 & 0.0479 & 0.0541 & 0.92 & -0.0018 & 0.0161 & 0.0177 & 0.98 \\
    $\widehat{Y}(0.25)$ & 0.2333 & 0.0006 & 0.0296 & 0.0388 & 0.94 & -0.0051 & 0.0340 & 0.0329 & 0.94 \\
    $\widehat{Y}(0.5)$ & 0.2508 & 0.0017 & 0.0187 & 0.0213 & 0.97 & -0.0168 & 0.0542 & 0.0510 & 0.94 \\
    $\widehat{Y}(0.75)$ & 0.2356 & 0.0007 & 0.0336 & 0.0387 & 0.91 & -0.0081 & 0.0295 & 0.0289 & 0.95 \\
    \bottomrule
    \end{tabular}%
  \label{tab:multinetwork}%
\end{table}%

\section{Evaluation of disseminated effects of community alerts in the Transmission Reduction Intervention Project}

 We applied the estimators proposed in Section 5.2 to estimate the causal effects of community alerts at baseline on report of risk behavior at the six-month visit. We assumed that TRIP was an undirected network because the links were defined by if two individuals engaged had sex or injected drugs together in the six months before the baseline interview as reported by at least one participant in the dyad.  This was an attempt to reduce the impact of possible missing links in the analysis due to stigma of sexual and drug use behaviors. The community alerts intervention status of the index participant and their neighbors was defined with respect to the baseline visit date of the index participant. The network structure in TRIP had 10 connected components with 216 participants and 362 shared connections (average degree is 3.35) after excluding isolates and 59 participants who were lost to follow-up before their six-month visit. Among the 216 participants in TRIP, 25 participants (11.6\%) received a community alert about the increased risk for HIV acquisition in close proximity in their network from the study team. We evaluated if information in the community alerts was disseminated to their the nearest neighbors and ultimately, if this resulted in a reduction in risk behavior among others in the network beyond those who were exposed to the alert themselves (Figure \ref{fig:trip_network}). Among participants with complete information on the questions related to sharing injection equipment, we considered the report of sharing injection equipment (or not) at the 6-month visit as the binary outcome, including sharing a syringe, cooker, filter or rinse water, or backloading to share injection drugs. The following baseline covariates were included in the adjusted models: HIV status, shared drug equipment (e.g., syringe) in last six months, the calendar date at first interview, education (primary school, high school, and post high school), and employment status (employed, unemployed/looking for a work, can't work because of health reason, and other). HIV status was ascertained in this study from a blood sample from each participant collected by a health program physician \citep{nikolopoulos2016network}. Given the study population included PWID in one geographic location, we assume that social desirability leading to possible reporting bias is comparable between the two exposure groups. Under this assumption, the reporting bias could be effectively eliminated when estimating contrasts between exposure groups. 
 
 We consider each of the assumptions in Section 5.1 in this setting. TRIP is an observational study with a nonrandomized intervention. Conditional exchangeability is required to identify causal effects. We also assume that if there are multiple versions of the community alerts that these different versions are irrelevant for causal contrasts of interest and this results in one version of exposure to the community alerts intervention and one version of no exposure to this intervention. TRIP recruited at least one wave of contact tracing for each participant of an HIV-infected seed; therefore, we expect that there could be dissemination or spillover between two individuals connected by a first-degree link (i.e., possible influence of their neighbors' intervention exposure on an individual's outcome). Based on the complex structure of the TRIP network resulting in possibly many different vectors of $A_{\N_i}$, a stratified interference may be more appropriate to ensure that the positivity assumption holds. For the reducible propensity score assumption, the neighbors's covariates were not significantly associated with their index individual's exposure and the neighbors's exposure was not significant associated with the index individual's covariates, so Assumption \ref{assump:reducible} used for IPW$_2$ may be plausible in this analysis (data not shown).

For the analysis, we reported the point estimates and corresponding Wald-type 95\% confidence intervals of each causal effect using both IPW$_1$ and IPW$_2$ estimators under allocation strategies 20\%, 30\%, 40\%, and 50\% because the most of individuals in the TRIP study had 20\% to 50\% of their nearest neighbors exposed to community alerts. The normality of random effects in IPW$_1$ was tested using a diagnostic test for mixed effects model in \cite{normality} and this resulted in a $p$-value $=0.012$ under the null hypothesis that the mixing distribution is normal. Due to this result and better finite-sample performance for IPW$_2$ with a smaller number of components (see Section 6), we recommend IPW$_2$ as a more appropriate estimator in this setting given the small number of components in the TRIP network. Based on the simulation scenario 4 results that showed better finite sample performance for 20 components, we used community detection to further divide the TRIP network into 20 components to possibly improve the finite-sample performance of the variance estimators. We report the variance estimates with and without dividing TRIP network 10 observed components to 20 components in Table \ref{tab:trip_full}. In addition to including the full set of covariates to adjust for measured confounding in the weight models, we conducted sensitivity analyses to evaluate the impact of different sets of covariates on the model results. We first considered univariate models; that is, adjustment for only one covariate at a time. Second, we estimated the effects using the full set of covariates, excluding one covariate at a time. Lastly, we estimated the effects without adjustment for any covariates. The results were largely robust to the set of measured covariates used to adjust for confounding. In addition, the results that used community detection to further divide TRIP into 20 components to estimate the asymptotic variances were comparable to an analysis that used the 10 observed components. All models results are summarized in Appendix D. The study protocol was reviewed and approved by the University of Rhode Island Institutional Review Board. All analyses were conducted using R (version 3.6.2), and R packages: igraph: Network Analysis and Visualization (version 1.3.4), lme4: Linear Mixed-Effects Models using 'Eigen' and S4 (version 1.1-30), and numDeriv: Accurate Numerical Derivatives (version 2016.8-1.1). 

%All models demonstrated an estimated protective disseminated effect, except for the scenario with the full set of covariates, excluding the calendar date at first interview, $\widehat{DE}_2(0.25)=0.0086$ when estimated using IPW$_2$; however, this did not achieve statistical significance. 

Direct, indirect, total, and overall effect estimates and their corresponding Wald-type 95\% confidence intervals of both estimators for different allocation strategies $\alpha=0.2, 0.3, 0.4$ and $0.5$ adjusting for all measured confounding variables are reported in Table \ref{tab:trip_full}. All estimates of the risk differences for both estimators IPW$_1$ and IPW$_2$ were protective, excluding the indirect effect under allocation strategy 30\% and 20\%, $\widehat{IE}_1(0.3, 0.2)=0.01$ using IPW$_1$ and $\widehat{IE}_2(0.3, 0.2)=0.00$ using IPW$_2$; however, these did not achieve statistical significance. These results suggest that the likelihood of reporting HIV risk behavior was reduced not only by an participant's exposure to community alerts, but also by the proportion of an participant's nearest neighbors exposed to community alerts from the study team. We report the confidence interval obtained using 10 components. Specifically, the estimated direct effect was $\widehat{DE}_1(0.5)=-0.18$ (95\% CI: $-0.49, 0.14$), estimated using IPW$_1$ and $\widehat{DE}_2(0.5)=-0.21$ (95\% CI: $-0.56, 0.15$) when estimated with IPW$_2$; that is, we expect 18  fewer reports of risk behavior per 100 participants if a participant receives the alert compared to if a participant
does not receive an alert with 50\% intervention coverage (i.e., 50\% of their neighbors receiving alerts) when estimated using IPW$_1$ (21 per 100 fewer using IPW$_2$).  The indirect effect is $\widehat{IE}_1(0.5, 0.2)=-0.03$ (95\% CI:$-0.07, 0.00$), estimated using IPW$_1$ under allocation strategies 20\% versus 50\% and $\widehat{IE}_2(0.5, 0.2)=-0.02$ (95\% CI:$-0.04, -0.01$) when estimated with IPW$_2$; in other words, we expect 4 fewer reports of risk behavior per 100 participants if a participant does not receive an alert with
50\% intervention coverage compared to only 20\% intervention coverage when estimated using IPW$_1$ (2 per 100 fewer using IPW$_2$). The total effects $\widehat{TE}_1(0.5, 0.2)=-0.21$ (95\% CI:$-0.53, 0.11$) estimated using IPW$_1$ and $\widehat{TE}_2(0.5, 0.2)=-0.23$ (95\% CI:$-0.58, 0.12$) estimated using IPW$_2$. We expect 21 fewer reports of risk behavior per 100 participants when estimated using IPW1 if a participant receives an alert with 50\% of their nearest neighbors alerted versus if a participant does not receive an alert and only 20\% of their nearest neighbors receive an alert (23 per 100 fewer using IPW$_2$). The overall effects, $\widehat{OE}_1(0.5, 0.2)=-0.11$ (95\% CI:$-0.27, 0.05$) using IPW$_1$ and $\widehat{OE}_2(0.5, 0.2)=-0.13$ (95\% CI:$-0.32, 0.06$) using IPW$_2$. When estimated using IPW$_1$, we expect 11 fewer reports of risk behavior per 100 participants if 50\% of the nearest neighbors and participant $i$ receive alerts compared to if only 20\% of the nearest neighbors and participant $i$ receive alerts (13 per 100 fewer using IPW$_2$). 

%Additionally, the effects under allocation strategies 75\% are likely better quantified using IPW$_1$. Besides the discussed allocation strategies above, we also reported the the effects with $\alpha=20\%, 30\%, 40\%, 50\%$ that are close to current exposure coverage in Appendix D, Table \ref{tab:20_50}.

% Table generated by Excel2LaTeX from sheet 'all'
\begin{table}[htbp]
  \centering
  \caption{The estimated risk differences and 95\% confidence intervals (CIs) estimated using the TRIP network with the original 10 network components, and 95\% CIs estimated by dividing TRIP network into 20 components, of the effects of community alerts at baseline on HIV risk behavior at 6 months adjusted for full set of measured confounding variables under allocation strategies 20\%, 30\%, 40\%, and 50\%}
    \begin{tabular}{lc|ccc|ccc}
    \toprule
         &      & \multicolumn{3}{c|}{IPW1} & \multicolumn{3}{c}{IPW2} \\
    \multicolumn{1}{c}{Effects} & Coverage & RD   & \multicolumn{2}{c|}{95\% CI} & RD   & \multicolumn{2}{c}{95\% CI} \\
         & $(\alpha, \alpha')$ &      & 10 components & 20 components &      & 10 components & 20 components \\
    \midrule
    \midrule
    \rowcolor[rgb]{ .949,  .949,  .949} Direct & (20\%, 20\%) & -0.06 & (-0.14, 0.01) & (-0.39, 0.26) & 0.01 & (-0.08, 0.10) & (-0.19, 0.21) \\
    \rowcolor[rgb]{ .949,  .949,  .949} Direct & (30\%, 30\%) & -0.10 & (-0.24, 0.04) & (-0.48, 0.29) & -0.09 & (-0.19, 0.02) & (-0.27, 0.10) \\
    \rowcolor[rgb]{ .949,  .949,  .949} Direct & (40\%, 40\%) & -0.14 & (-0.36, 0.09) & (-0.52, 0.24) & -0.16 & (-0.40, 0.09) & (-0.38, 0.07) \\
    \rowcolor[rgb]{ .949,  .949,  .949} Direct & (50\%, 50\%) & -0.18 & (-0.49, 0.14) & (-0.52, 0.17) & -0.21 & (-0.56, 0.15) & (-0.47, 0.06) \\
    Indirect & (30\%, 20\%) & 0.01 & (-0.02, 0.03) & (-0.03, 0.04) & 0.00 & (-0.01, 0.02) & (-0.04, 0.05) \\
    Indirect & (40\%, 20\%) & -0.01 & (-0.03, 0.01) & (-0.06, 0.04) & -0.01 & (-0.02, 0.01) & (-0.09, 0.07) \\
    Indirect & (50\%, 20\%) & -0.03 & (-0.07, 0.00) & (-0.10, 0.03) & -0.02 & (-0.04, -0.01) & (-0.14, 0.10) \\
    Indirect & (40\%, 30\%) & -0.01 & (-0.03, -0.00) & (-0.04, 0.01) & -0.01 & (-0.02, -0.00) & (-0.05, 0.03) \\
    Indirect & (50\%, 40\%) & -0.03 & (-0.05, 0.00) & (-0.06, 0.01) & -0.01 & (-0.03, -0.00) & (-0.06, 0.03) \\
    Indirect & (50\%, 30\%) & -0.04 & (-0.08, 0.00) & (-0.09, 0.01) & -0.02 & (-0.04, -0.01) & (-0.11, 0.06) \\
    \rowcolor[rgb]{ .949,  .949,  .949} Total & (30\%, 20\%) & -0.09 & (-0.22, 0.04) & (-0.49, 0.31) & -0.08 & (-0.17, 0.01) & (-0.28, 0.12) \\
    \rowcolor[rgb]{ .949,  .949,  .949} Total & (40\%, 20\%) & -0.15 & (-0.36, 0.07) & (-0.53, 0.24) & -0.16 & (-0.40, 0.07) & (-0.37, 0.05) \\
    \rowcolor[rgb]{ .949,  .949,  .949} Total & (50\%, 20\%) & -0.21 & (-0.53, 0.11) & (-0.54, 0.11) & -0.23 & (-0.58, 0.12) & (-0.43, -0.02) \\
    \rowcolor[rgb]{ .949,  .949,  .949} Total & (40\%, 30\%) & -0.15 & (-0.38, 0.08) & (-0.52, 0.22) & -0.17 & (-0.41, 0.08) & (-0.37, 0.04) \\
    \rowcolor[rgb]{ .949,  .949,  .949} Total & (50\%, 40\%) & -0.20 & (-0.53, 0.13) & (-0.52, 0.12) & -0.22 & (-0.58, 0.14) & (-0.45, 0.01) \\
    \rowcolor[rgb]{ .949,  .949,  .949} Total & (50\%, 30\%) & -0.22 & (-0.56, 0.12) & (-0.53, 0.09) & -0.23 & (-0.59, 0.13) & (-0.44, -0.02) \\
    Overall & (30\%, 20\%) & -0.01 & (-0.03, 0.01) & (-0.08, 0.06) & -0.03 & (-0.06, 0.01) & (-0.07, 0.02) \\
    Overall & (40\%, 20\%) & -0.05 & (-0.12, 0.02) & (-0.15, 0.05) & -0.07 & (-0.18, 0.03) & (-0.15, 0.00) \\
    Overall & (50\%, 20\%) & -0.11 & (-0.27, 0.05) & (-0.21, -0.01) & -0.13 & (-0.32, 0.06) & (-0.21, -0.04) \\
    Overall & (40\%, 30\%) & -0.04 & (-0.10, 0.02) & (-0.07, -0.01) & -0.05 & (-0.12, 0.02) & (-0.08, -0.01) \\
    Overall & (50\%, 40\%) & -0.06 & (-0.15, 0.03) & (-0.09, -0.03) & -0.05 & (-0.14, 0.03) & (-0.08, -0.03) \\
    Overall & (50\%, 30\%) & -0.10 & (-0.25, 0.05) & (-0.14, -0.06) & -0.10 & (-0.26, 0.05) & (-0.16, -0.05) \\
    \bottomrule
    \end{tabular}%
  \label{tab:trip_full}%
\end{table}%

\begin{comment}
\begin{figure}
\centering
\caption{The point estimates of risk differences and the Wald 95\% CI of the effect of community alert and HIV risk behavior on TRIP using IPW$_1$ and IPW$_2$.}
\includegraphics[scale=0.53]{TRIP_figure_0.2_0.5.eps}
\label{fig:TRIP}
\end{figure}
\end{comment}

\section{Discussion}
In this paper, methods for evaluating disseminated effects were developed for the setting of network-based studies by leveraging a nearest neighbor interference set. The proposed approach uses connections (i.e. links) between individuals in a network and allows for overlapping interference sets within each component of the network. The two proposed estimators were shown to be consistent and asymptotically normal. Importantly, a consistent, closed-form estimator of the asymptotic variance was derived. The simulation study demonstrated that the two IPW estimators had reasonable finite-sample performance in terms of consistency and empirical coverage for a large number ($>100$) of components in the observed network. The proposed variance estimators incorporate the observed network structure by assigning each individual a unique propensity score defined by their own nearest neighbors in which the nearest neighbors for individuals can overlap. We compared the performance of our variance estimators to the estimators for the asymptotic variances that assume partial interference with component-level propensity scores \citep{liu2016inverse} by using the observed network components as partial interference sets. In Figure \ref{fig:liuasymp}, our variance estimators were more efficient and closer to the empirical standard error by utilizing the network structure in a nearest neighbors level propensity score as compared to Liu's estimator. In the additional simulation scenario 4, the empirical coverage probabilities were above the nominal level (97\%-100\%) when using the TRIP network with only 10 components. This may be a result of the uncertainty due to the imbalanced component size observed in TRIP, where the total number of nodes was 216 and the largest component had size 186. After using community detection to further divide the network into a larger number of components ($m=20$), the coverage level then decreased to average of 93\%. 

Based on the simulation results, both estimators performed well in terms of finite sample bias. IPW$_2$ demonstrated better performance for variance estimation (i.e., ASE was closer to ESE) when the number of network components was small ($<50$), while IPW$_1$ had lower coverage for the confidence intervals. When the number of network components was large ($\geq100$), the estimated average standard error for IPW$_1$ resulted in confidence intervals with coverage around the nominal level, while IPW$_2$ tended to have coverage above the nominal level. Based on these findings, the estimation of these effects in TRIP network using IPW$_2$ may be preferred over IPW$_1$ due to the small number of network components with the caveat that these recommendations may be sensitive to specification in the simulation scenarios, including features of the study design. In addition, we explored adding additional covariates with larger parameter values in the exposure generating model (Table 5). In this case, the estimators had slightly larger bias compared to Table 4 and ECP somewhat below the nominal, while IPW$_2$ had slightly higher coverage than IPW$_1$. As reported in Table 3, a violation of the stratified interference assumption when the exposure mechanism is misspecified resulted in deviations from the nominal coverage level for both estimators.

With these methods, we now have an approach to quantify the social and biological influence on the determinants of risk and HIV transmission in HIV risk networks of PWIDs \citep{friedmanSocialNet2001} when evaluating the impact of interventions, such as TasP, or how interventions permeate a risk network \citep{nikolopoulos2016network,friedman2014socially}. These new methodologies will improve the identification of best preventive practices for PWID and provide evidence to expedite policy changes to improve access to HIV treatment and risk reduction interventions in subpopulations of high-risk drug users. In the TRIP study, these methods allowed for quantification of the extent to which the community alerts intervention reduced onward transmission to others in the community by tracking incident infections in the risk networks as measured through the proxy of self-reported HIV risk behaviors. Correctly conducted and analyzed studies among PWID will improve existing interventions, inform new interventions, and has the potential to reduce incident HIV infections in this subpopulation.

Studies of network effects among PWID are rich with future methodological problems. The simulation study indicated that the asymptotic variance estimators of the IPW estimators had coverage below the nominal level when the number of components in the network is limited ($<$50), while both IPW estimators were unbiased in finite samples ($<$5\%; bias/true value). Finite sample correction for estimating asymptotic variances is needed when the network has small number of components. As the approach in this paper used components as independent units for the variance estimation, developing methodologies with heterogeneous correlation structures within a large size component should be included in future work. Furthermore, the outcome of interest may be missing due to participant loss to follow-up in some intervention-based studies when outcomes are ascertained post-intervention. For example, 21\% of TRIP participants were lost to follow-up by the six-month visit. Future work should include development of censoring methods to evaluate the IPW outcomes in the presence of missing outcomes or alternative methods to also address missing links in the network. With regard to real data application, the impact of unmeasured confounding is important because this would violate the conditional exchangeability assumption; however, sensitivity analyses in the presence of interference currently only exist for two-stage randomized trials with clustering features \citep{tylersensitivity2014}. Designing sensitivity analyses to assess the bias of unmeasured confounding in network-based studies should be included in future research. In addition, if the spillover set actually included two-degree neighbors or other sets of individuals in the network, the nearest neighbors interference assumption would not be valid. We recommend the development of future methods that consider alternative definitions of the spillover set in the network. For example, we could also have a violation of the stratified interference assumption if in fact one of the neighbors was a closer contact or more important to the index participant. We recommend for future work the incorporation of edge weights into this method to reflect variations in the strength of connections relevant for spillover.  With these improved inferential methods, investigators will be able to answer questions they were previously unable to address in network-based studies, leading to more effective intervention implementation and far-reaching policy change to prevent HIV infection, reduce risk behavior, ultimately, improve HIV treatment and care among PWID. In addition to study HIV transmission among PWID, this method can also be applied in a wider context to study sexually transmitted infection diseases such as genital herpes and trichomoniasis among adolescents and young adults, men who have sex with men, or pregnant women.

%%%%%% include this section if you wish to acknowledge people,
%%%%%% grant support, etc.

\section*{Acknowledgements}
These findings are presented on behalf of the Transmission Reduction Intervention Project (TRIP). We would like to thank all of the TRIP investigators, data management teams, and participants who contributed to this project. We thank Ke Zhang for her editorial comments. The project described was supported by the Avenir Award Program for Research on Substance Abuse and HIV/AIDS (DP2) from National Institute on Drug Abuse of the National Institutes of Health Award Number DP2DA046856. Dr. Samuel Friedman was partially supported by the National Institute on Drug Abuse of the National Institutes of Health Award Number DP1DA034989, which funded Preventing HIV Transmission by Recently-Infected Drug Users (TRIP), and the National Institute on Drug Abuse of the National Institutes of Health Award Number P30DA011041, which supported the Center for Drug Use and HIV Research at New York University. Dr. M. Elizabeth Halloran was partially supported by the National Institute of Allergy and Infectious Diseases of the National Institutes of Health Award Number R01AI085073 titled Causal Inference in Infectious Disease Prevention Studies. The content is solely the responsibility of the authors and does not necessarily represent the official views of the National Institutes of Health.

\begin{supplement}
\noindent \stitle{Dataset and codes}
\sdescription{The TRIP datasets are available upon reasonable request to the corresponding author subject to approval by the TRIP investigators. The simulation code and datasets are available from the corresponding author on reasonable request. Codes and a sample dataset can be found on github: https://github.com/uri-ncipher/Nearest-Neighbor-estimators.}
\end{supplement}

\bibliographystyle{imsart-nameyear} \bibliography{nearestneighbors}
%\begin{comment}
\newpage
\setcounter{page}{1}
\clearpage
\appendix

\section{Proof of Proposition 1}
\noindent To show $\widehat{Y}^{IPW_1}(a, \alpha)$ is unbiased with known propensity score, see the following:
\begin{align}
    E[\widehat{Y}^{IPW_1}(a, \alpha)]&=\frac{1}{n}\sum_{i=1}^n E\big[\frac{y_i(A_i, A_{\N_i})I(A_i=a)\pi(A_{\N_i};\alpha)}{f_1(A_i, A_{\N_i}|Z_i, Z_{\N_i})}\big]\notag\\
    &=\frac{1}{n}\sum_{i=1}^n \sum_{a_i, a_{\N_i}} \frac{y_i(a_i, a_{\N_i})I(a_i=a)\pi(a_{\N_i};\alpha)}{f_1(a_i, a_{\N_i}|Z_i, Z_{\N_i})}f_1(a_i, a_{\N_i}|Z_i, Z_{\N_i})\notag\\
    &=\frac{1}{n}\sum_{i=1}^n \sum_{ a_{\N_i}}y_i(a_i=a, a_{\N_i})\pi(a_{\N_i};\alpha)\notag\\
    &=\bar{y}(a, \alpha)\notag
\end{align}
The unbiasedness of the marginal inverse probability weighted estimator, $\widehat{Y}^{IPW_1}(\alpha)$, can be proved similarly. 

Under the assumption that $y(a_i, a_{\N_i})=y(a_i, s_i)$, IPW$_2$ is also unbiased with a known propensity score
\begin{align}
    E[\widehat{Y}^{IPW_2}(a,\alpha)]&=\frac{1}{n}\sum_{i=1}^n E\big[\frac{y_i(A_i, S_i)I(A_i=a)\pi(A_{\N_i};\alpha)}{f_2(A_i, S_i|Z_i, Z_{\N_i})}\big]\notag\\
    &=\frac{1}{n}\sum_{i=1}^n \sum_{a_i, s_i} \frac{y_i(a_i, s_i)I(a_i=a)\pi(a_{\N_i};\alpha)}{f_2(a_i, s_i|Z_i, Z_{\N_i})}f_2(a_i, s_i|Z_i, Z_{\N_i})\notag\\
    &=\frac{1}{n}\sum_{i=1}^n \sum_{j=0}^{d_i}{d_i \choose j}y_i(a_i=a, j)\alpha^j(1-\alpha)^{d_i-j}\notag\\
    &=\frac{1}{n}\sum_{i=1}^n \sum_{a_{\N_i}}y_i(a_i=a, a_{\N_i})\pi(a_{\N_i};\alpha)\notag\\
    &=\bar{y}(a, \alpha)\notag
\end{align}

\newpage
\section{Proposition 2 and Sandwich-Type Estimators of the Variance}
\noindent Following \citet{mestimator2013}, to estimate the parameters in the exposure propensity score model $\hat{\Theta}$, we
let $$\psi_{\eta}(Y_{\nu},A_{\nu},Z_{\nu}; \theta)=\frac{1}{k}\sum_{i\in V(C_\nu)}\frac{\partial \log f_r(A_{\nu i}, A_{\N_{\nu i}}|Z_{\nu i}, Z_{\N_{\nu i}})}{\partial \eta}, \eta\in\Theta.$$ Estimates $\hat{\eta}$ that maximize the $\log$ likelihood are solutions to the score equation $$\sum_{\nu=1}^m\psi_{\eta}(Y_{\nu},A_{\nu},Z_{\nu}; \theta)=\frac{1}{k}\sum_{\nu=1}^m \sum_{i\in V(C_\nu)}\frac{\partial \log f_r(A_{\nu i}, A_{\N_{\nu i}}|Z_{\nu i}, Z_{\N_{\nu i}})}{\partial \eta}=0.$$ 
The estimating equations for the remaining parameters $\hat{\theta}_{0\alpha}, \hat{\theta}_{1\alpha}, \hat{\theta}_{\alpha}$ are described in Section 5.3. Let 
\begin{align*}
\psi_{\nu}(Y_\nu, A_\nu, Z_\nu; \theta)=\begin{pmatrix}\psi_{\eta}(Y_\nu, A_\nu, Z_\nu; \theta) \\ \psi_0(Y_\nu, A_\nu, Z_\nu; \theta; \alpha) \\ \psi_1(Y_\nu, A_\nu, Z_\nu; \theta; \alpha) \\ \psi_2(Y_\nu, A_\nu, Z_\nu; \theta; \alpha)\end{pmatrix}_{\eta\in\Theta}.\end{align*} Therefore, $\displaystyle \sum_{i=1}^m\psi_{\nu}(Y_{\nu},A_{\nu},Z_{\nu}; \hat{\theta})=0$. Under suitable regularity conditions and due to the unbiased estimating equations, as $m\rightarrow \infty$, $\hat{\theta}$ converges in probability to $\theta$ and
$\sqrt{m}(\hat{\theta}-\theta)$ converges in distribution to a multivariate normal $N(0, \Sigma)$, where 
$$\Sigma=\frac{1}{m}A^{-1}(\theta)B(\theta)A(\theta)^{-T}$$ with $$A(\theta)=E[-\dot{\psi}_{\nu}(Y_{\nu},A_{\nu},Z_{\nu}; \theta)]=E[-\partial\psi_{\nu}(Y_{\nu},A_{\nu},Z_{\nu}; \theta)/\partial \theta^T]$$ and $$B(\theta)=E[\psi_{\nu}(Y_{\nu},A_{\nu},Z_{\nu}; \theta)\psi_{\nu}(Y_{\nu},A_{\nu},Z_{\nu}; \theta)^T].$$ The true parameter $\theta$ is defined as the solution to the equation
$$\int \psi(Y_{\nu},A_{\nu},Z_{\nu}; \theta)dF_{\nu}(Y_{\nu},A_{\nu},Z_{\nu})=0$$ where $F_{\nu}$ is the cumulative distribution function of $(Y_{\nu},A_{\nu},Z_{\nu})$. 

The empirical sandwich-type estimator can be used to estimate the asymptotic variance for the direct, disseminated, composite and overall estimators. Replacing $A(\theta)$ and $B(\theta)$ with empirical estimators in Proposition 2 yields a consistent sandwich estimator of the asymptotic variance $\Sigma$ $$\hat{\Sigma}_m=\frac{1}{m}A_m(\hat{\theta})^{-1}B_m(\hat{\theta})A_m(\hat{\theta}))^{-T}$$ where 

$$A_m(\hat{\theta})=\frac{1}{m}\sum_{\nu=1}^m-\dot\psi_{\nu}(Y_{\nu},A_{\nu},Z_{\nu};\hat{\theta}; \alpha)=-\frac{1}{m}\sum_{\nu=1}^m\begin{pmatrix} A_{11}(Y_{\nu},A_{\nu},Z_{\nu};\hat{\theta}) & 0\\
A_{2\cdot}(Y_{\nu},A_{\nu},Z_{\nu};\hat{\theta}; \alpha) & -I_{3\times 3}\end{pmatrix}$$
in which $$A_{11}(Y_{\nu},A_{\nu},Z_{\nu};\hat{\theta})=\biggl(\frac{\partial \psi_\eta(Y_{\nu},A_{\nu},Z_{\nu};\hat{\theta})}{\partial \eta'}\biggl)_{\eta, \eta'\in \Theta}$$ $$A_{21}(Y_{\nu},A_{\nu},Z_{\nu};\hat{\theta}; \alpha)=\biggl(\frac{\partial \psi_0(Y_{\nu},A_{\nu},Z_{\nu};\hat{\theta}; \alpha)}{\partial \eta'}\biggl)_{\eta'\in \Theta}$$ $$A_{31}(Y_{\nu},A_{\nu},Z_{\nu};\hat{\theta}; \alpha)=\biggl(\frac{\partial \psi_1(Y_{\nu},A_{\nu},Z_{\nu};\hat{\theta}; \alpha)}{\partial \eta'}\biggl)_{\eta'\in \Theta}$$ and $$A_{41}(Y_{\nu},A_{\nu},Z_{\nu};\hat{\theta}; \alpha)=\biggl(\frac{\partial \psi_2(Y_{\nu},A_{\nu},Z_{\nu};\hat{\theta}; \alpha)}{\partial \eta'}\biggl)_{\eta'\in \Theta}.$$ Let $A_{2\cdot}(Y_{\nu},A_{\nu},Z_{\nu};\hat{\theta}; \alpha)=\begin{pmatrix} A_{21}(Y_{\nu},A_{\nu},Z_{\nu};\hat{\theta}; \alpha) & A_{31}(Y_{\nu},A_{\nu},Z_{\nu};\hat{\theta}; \alpha)& A_{41}(Y_{\nu},A_{\nu},Z_{\nu};\hat{\theta}; \alpha)\end{pmatrix}^T$

\begin{align*}B_m(\hat{\theta})&=\frac{1}{m}\sum_{\nu=1}^m \psi_{\nu}(Y_{\nu},A_{\nu},Z_{\nu};\hat{\theta})\psi_{\nu}(Y_{\nu},A_{\nu},Z_{\nu};\hat{\theta})^T\end{align*}

\noindent That is, $\hat{\Sigma}_m$ is a consistent estimator of $\Sigma$. We provide a sandwich estimator of the variance for the disseminated effect. An analogous procedure can be used to obtain the sandwich variance of the variance for the estimators of the direct, overall, and total effects. Let 
\begin{align*}
\psi_{\nu}(Y_\nu, A_\nu, Z_\nu; \theta)=\begin{pmatrix}\psi_{\eta}(Y_\nu, A_\nu, Z_\nu; \theta) \\ \psi_0(Y_\nu, A_\nu, Z_\nu; \theta; \alpha_1) \\ \psi_0(Y_\nu, A_\nu, Z_\nu; \theta; \alpha_0) \\ \psi_1(Y_\nu, A_\nu, Z_\nu; \theta; \alpha_1) \\ \psi_1(Y_\nu, A_\nu, Z_\nu; \theta; \alpha_0) \\ \psi_2(Y_\nu, A_\nu, Z_\nu; \theta; \alpha_1)\\\psi_2(Y_\nu, A_\nu, Z_\nu; \theta; \alpha_0)\end{pmatrix}_{\eta\in\Theta}.\end{align*} Replacing $A(\theta)$ and $B(\theta)$ with empirical estimators in Proposition 2 yields a consistent sandwich estimator of the asymptotic variance of $\overline{IE}(\alpha_1,\alpha_0)$ is $$\hat{\Sigma}_{IE} = \frac{1}{m}\lambda^T A_m(\hat{\theta})^{-1}B_m(\hat{\theta})A_m(\hat{\theta})^{-T} \lambda$$ with $\lambda = (0_{1 \times p}, 1, -1, 0, 0, 0, 0)^T$. The estimated standard error (se) is $\widehat{\text{se}}(\widehat{IE}_r(\alpha_1,\alpha_0)) = \sqrt{\hat{\Sigma}_{IE}}$.

\begin{comment}
\begin{align}
B_m(\hat{\theta})&=\frac{1}{m}\sum_{i=1}^m \psi(O_i)\psi(O_i)^T\notag\\
&=\frac{1}{m}\sum_{i=1}^m\scriptsize\begin{pmatrix}
\psi_{\gamma_1}(O_i)\psi_{\gamma_1}(O_i) & \cdots & \psi_{\gamma_1}(O_i)\psi_{\gamma_p}(O_i) &\psi_{\gamma_1}(O_i)\psi_{p+1}(O_i) & \psi_{\gamma_1}(O_i)\psi_{p+2}(O_i) & \psi_{\gamma_1}(O_i)\psi_{p+3}(O_i)\\
\vdots & \ddots & \vdots &\vdots & \vdots &\vdots \\
\psi_{\gamma_p}(O_i)\psi_{\gamma_1}(O_i) & \cdots & \psi_{\gamma_p}(O_i)\psi_{\gamma_p}(O_i) &\psi_{\gamma_p}(O_i)\psi_{p+1}(O_i) & \psi_{\gamma_p}(O_i)\psi_{p+2}(O_i) & \psi_{\gamma_p}(O_i)\psi_{p+3}(O_i) \\
\psi_{p+1}(O_i)\psi_{\gamma_1}(O_i) & \cdots & \psi_{p+1}(O_i)\psi_{\gamma_p}(O_i) &\psi_{p+1}(O_i)\psi_{p+1}(O_i) & \psi_{p+1}(O_i)\psi_{p+2}(O_i) & \psi_{p+1}(O_i)\psi_{p+3}(O_i)\\
\psi_{p+2}(O_i)\psi_{\gamma_1}(O_i) & \cdots & \psi_{p+2}(O_i)\psi_{\gamma_p}(O_i) &\psi_{p+2}(O_i)\psi_{p+1}(O_i) & \psi_{p+2}(O_i)\psi_{p+2}(O_i) & \psi_{p+2}(O_i)\psi_{p+3}(O_i)\\
\psi_{p+3}(O_i)\psi_{\gamma_1}(O_i) & \cdots & \psi_{p+3}(O_i)\psi_{\gamma_p}(O_i) &\psi_{p+3}(O_i)\psi_{p+1}(O_i) & \psi_{p+3}(O_i)\psi_{p+2}(O_i) & \psi_{p+3}(O_i)\psi_{p+3}(O_i)
\end{pmatrix}\notag
\end{align}
\end{comment}

\newpage
\section{Simulation Results}
\label{s:tables}
In this section, we include the simulation results from 1000 simulation data set on networks with 10, 50, 100, 150, 200 components. The simulation results include the true values of the average potential outcomes, and bias on estimations of inverse probability weighted estimators (IPW$_1$ and IPW$_2$) for exposed ($a=1$) and not exposed ($a=0$), and marginal estimators under allocation strategies 25\%, 50\%, and 75\%, the asymptotic standard errors (ASE), and empirical coverage probabilities (ECP) (Table A1 to A5). Additionally, we used a different exposure generating model given by $A_i={\rm Bern}(p={\rm logit}^{-1}(-0.5-1.5\cdot Z_i+b_\nu))$ in data simulating on a network with 100 components (Table A6). 

\setcounter{table}{0}
\renewcommand{\thetable}{A\arabic{table}}

% Table generated by Excel2LaTeX from sheet 'Sheet1'
\begin{table}[htbp]
  \centering
  \caption{Simulation results of IPW$_1$ (left) and IPW$_2$ (right) on network with 10 components.}
    \begin{tabular}{lr|rrrr|rrrr}
    \toprule
    \rowcolor[rgb]{ .949,  .949,  .949}      &      & \multicolumn{4}{c|}{IPW$_1$} & \multicolumn{4}{c}{IPW$_2$} \\
    \rowcolor[rgb]{ .949,  .949,  .949}      & \multicolumn{1}{c|}{True} & \multicolumn{1}{c}{Bias} & \multicolumn{1}{c}{ESE} & \multicolumn{1}{c}{ASE} & \multicolumn{1}{c|}{ECP} & \multicolumn{1}{c}{Bias} & \multicolumn{1}{c}{ESE} & \multicolumn{1}{c}{ASE} & \multicolumn{1}{c}{ECP} \\
    \midrule
    \midrule
    $\widehat{Y}(1, 0.25)$ & 0.247 & 0.004 & 0.158 & 0.110 & 0.73 & -0.013 & 0.285 & 0.122 & 0.77 \\
    $\widehat{Y}(1, 0.5)$ & 0.226 & 0.005 & 0.098 & 0.076 & 0.83 & -0.001 & 0.091 & 0.077 & 0.90 \\
    $\widehat{Y}(1, 0.75)$ & 0.205 & <0.001 & 0.124 & 0.083 & 0.75 & 0.016 & 0.095 & 0.074 & 0.79 \\
    $\widehat{Y}(0, 0.25)$ & 0.227 & 0.004 & 0.111 & 0.082 & 0.80 & 0.013 & 0.115 & 0.079 & 0.80 \\
    $\widehat{Y}(0, 0.5)$ & 0.274 & 0.010 & 0.098 & 0.082 & 0.83 & 0.004 & 0.088 & 0.081 & 0.91 \\
    $\widehat{Y}(0, 0.75)$ & 0.324 & 0.016 & 0.171 & 0.128 & 0.76 & 0.011 & 0.189 & 0.127 & 0.78 \\
    $\widehat{Y}(0.25)$ & 0.232 & 0.004 & 0.093 & 0.073 & 0.82 & 0.006 & 0.114 & 0.073 & 0.83 \\
    $\widehat{Y}(0.5)$ & 0.250 & 0.007 & 0.070 & 0.061 & 0.85 & 0.001 & 0.064 & 0.060 & 0.92 \\
    $\widehat{Y}(0.75)$ & 0.235 & 0.004 & 0.103 & 0.076 & 0.79 & 0.015 & 0.087 & 0.071 & 0.81 \\
    \bottomrule
    \end{tabular}%
  \label{tab:addlabel}%
\end{table}%

% Table generated by Excel2LaTeX from sheet 'Sheet1'
\begin{table}[H]
  \centering
  \caption{Simulation results of IPW$_1$ (left) and IPW$_2$ (right) on network with 50 components}
    \begin{tabular}{lr|rrrr|rrrr}
    \toprule
    \rowcolor[rgb]{ .949,  .949,  .949}      &      & \multicolumn{4}{c|}{IPW$_1$} & \multicolumn{4}{c}{IPW$_2$} \\
    \rowcolor[rgb]{ .949,  .949,  .949}      & \multicolumn{1}{c|}{True} & \multicolumn{1}{c}{Bias} & \multicolumn{1}{c}{ESE} & \multicolumn{1}{c}{ASE} & \multicolumn{1}{c|}{ECP} & \multicolumn{1}{c}{Bias} & \multicolumn{1}{c}{ESE} & \multicolumn{1}{c}{ASE} & \multicolumn{1}{c}{ECP} \\
    \midrule
    \midrule
    $\widehat{Y}(1, 0.25)$ & 0.249 & 0.004 & 0.066 & 0.060 & 0.91 & -0.008 & 0.077 & 0.062 & 0.89 \\
    $\widehat{Y}(1, 0.5)$ & 0.227 & 0.002 & 0.040 & 0.038 & 0.94 & -0.001 & 0.033 & 0.034 & 0.96 \\
    $\widehat{Y}(1, 0.75)$ & 0.206 & -0.002 & 0.053 & 0.048 & 0.91 & -0.001 & 0.052 & 0.044 & 0.91 \\
    $\widehat{Y}(0, 0.25)$ & 0.227 & -0.002 & 0.049 & 0.048 & 0.94 & -0.002 & 0.048 & 0.043 & 0.92 \\
    $\widehat{Y}(0, 0.5)$ & 0.274 & 0.003 & 0.038 & 0.040 & 0.97 & -0.002 & 0.033 & 0.036 & 0.97 \\
    $\widehat{Y}(0, 0.75)$ & 0.325 & 0.002 & 0.071 & 0.067 & 0.93 & -0.014 & 0.081 & 0.071 & 0.93 \\
    $\widehat{Y}(0.25)$ & 0.233 & -0.001 & 0.040 & 0.039 & 0.94 & -0.003 & 0.040 & 0.038 & 0.93 \\
    $\widehat{Y}(0.5)$ & 0.250 & 0.002 & 0.028 & 0.029 & 0.97 & -0.001 & 0.023 & 0.026 & 0.98 \\
    $\widehat{Y}(0.75)$ & 0.235 & -0.001 & 0.042 & 0.041 & 0.94 & -0.006 & 0.044 & 0.040 & 0.93 \\
    \bottomrule
    \end{tabular}%
  \label{tab:addlabel}%
\end{table}%

% Table generated by Excel2LaTeX from sheet 'Sheet1'
\begin{table}[H]
  \centering
  \caption{Simulation results of IPW$_1$ (left) and IPW$_2$ (right) on network with 100 components}
    \begin{tabular}{lr|rrrr|rrrr}
    \toprule
    \rowcolor[rgb]{ .949,  .949,  .949}      &      & \multicolumn{4}{c|}{IPW$_1$} & \multicolumn{4}{c}{IPW$_2$} \\
    \rowcolor[rgb]{ .949,  .949,  .949}      & \multicolumn{1}{c|}{True} & \multicolumn{1}{c}{Bias} & \multicolumn{1}{c}{ESE} & \multicolumn{1}{c}{ASE} & \multicolumn{1}{c|}{ECP} & \multicolumn{1}{c}{Bias} & \multicolumn{1}{c}{ESE} & \multicolumn{1}{c}{ASE} & \multicolumn{1}{c}{ECP} \\
    \midrule
    \midrule
    $\widehat{Y}(1, 0.25)$ & 0.249 & <0.001 & 0.047 & 0.044 & 0.92 & -0.010 & 0.051 & 0.047 & 0.92 \\
    $\widehat{Y}(1, 0.5)$ & 0.227 & -0.001 & 0.029 & 0.027 & 0.91 & -0.001 & 0.023 & 0.024 & 0.96 \\
    $\widehat{Y}(1, 0.75)$ & 0.206 & -0.005 & 0.041 & 0.035 & 0.91 & -0.005 & 0.036 & 0.033 & 0.94 \\
    $\widehat{Y}(0, 0.25)$ & 0.227 & -0.006 & 0.033 & 0.034 & 0.92 & -0.002 & 0.032 & 0.032 & 0.95 \\
    $\widehat{Y}(0, 0.5)$ & 0.274 & 0.003 & 0.025 & 0.029 & 0.95 & -0.002 & 0.023 & 0.025 & 0.98 \\
    $\widehat{Y}(0, 0.75)$ & 0.325 & 0.006 & 0.047 & 0.048 & 0.89 & -0.016 & 0.058 & 0.053 & 0.94 \\
    $\widehat{Y}(0.25)$ & 0.233 & -0.004 & 0.028 & 0.028 & 0.95 & -0.004 & 0.028 & 0.028 & 0.97 \\
    $\widehat{Y}(0.5)$ & 0.250 & 0.001 & 0.019 & 0.021 & 0.95 & -0.002 & 0.016 & 0.018 & 0.99 \\
    $\widehat{Y}(0.75)$ & 0.235 & -0.002 & 0.033 & 0.030 & 0.92 & -0.008 & 0.031 & 0.030 & 0.96 \\
    \bottomrule
    \end{tabular}%
  \label{tab:100comp}%
\end{table}%

% Table generated by Excel2LaTeX from sheet 'Sheet1'
\begin{table}[H]
  \centering
  \caption{Simulation results of IPW$_1$ (left) and IPW$_2$ (right) on network with 150 components}
    \begin{tabular}{lr|rrrr|rrrr}
    \toprule
    \rowcolor[rgb]{ .949,  .949,  .949}      &      & \multicolumn{4}{c|}{IPW$_1$} & \multicolumn{4}{c}{IPW$_2$} \\
    \rowcolor[rgb]{ .949,  .949,  .949}      & \multicolumn{1}{c|}{True} & \multicolumn{1}{c}{Bias} & \multicolumn{1}{c}{ESE} & \multicolumn{1}{c}{ASE} & \multicolumn{1}{c|}{ECP} & \multicolumn{1}{c}{Bias} & \multicolumn{1}{c}{ESE} & \multicolumn{1}{c}{ASE} & \multicolumn{1}{c}{ECP} \\
    \midrule
    \midrule
    $\widehat{Y}(1, 0.25)$ & 0.249 & 0.001 & 0.036 & 0.037 & 0.93 & -0.012 & 0.041 & 0.039 & 0.95 \\
    $\widehat{Y}(1, 0.5)$ & 0.226 & 0.001 & 0.021 & 0.022 & 0.96 & -0.001 & 0.017 & 0.020 & 0.98 \\
    $\widehat{Y}(1, 0.75)$ & 0.205 & -0.003 & 0.028 & 0.029 & 0.95 & -0.004 & 0.027 & 0.028 & 0.96 \\
    $\widehat{Y}(0, 0.25)$ & 0.228 & -0.002 & 0.027 & 0.028 & 0.95 & -0.004 & 0.026 & 0.027 & 0.97 \\
    $\widehat{Y}(0, 0.5)$ & 0.274 & 0.001 & 0.022 & 0.023 & 0.96 & -0.002 & 0.018 & 0.020 & 0.98 \\
    $\widehat{Y}(0, 0.75)$ & 0.325 & 0.001 & 0.040 & 0.041 & 0.94 & -0.017 & 0.044 & 0.044 & 0.96 \\
    $\widehat{Y}(0.25)$ & 0.233 & -0.001 & 0.023 & 0.023 & 0.94 & -0.006 & 0.022 & 0.023 & 0.97 \\
    $\widehat{Y}(0.5)$ & 0.250 & 0.001 & 0.015 & 0.017 & 0.97 & -0.002 & 0.012 & 0.015 & 0.99 \\
    $\widehat{Y}(0.75)$ & 0.235 & -0.002 & 0.023 & 0.024 & 0.96 & -0.007 & 0.023 & 0.024 & 0.97 \\
    \bottomrule
    \end{tabular}%
  \label{tab:addlabel}%
\end{table}%

% Table generated by Excel2LaTeX from sheet 'Sheet1'
\begin{table}[H]
  \centering
  \caption{Simulation results of IPW$_1$ (left) and IPW$_2$ (right) on network with 200 components}
    \begin{tabular}{lr|rrrr|rrrr}
    \toprule
    \rowcolor[rgb]{ .949,  .949,  .949}      &      & \multicolumn{4}{c|}{IPW$_1$} & \multicolumn{4}{c}{IPW$_2$} \\
    \rowcolor[rgb]{ .949,  .949,  .949}      & \multicolumn{1}{c|}{True} & \multicolumn{1}{c}{Bias} & \multicolumn{1}{c}{ESE} & \multicolumn{1}{c}{ASE} & \multicolumn{1}{c|}{ECP} & \multicolumn{1}{c}{Bias} & \multicolumn{1}{c}{ESE} & \multicolumn{1}{c}{ASE} & \multicolumn{1}{c}{ECP} \\
    \midrule
    \midrule
    $\widehat{Y}(1, 0.25)$ & 0.248 & 0.001 & 0.031 & 0.032 & 0.94 & -0.011 & 0.036 & 0.034 & 0.95 \\
    $\widehat{Y}(1, 0.5)$ & 0.226 & 0.001 & 0.019 & 0.019 & 0.94 & <0.001 & 0.016 & 0.017 & 0.97 \\
    $\widehat{Y}(1, 0.75)$ & 0.205 & -0.002 & 0.025 & 0.025 & 0.94 & -0.004 & 0.024 & 0.024 & 0.95 \\
    $\widehat{Y}(0, 0.25)$ & 0.228 & -0.003 & 0.023 & 0.024 & 0.95 & -0.007 & 0.024 & 0.024 & 0.95 \\
    $\widehat{Y}(0, 0.5)$ & 0.274 & 0.001 & 0.019 & 0.020 & 0.96 & -0.002 & 0.016 & 0.018 & 0.97 \\
    $\widehat{Y}(0, 0.75)$ & 0.325 & -0.001 & 0.036 & 0.035 & 0.94 & -0.016 & 0.041 & 0.038 & 0.95 \\
    $\widehat{Y}(0.25)$ & 0.233 & -0.002 & 0.019 & 0.020 & 0.97 & -0.007 & 0.021 & 0.021 & 0.97 \\
    $\widehat{Y}(0.5)$ & 0.250 & 0.001 & 0.014 & 0.014 & 0.96 & -0.001 & 0.011 & 0.013 & 0.98 \\
    $\widehat{Y}(0.75)$ & 0.235 & -0.002 & 0.021 & 0.021 & 0.95 & -0.008 & 0.020 & 0.020 & 0.97 \\
    \bottomrule
    \end{tabular}%
  \label{tab:200comp}%
\end{table}%

% Table generated by Excel2LaTeX from sheet 'Sheet1'
\begin{table}[H]
  \centering
  \caption{Simulation results of IPW$_1$ (left) and IPW$_2$ (right) using propensity score model $A_i={\rm Bern}({\rm logit}^{-1}(-0.5-1.5\cdot Z_i+b_\nu)$ on a network with 100 components.}
    \begin{tabular}{lr|rrrr|rrrr}
    \toprule
    \rowcolor[rgb]{ .949,  .949,  .949}       &       & \multicolumn{4}{c|}{IPW$_1$}    & \multicolumn{4}{c}{IPW$_2$} \\
    \rowcolor[rgb]{ .949,  .949,  .949}       & \multicolumn{1}{c|}{True} & \multicolumn{1}{c}{Bias} & \multicolumn{1}{c}{ESE} & \multicolumn{1}{c}{ASE} & \multicolumn{1}{c|}{ECP} & \multicolumn{1}{c}{Bias} & \multicolumn{1}{c}{ESE} & \multicolumn{1}{c}{ASE} & \multicolumn{1}{c}{ECP} \\
    \midrule
    \midrule
    $\widehat{Y}(1, 0.25)$ & 0.2482 & 0.0017 & 0.041 & 0.044 & 0.94  & 0.0017 & 0.035 & 0.038 & 0.97 \\
    $\widehat{Y}(1, 0.5)$ & 0.2260 & 0.0006 & 0.068 & 0.074 & 0.86  & 0.0011 & 0.114 & 0.055 & 0.92 \\
    $\widehat{Y}(1, 0.75)$ & 0.2051 & -0.0005 & 0.171 & 0.183 & 0.68  & -0.0079 & 0.540  & 0.099 & 0.71 \\
     $\widehat{Y}(0, 0.25)$ & 0.2278 & 0.0011 & 0.020  & 0.022 & 0.96  & -0.0002 & 0.017 & 0.019 & 0.99 \\
     $\widehat{Y}(0, 0.5)$ & 0.2743 & 0.0003 & 0.050  & 0.045 & 0.90   & -0.0093 & 0.052 & 0.045 & 0.95 \\
     $\widehat{Y}(0, 0.75)$ & 0.3325 & 0.0021 & 0.172 & 0.124 & 0.79  & -0.0348 & 0.208 & 0.130  & 0.81 \\
    $\widehat{Y}(0.25)$ & 0.2329 & 0.0012 & 0.018 & 0.021 & 0.97  & 0.0002 & 0.015 & 0.018 & 0.98 \\
    $\widehat{Y}(0.5)$ & 0.2502 & 0.0005 & 0.042 & 0.047 & 0.89  & -0.0041 & 0.063 & 0.040  & 0.95 \\
    $\widehat{Y}(0.75)$ & 0.2350& -0.0018 & 0.133 & 0.150  & 0.73  & -0.0146 & 0.408 & 0.091 & 0.78 \\
    \hline
    \end{tabular}%
  \label{tab:trtdist}%
\end{table}%

\newpage
\section{Community Alerts and HIV Risk Behavior in TRIP at 6 months}
In this section, we report the point estimates for direct, indirect, total, and overall effects under allocation strategies 20\%, 30\%, 40\% and 50\% and their corresponding 95\% confidence intervals for the effect of community alerts at baseline on HIV risk behavior at 6-month follow up in TRIP using different set of measured confounding variables. The full set of confounding variables are HIV status, shared drug equipment (e.g. syringe) in last six months, the calendar date at first interview, education (primary school, first 3 years of high school, last 3 years of high school, and post high school), and employment status (employed, unemployed/looking for a work, can’t work because of health reason, and other). We first considered univariate models; that is, only use one confounding variable at a time (Tables \ref{tab:TRIP_HIV}, \ref{tab:TRIP_date}, \ref{tab:TRIP_share}, \ref{tab:TRIP_edu}, \ref{tab:TRIP_employ}), and a model not adjusted for any confounders (Table \ref{tab:TRIP_noadj}). Second, we estimated the effects using all of the variables, but excluding one at a time (Tables \ref{tab:TRIP_noHIV}, \ref{tab:TRIP_noDate}, \ref{tab:TRIP_noShare}, \ref{tab:TRIP_noEdu}, \ref{tab:TRIP_noemploy}). The 95\% CI (10) were estimated by original 10 components TRIP network, and 95\% CI (20) were estimated by dividing TRIP network into 20 components using community detection. 

% Table generated by Excel2LaTeX from sheet 'HIV'
\begin{table}[htbp]
  \centering
  \caption{Estimated risk differences and 95\% confidence intervals (CIs) of the effects of community alerts on HIV risk behavior at 6 months in TRIP adjusted only for HIV status.}
    \begin{tabular}{ll|rrrrrr}
    \toprule
    \multicolumn{1}{c}{Effects} & Coverage & \multicolumn{3}{c}{IPW$_1$} & \multicolumn{3}{c}{IPW$_2$} \\
         & $(\alpha, \alpha')$ & \multicolumn{1}{c}{RD} & \multicolumn{1}{c}{95\% CI (10)} & \multicolumn{1}{c}{95\% CI (20)} & \multicolumn{1}{c}{RD} & \multicolumn{1}{c}{95\% CI (10)} & \multicolumn{1}{c}{95\% CI (20)} \\
    \midrule
    \midrule
    \rowcolor[rgb]{ .949,  .949,  .949} Direct & (20\%, 20\%) & -0.0803 & (-0.154,-0.007) & (-0.246, 0.085) & -0.0237 & (-0.070, 0.022) & (-0.219, 0.171) \\
    \rowcolor[rgb]{ .949,  .949,  .949} Direct & (30\%, 30\%) & -0.1301 & (-0.285, 0.025) & (-0.327, 0.067) & -0.1196 & (-0.272, 0.033) & (-0.312, 0.073) \\
    \rowcolor[rgb]{ .949,  .949,  .949} Direct & (40\%, 40\%) & -0.1808 & (-0.439, 0.078) & (-0.408, 0.046) & -0.1913 & (-0.481, 0.099) & (-0.430, 0.048) \\
    \rowcolor[rgb]{ .949,  .949,  .949} Direct & (50\%, 50\%) & -0.2256 & (-0.587, 0.136) & (-0.482, 0.031) & -0.2366 & (-0.622, 0.149) & (-0.517, 0.044) \\
    Indirect & (30\%, 20\%) & 0.0055 & (-0.011, 0.022) & (-0.020, 0.031) & 0.0063 & (-0.010, 0.022) & (-0.050, 0.063) \\
    Indirect & (40\%, 20\%) & -0.0057 & (-0.026, 0.015) & (-0.063, 0.052) & -0.0058 & (-0.016, 0.005) & (-0.103, 0.091) \\
    Indirect & (50\%, 20\%) & -0.0240 & (-0.059, 0.011) & (-0.128, 0.080) & -0.0269 & (-0.052,-0.002) & (-0.158, 0.105) \\
    Indirect & (40\%, 30\%) & -0.0112 & (-0.025, 0.002) & (-0.046, 0.023) & -0.0120 & (-0.025, 0.001) & (-0.055, 0.031) \\
    Indirect & (50\%, 40\%) & -0.0184 & (-0.040, 0.003) & (-0.068, 0.032) & -0.0211 & (-0.046, 0.004) & (-0.060, 0.018) \\
    Indirect & (50\%, 30\%) & -0.0295 & (-0.064, 0.005) & (-0.113, 0.054) & -0.0331 & (-0.071, 0.004) & (-0.113, 0.047) \\
    \rowcolor[rgb]{ .949,  .949,  .949} Total & (30\%, 20\%) & -0.1246 & (-0.268, 0.018) & (-0.325, 0.076) & -0.1134 & (-0.252, 0.025) & (-0.311, 0.084) \\
    \rowcolor[rgb]{ .949,  .949,  .949} Total & (40\%, 20\%) & -0.1865 & (-0.439, 0.066) & (-0.398, 0.025) & -0.1970 & (-0.484, 0.090) & (-0.407, 0.013) \\
    \rowcolor[rgb]{ .949,  .949,  .949} Total & (50\%, 20\%) & -0.2496 & (-0.617, 0.118) & (-0.455,-0.045) & -0.2635 & (-0.671, 0.144) & (-0.476,-0.051) \\
    \rowcolor[rgb]{ .949,  .949,  .949} Total & (40\%, 30\%) & -0.1920 & (-0.458, 0.074) & (-0.403, 0.019) & -0.2033 & (-0.506, 0.099) & (-0.423, 0.016) \\
    \rowcolor[rgb]{ .949,  .949,  .949} Total & (50\%, 40\%) & -0.2440 & (-0.618, 0.130) & (-0.466,-0.022) & -0.2577 & (-0.668, 0.152) & (-0.512,-0.004) \\
    \rowcolor[rgb]{ .949,  .949,  .949} Total & (50\%, 30\%) & -0.2552 & (-0.636, 0.126) & (-0.461,-0.050) & -0.2697 & (-0.692, 0.152) & (-0.501,-0.039) \\
    Overall & (30\%, 20\%) & -0.0174 & (-0.039, 0.004) & (-0.051, 0.017) & -0.0249 & (-0.061, 0.011) & (-0.072, 0.022) \\
    Overall & (40\%, 20\%) & -0.0619 & (-0.148, 0.024) & (-0.118,-0.006) & -0.0775 & (-0.196, 0.041) & (-0.151,-0.004) \\
    Overall & (50\%, 20\%) & -0.1208 & (-0.297, 0.056) & (-0.194,-0.048) & -0.1404 & (-0.360, 0.079) & (-0.230,-0.051) \\
    Overall & (40\%, 30\%) & -0.0445 & (-0.110, 0.021) & (-0.069,-0.020) & -0.0526 & (-0.135, 0.030) & (-0.087,-0.018) \\
    Overall & (50\%, 40\%) & -0.0588 & (-0.149, 0.032) & (-0.086,-0.032) & -0.0629 & (-0.164, 0.038) & (-0.096,-0.030) \\
    Overall & (50\%, 30\%) & -0.1033 & (-0.259, 0.053) & (-0.152,-0.055) & -0.1155 & (-0.299, 0.068) & (-0.178,-0.053) \\
    \bottomrule
    \end{tabular}%
  \label{tab:TRIP_HIV}%
\end{table}%

% Table generated by Excel2LaTeX from sheet 'date'
\begin{table}[h]
  \centering
  \caption{Estimated risk differences and 95\% confidence intervals (CIs) of the effects of community alerts on HIV risk behavior at 6 months in TRIP adjusted only for the calendar date at first interview.}
    \begin{tabular}{lc|rrrrrr}
    \toprule
    \multicolumn{1}{c}{Effects} & Coverage & \multicolumn{3}{c}{IPW$_1$} & \multicolumn{3}{c}{IPW$_2$} \\
         & $(\alpha, \alpha')$ & \multicolumn{1}{c}{RD} & \multicolumn{1}{c}{95\% CI (10)} & \multicolumn{1}{c}{95\% CI (20)} & \multicolumn{1}{c}{RD} & \multicolumn{1}{c}{95\% CI (10)} & \multicolumn{1}{c}{95\% CI (20)} \\
    \midrule
    \midrule
    \rowcolor[rgb]{ .949,  .949,  .949} Direct & (20\%, 20\%) & -0.0688 & (-0.132,-0.005) & (-0.270,0.133) & 0.0052 & (-0.100, 0.111) & (-0.212, 0.223) \\
    \rowcolor[rgb]{ .949,  .949,  .949} Direct & (30\%, 30\%) & -0.1066 & (-0.215, 0.002) & (-0.333, 0.120) & -0.0739 & (-0.143,-0.005) & (-0.304, 0.156) \\
    \rowcolor[rgb]{ .949,  .949,  .949} Direct & (40\%, 40\%) & -0.1515 & (-0.349, 0.046) & (-0.386, 0.083) & -0.1454 & (-0.342, 0.051) & (-0.406, 0.116) \\
    \rowcolor[rgb]{ .949,  .949,  .949} Direct & (50\%, 50\%) & -0.1947 & (-0.492, 0.103) & (-0.437, 0.048) & -0.2039 & (-0.521, 0.113) & (-0.501, 0.093) \\
    Indirect & (30\%, 20\%) & 0.0040 & (-0.011, 0.019) & (-0.019, 0.026) & 0.0040 & (-0.008, 0.016) & (-0.046, 0.054) \\
    Indirect & (40\%, 20\%) & -0.0103 & (-0.029, 0.009) & (-0.054, 0.033) & -0.0044 & (-0.015, 0.007) & (-0.096, 0.087) \\
    Indirect & (50\%, 20\%) & -0.0333 & (-0.072, 0.006) & (-0.106, 0.039) & -0.0174 & (-0.032,-0.003) & (-0.153, 0.119) \\
    Indirect & (40\%, 30\%) & -0.0142 & (-0.030, 0.002) & (-0.039, 0.010) & -0.0084 & (-0.015,-0.001) & (-0.053, 0.036) \\
    Indirect & (50\%, 40\%) & -0.0230 & (-0.050, 0.004) & (-0.057, 0.011) & -0.0131 & (-0.025,-0.001) & (-0.063, 0.037) \\
    Indirect & (50\%, 30\%) & -0.0373 & (-0.080, 0.005) & (-0.095, 0.020) & -0.0214 & (-0.040,-0.003) & (-0.114, 0.071) \\
    \rowcolor[rgb]{ .949,  .949,  .949} Total & (30\%, 20\%) & -0.1027 & (-0.204,-0.001) & (-0.335, 0.129) & -0.0699 & (-0.131,-0.009) & (-0.309, 0.170) \\
    \rowcolor[rgb]{ .949,  .949,  .949} Total & (40\%, 20\%) & -0.1618 & (-0.361, 0.038) & (-0.395, 0.072) & -0.1498 & (-0.342, 0.042) & (-0.392, 0.092) \\
    \rowcolor[rgb]{ .949,  .949,  .949} Total & (50\%, 20\%) & -0.2280 & (-0.547, 0.091) & (-0.450,-0.006) & -0.2213 & (-0.542, 0.100) & (-0.447, 0.004) \\
    \rowcolor[rgb]{ .949,  .949,  .949} Total & (40\%, 30\%) & -0.1658 & (-0.376, 0.044) & (-0.395, 0.063) & -0.1538 & (-0.356, 0.048) & (-0.395, 0.087) \\
    \rowcolor[rgb]{ .949,  .949,  .949} Total & (50\%, 40\%) & -0.2177 & (-0.537, 0.101) & (-0.444, 0.009) & -0.2169 & (-0.542, 0.109) & (-0.474, 0.040) \\
    \rowcolor[rgb]{ .949,  .949,  .949} Total & (50\%, 30\%) & -0.2320 & (-0.563, 0.099) & (-0.451,-0.012) & -0.2253 & (-0.556, 0.106) & (-0.458, 0.007) \\
    Overall & (30\%, 20\%) & -0.0143 & (-0.030, 0.002) & (-0.051, 0.023) & -0.0192 & (-0.044, 0.006) & (-0.070, 0.031) \\
    Overall & (40\%, 20\%) & -0.0571 & (-0.132, 0.017) & (-0.118, 0.004) & -0.0636 & (-0.155, 0.028) & (-0.140, 0.013) \\
    Overall & (50\%, 20\%) & -0.1169 & (-0.282, 0.049) & (-0.195,-0.038) & -0.1204 & (-0.301, 0.060) & (-0.204,-0.037) \\
    Overall & (40\%, 30\%) & -0.0429 & (-0.104, 0.018) & (-0.070,-0.016) & -0.0443 & (-0.111, 0.022) & (-0.074,-0.015) \\
    Overall & (50\%, 40\%) & -0.0598 & (-0.152, 0.032) & (-0.085,-0.034) & -0.0568 & (-0.146, 0.032) & (-0.084,-0.030) \\
    Overall & (50\%, 30\%) & -0.1026 & (-0.256, 0.050) & (-0.153,-0.053) & -0.1012 & (-0.257, 0.054) & (-0.150,-0.053) \\
    \bottomrule
    \end{tabular}%
  \label{tab:TRIP_date}%
\end{table}%

% Table generated by Excel2LaTeX from sheet 'share.x'
\begin{table}[htbp]
  \centering
  \caption{Estimated risk differences and 95\% confidence intervals (CIs) of the effects of community alerts on HIV risk behavior at 6 months in TRIP adjusted only for shared drug equipment (e.g. syringe) in last six months.}
    \begin{tabular}{lc|rllrll}
    \toprule
    \multicolumn{1}{c}{Effects} & Coverage & \multicolumn{3}{c}{IPW$_1$} & \multicolumn{3}{c}{IPW$_2$} \\
         & $(\alpha, \alpha')$ & \multicolumn{1}{c}{RD} & \multicolumn{1}{c}{95\% CI (10)} & \multicolumn{1}{c}{95\% CI (20)} & \multicolumn{1}{c}{RD} & \multicolumn{1}{c}{95\% CI (10)} & \multicolumn{1}{c}{95\% CI (20)} \\
    \midrule
    \midrule
    \rowcolor[rgb]{ .949,  .949,  .949} Direct & (20\%, 20\%) & -0.0889 & (-0.173,-0.004) & (-0.324, 0.146) & 0.0253 & (-0.121, 0.172) & (-0.222, 0.273) \\
    \rowcolor[rgb]{ .949,  .949,  .949} Direct & (30\%, 30\%) & -0.1337 & (-0.288, 0.020) & (-0.417, 0.15) & -0.0609 & (-0.107,-0.015) & (-0.335, 0.213) \\
    \rowcolor[rgb]{ .949,  .949,  .949} Direct & (40\%, 40\%) & -0.1792 & (-0.423, 0.065) & (-0.489, 0.131) & -0.1386 & (-0.318, 0.041) & (-0.452, 0.175) \\
    \rowcolor[rgb]{ .949,  .949,  .949} Direct & (50\%, 50\%) & -0.2186 & (-0.552, 0.115) & (-0.548, 0.110) & -0.2019 & (-0.514, 0.110) & (-0.556, 0.152) \\
    Indirect & (30\%, 20\%) & 0.0057 & (-0.012, 0.024) & (-0.019, 0.031) & 0.0104 & (-0.015, 0.035) & (-0.048, 0.068) \\
    Indirect & (40\%, 20\%) & -0.0057 & (-0.025, 0.013) & (-0.065, 0.053) & 0.0074 & (-0.025, 0.040) & (-0.104, 0.119) \\
    Indirect & (50\%, 20\%) & -0.0259 & (-0.057, 0.006) & (-0.137, 0.085) & -0.0024 & (-0.037, 0.032) & (-0.169, 0.165) \\
    Indirect & (40\%, 30\%) & -0.0114 & (-0.024, 0.001) & (-0.049, 0.027) & -0.0030 & (-0.012, 0.006) & (-0.059, 0.053) \\
    Indirect & (50\%, 40\%) & -0.0202 & (-0.044, 0.003) & (-0.077, 0.036) & -0.0097 & (-0.018,-0.001) & (-0.071, 0.051) \\
    Indirect & (50\%, 30\%) & -0.0316 & (-0.067,0.004) & (-0.125, 0.062) & -0.0127 & (-0.027, 0.002) & (-0.128, 0.102) \\
    \rowcolor[rgb]{ .949,  .949,  .949} Total & (30\%, 20\%) & -0.1280 & (-0.270, 0.014) & (-0.411, 0.155) & -0.0506 & (-0.091,-0.010) & (-0.329, 0.228) \\
    \rowcolor[rgb]{ .949,  .949,  .949} Total & (40\%, 20\%) & -0.1848 & (-0.424, 0.054) & (-0.470, 0.100) & -0.1312 & (-0.282, 0.019) & (-0.410, 0.148) \\
    \rowcolor[rgb]{ .949,  .949,  .949} Total & (50\%, 20\%) & -0.2445 & (-0.592, 0.103) & (-0.505, 0.016) & -0.2043 & (-0.487, 0.079) & (-0.459, 0.050) \\
    \rowcolor[rgb]{ .949,  .949,  .949} Total & (40\%, 30\%) & -0.1906 & (-0.444, 0.063) & (-0.477, 0.096) & -0.1416 & (-0.316, 0.032) & (-0.425, 0.142) \\
    \rowcolor[rgb]{ .949,  .949,  .949} Total & (50\%, 40\%) & -0.2388 & (-0.592, 0.115) & (-0.525, 0.047) & -0.2117 & (-0.525, 0.101) & (-0.514, 0.091) \\
    \rowcolor[rgb]{ .949,  .949,  .949} Total & (50\%, 30\%) & -0.2502 & (-0.612, 0.112) & (-0.513, 0.013) & -0.2147 & (-0.521, 0.092) & (-0.482, 0.053) \\
    Overall & (30\%, 20\%) & -0.0166 & (-0.035, 0.002) & (-0.058, 0.025) & -0.0130 & (-0.025,-0.001) & (-0.068, 0.042) \\
    Overall & (40\%, 20\%) & -0.0595 & (-0.138, 0.019) & (-0.124, 0.005) & -0.0531 & (-0.122, 0.016) & (-0.135, 0.029) \\
    Overall & (50\%, 20\%) & -0.1174 & (-0.285, 0.050) & (-0.194,-0.041) & -0.1084 & (-0.263, 0.046) & (-0.195,-0.021) \\
    Overall & (40\%, 30\%) & -0.0429 & (-0.104, 0.019) & (-0.069,-0.017) & -0.0401 & (-0.098, 0.018) & (-0.070,-0.010) \\
    Overall & (50\%, 40\%) & -0.0578 & (-0.147, 0.032) & (-0.086,-0.030) & -0.0553 & (-0.141, 0.030) & (-0.081,-0.029) \\
    Overall & (50\%, 30\%) & -0.1008 & (-0.252, 0.050) & (-0.149,-0.052) & -0.0954 & (-0.239, 0.048) & (-0.142,-0.048) \\
    \bottomrule
    \end{tabular}%
  \label{tab:TRIP_share}%
\end{table}%

% Table generated by Excel2LaTeX from sheet 'education'
\begin{table}[htbp]
  \centering
  \caption{Estimated risk differences and 95\% confidence intervals (CIs) of the effects of community alerts on HIV risk behavior at 6 months in TRIP adjusted only for education (primary school, high school, and post high school).}
    \begin{tabular}{lc|rllrll}
    \toprule
    \multicolumn{1}{c}{Effects} & Coverage & \multicolumn{3}{c}{IPW$_1$} & \multicolumn{3}{c}{IPW$_2$} \\
         & $(\alpha, \alpha')$ & \multicolumn{1}{c}{RD} & \multicolumn{1}{c}{95\% CI (10)} & \multicolumn{1}{c}{95\% CI (20)} & \multicolumn{1}{c}{RD} & \multicolumn{1}{c}{95\% CI (10)} & \multicolumn{1}{c}{95\% CI (20)} \\
    \midrule
    \midrule
    \rowcolor[rgb]{ .949,  .949,  .949} Direct & (20\%, 20\%) & -0.0799 & (-0.156,-0.004) & (-0.266, 0.106) & 0.0049 & (-0.093, 0.102) & (-0.212, 0.222) \\
    \rowcolor[rgb]{ .949,  .949,  .949} Direct & (30\%, 30\%) & -0.1294 & (-0.294, 0.035) & (-0.347, 0.088) & -0.0827 & (-0.172, 0.006) & (-0.313, 0.148) \\
    \rowcolor[rgb]{ .949,  .949,  .949} Direct & (40\%, 40\%) & -0.1799 & (-0.456, 0.096) & (-0.418, 0.058) & -0.1650 & (-0.414, 0.084) & (-0.453, 0.123) \\
    \rowcolor[rgb]{ .949,  .949,  .949} Direct & (50\%, 50\%) & -0.2246 & (-0.608, 0.159) & (-0.481, 0.032) & -0.2347 & (-0.631, 0.162) & (-0.592, 0.122) \\
    Indirect & (30\%, 20\%) & 0.0054 & (-0.016, 0.027) & (-0.014, 0.025) & 0.0106 & (-0.017, 0.038) & (-0.048, 0.069) \\
    Indirect & (40\%, 20\%) & -0.0060 & (-0.037, 0.025) & (-0.047, 0.035) & 0.0136 & (-0.037, 0.064) & (-0.108, 0.135) \\
    Indirect & (50\%, 20\%) & -0.0245 & (-0.069, 0.020) & (-0.104, 0.054) & 0.0139 & (-0.062, 0.090) & (-0.179, 0.207) \\
    Indirect & (40\%, 30\%) & -0.0114 & (-0.026, 0.003) & (-0.037, 0.014) & 0.0031 & (-0.020, 0.027) & (-0.062, 0.068) \\
    Indirect & (50\%, 40\%) & -0.0185 & (-0.038, 0.001) & (-0.061, 0.024) & 0.0003 & (-0.026, 0.026) & (-0.076, 0.076) \\
    Indirect & (50\%, 30\%) & -0.0299 & (-0.064, 0.004) & (-0.097, 0.037) & 0.0034 & (-0.046, 0.053) & (-0.137, 0.143) \\
    \rowcolor[rgb]{ .949,  .949,  .949} Total & (30\%, 20\%) & -0.1240 & (-0.273, 0.025) & (-0.350, 0.102) & -0.0722 & (-0.139,-0.005) & (-0.309, 0.164) \\
    \rowcolor[rgb]{ .949,  .949,  .949} Total & (40\%, 20\%) & -0.1859 & (-0.446, 0.074) & (-0.422, 0.050) & -0.1514 & (-0.353, 0.050) & (-0.395, 0.092) \\
    \rowcolor[rgb]{ .949,  .949,  .949} Total & (50\%, 20\%) & -0.2491 & (-0.622, 0.123) & (-0.475,-0.023) & -0.2207 & (-0.547, 0.105) & (-0.452, 0.011) \\
    \rowcolor[rgb]{ .949,  .949,  .949} Total & (40\%, 30\%) & -0.1913 & (-0.469, 0.086) & (-0.421, 0.038) & -0.1619 & (-0.390, 0.066) & (-0.412, 0.088) \\
    \rowcolor[rgb]{ .949,  .949,  .949} Total & (50\%, 40\%) & -0.2431 & (-0.634, 0.147) & (-0.472,-0.014) & -0.2344 & (-0.609, 0.140) & (-0.526, 0.057) \\
    \rowcolor[rgb]{ .949,  .949,  .949} Total & (50\%, 30\%) & -0.2545 & (-0.646, 0.137) & (-0.475,-0.034) & -0.2313 & (-0.584, 0.121) & (-0.479, 0.016) \\
    Overall & (30\%, 20\%) & -0.0174 & (-0.039, 0.004) & (-0.056, 0.021) & -0.0152 & (-0.031,0.0005) & (-0.066, 0.036) \\
    Overall & (40\%, 20\%) & -0.0620 & (-0.145, 0.021) & (-0.125, 0.001) & -0.0533 & (-0.122, 0.015) & (-0.130, 0.023) \\
    Overall & (50\%, 20\%) & -0.1208 & (-0.292, 0.050) & (-0.199,-0.043) & -0.1044 & (-0.249, 0.040) & (-0.190,-0.019) \\
    Overall & (40\%, 30\%) & -0.0445 & (-0.108, 0.019) & (-0.071,-0.018) & -0.0381 & (-0.091, 0.015) & (-0.067,-0.009) \\
    Overall & (50\%, 40\%) & -0.0589 & (-0.147, 0.030) & (-0.085,-0.032) & -0.0510 & (-0.128, 0.026) & (-0.077,-0.025) \\
    Overall & (50\%, 30\%) & -0.1034 & (-0.255, 0.048) & (-0.153,-0.054) & -0.0891 & (-0.219, 0.041) & (-0.137,-0.042) \\
    \bottomrule
    \end{tabular}%
  \label{tab:TRIP_edu}%
\end{table}%

% Table generated by Excel2LaTeX from sheet 'employ'
\begin{table}[htbp]
  \centering
  \caption{Estimated risk differences and 95\% confidence intervals (CIs) of the effects of community alerts on HIV risk behavior at 6 months in TRIP adjusted only for employment status (employed, unemployed/looking for a work, can’t work because of health reason, and other).}
    \begin{tabular}{lc|rllrll}
    \toprule
    \multicolumn{1}{c}{Effects} & Coverage & \multicolumn{3}{c}{IPW$_1$} & \multicolumn{3}{c}{IPW$_2$} \\
         & $(\alpha, \alpha')$ & \multicolumn{1}{c}{RD} & \multicolumn{1}{c}{95\% CI (10)} & \multicolumn{1}{c}{95\% CI (20)} & \multicolumn{1}{c}{RD} & \multicolumn{1}{c}{95\% CI (10)} & \multicolumn{1}{c}{95\% CI (20)} \\
    \midrule
    \midrule
    \rowcolor[rgb]{ .949,  .949,  .949} Direct & (20\%, 20\%) & -0.0821 & (-0.158,-0.006) & (-0.282, 0.118) & 0.0185 & (-0.107, 0.144) & (-0.198, 0.235) \\
    \rowcolor[rgb]{ .949,  .949,  .949} Direct & (30\%, 30\%) & -0.1309 & (-0.279, 0.017) & (-0.385, 0.123) & -0.0481 & (-0.094,-0.002) & (-0.287, 0.191) \\
    \rowcolor[rgb]{ .949,  .949,  .949} Direct & (40\%, 40\%) & -0.1807 & (-0.427, 0.066) & (-0.474, 0.113) & -0.1076 & (-0.241, 0.025) & (-0.371, 0.156) \\
    \rowcolor[rgb]{ .949,  .949,  .949} Direct & (50\%, 50\%) & -0.2253 & (-0.573, 0.123) & (-0.548, 0.098) & -0.1578 & (-0.394, 0.079) & (-0.444, 0.129) \\
    Indirect & (30\%, 20\%) & 0.0051 & (-0.011, 0.021) & (-0.024, 0.034) & -0.0060 & (-0.015, 0.003) & (-0.046, 0.034) \\
    Indirect & (40\%, 20\%) & -0.0069 & (-0.026, 0.012) & (-0.077, 0.063) & -0.0247 & (-0.057, 0.007) & (-0.097, 0.047) \\
    Indirect & (50\%, 20\%) & -0.0255 & (-0.059, 0.008) & (-0.151, 0.100) & -0.0462 & (-0.102, 0.010) & (-0.154, 0.062) \\
    Indirect & (40\%, 30\%) & -0.0120 & (-0.026, 0.002) & (-0.055, 0.031) & -0.0188 & (-0.043, 0.005) & (-0.053, 0.016) \\
    Indirect & (50\%, 40\%) & -0.0186 & (-0.039, 0.002) & (-0.077, 0.040) & -0.0215 & (-0.046, 0.003) & (-0.064, 0.021) \\
    Indirect & (50\%, 30\%) & -0.0305 & (-0.065, 0.004) & (-0.131, 0.070) & -0.0402 & (-0.089, 0.009) & (-0.115, 0.035) \\
    \rowcolor[rgb]{ .949,  .949,  .949} Total & (30\%, 20\%) & -0.1258 & (-0.264, 0.012) & (-0.378, 0.126) & -0.0540 & (-0.101,-0.007) & (-0.302, 0.194) \\
    \rowcolor[rgb]{ .949,  .949,  .949} Total & (40\%, 20\%) & -0.1876 & (-0.432, 0.057) & (-0.451, 0.076) & -0.1324 & (-0.295, 0.030) & (-0.387, 0.123) \\
    \rowcolor[rgb]{ .949,  .949,  .949} Total & (50\%, 20\%) & -0.2508 & (-0.610, 0.108) & (-0.498,-0.004) & -0.2041 & (-0.495, 0.087) & (-0.443, 0.035) \\
    \rowcolor[rgb]{ .949,  .949,  .949} Total & (40\%, 30\%) & -0.1927 & (-0.449, 0.064) & (-0.461, 0.075) & -0.1264 & (-0.283, 0.030) & (-0.377, 0.124) \\
    \rowcolor[rgb]{ .949,  .949,  .949} Total & (50\%, 40\%) & -0.2439 & (-0.607, 0.119) & (-0.521, 0.034) & -0.1793 & (-0.440, 0.081) & (-0.434, 0.076) \\
    \rowcolor[rgb]{ .949,  .949,  .949} Total & (50\%, 30\%) & -0.2558 & (-0.628, 0.116) & (-0.508,-0.004) & -0.1981 & (-0.482, 0.086) & (-0.438, 0.042) \\
    Overall & (30\%, 20\%) & -0.0178 & (-0.040, 0.004) & (-0.057, 0.021) & -0.0241 & (-0.057, 0.009) & (-0.074, 0.026) \\
    Overall & (40\%, 20\%) & -0.0628 & (-0.148, 0.023) & (-0.126,0.0003) & -0.0715 & (-0.176, 0.033) & (-0.150, 0.007) \\
    Overall & (50\%, 20\%) & -0.1217 & (-0.297, 0.054) & (-0.200,-0.043) & -0.1288 & (-0.324, 0.066) & (-0.219,-0.038) \\
    Overall & (40\%, 30\%) & -0.0450 & (-0.110, 0.020) & (-0.072,-0.018) & -0.0474 & (-0.119, 0.025) & (-0.081,-0.014) \\
    Overall & (50\%, 40\%) & -0.0589 & (-0.150, 0.032) & (-0.087,-0.031) & -0.0574 & (-0.148, 0.033) & (-0.085,-0.029) \\
    Overall & (50\%, 30\%) & -0.1039 & (-0.260, 0.052) & (-0.154,-0.054) & -0.1047 & (-0.267, 0.057) & (-0.159,-0.050) \\
    \bottomrule
    \end{tabular}%
  \label{tab:TRIP_employ}%
\end{table}%

% Table generated by Excel2LaTeX from sheet 'no_HIV'
\begin{table}[htbp]
  \centering
  \caption{Estimated risk differences and 95\% confidence intervals (CIs) of the effects of community alerts on HIV risk behavior at 6 months in TRIP adjusted for shared drug equipment (e.g. syringe) in last six months, the calendar date at first interview, education (primary school, high school, and post high school), and employment status (employed, unemployed/looking for a work, can’t work because of health reason, and other).}
    \begin{tabular}{lc|rllrll}
    \toprule
    \multicolumn{1}{c}{Effects} & Coverage & \multicolumn{3}{c}{IPW$_1$} & \multicolumn{3}{c}{IPW$_2$} \\
         & $(\alpha, \alpha')$ & \multicolumn{1}{c}{RD} & \multicolumn{1}{c}{95\% CI (10)} & \multicolumn{1}{c}{95\% CI (20)} & \multicolumn{1}{c}{RD} & \multicolumn{1}{c}{95\% CI (10)} & \multicolumn{1}{c}{95\% CI (20)} \\
    \midrule
    \midrule
    \rowcolor[rgb]{ .949,  .949,  .949} Direct & (20\%, 20\%) & -0.0671 & (-0.125,-0.010) & (-0.317, 0.183) & 0.0343 & (-0.108, 0.176) & (-0.191, 0.260) \\
    \rowcolor[rgb]{ .949,  .949,  .949} Direct & (30\%, 30\%) & -0.0992 & (-0.198,-0.00004) & (-0.395, 0.196) & -0.0420 & (-0.072,-0.012) & (-0.290, 0.206) \\
    \rowcolor[rgb]{ .949,  .949,  .949} Direct & (40\%, 40\%) & -0.1396 & (-0.321, 0.042) & (-0.442, 0.163) & -0.1120 & (-0.272, 0.048) & (-0.397, 0.173) \\
    \rowcolor[rgb]{ .949,  .949,  .949} Direct & (50\%, 50\%) & -0.1791 & (-0.451, 0.093) & (-0.469, 0.111) & -0.1735 & (-0.467, 0.120) & (-0.501, 0.154) \\
    Indirect & (30\%, 20\%) & 0.0053 & (-0.014, 0.024) & (-0.021, 0.032) & 0.0036 & (-0.013, 0.021) & (-0.043, 0.050) \\
    Indirect & (40\%, 20\%) & -0.0093 & (-0.028, 0.009) & (-0.055, 0.037) & -0.0013 & (-0.028, 0.025) & (-0.094, 0.091) \\
    Indirect & (50\%, 20\%) & -0.0350 & (-0.075, 0.005) & (-0.104, 0.034) & -0.0079 & (-0.048, 0.032) & (-0.156, 0.141) \\
    Indirect & (40\%, 30\%) & -0.0146 & (-0.031, 0.002) & (-0.038, 0.009) & -0.0049 & (-0.016, 0.006) & (-0.054, 0.044) \\
    Indirect & (50\%, 40\%) & -0.0257 & (-0.059, 0.008) & (-0.055, 0.003) & -0.0065 & (-0.021, 0.008) & (-0.068, 0.054) \\
    Indirect & (50\%, 30\%) & -0.0403 & (-0.090, 0.009) & (-0.092, 0.011) & -0.0114 & (-0.037, 0.014) & (-0.120, 0.097) \\
    \rowcolor[rgb]{ .949,  .949,  .949} Total & (30\%, 20\%) & -0.0939 & (-0.182,-0.005) & (-0.393, 0.205) & -0.0384 & (-0.069,-0.008) & (-0.296, 0.219) \\
    \rowcolor[rgb]{ .949,  .949,  .949} Total & (40\%, 20\%) & -0.1489 & (-0.328, 0.030) & (-0.442, 0.144) & -0.1133 & (-0.251, 0.024) & (-0.377, 0.151) \\
    \rowcolor[rgb]{ .949,  .949,  .949} Total & (50\%, 20\%) & -0.2140 & (-0.510, 0.082) & (-0.474, 0.046) & -0.1813 & (-0.441, 0.079) & (-0.427, 0.064) \\
    \rowcolor[rgb]{ .949,  .949,  .949} Total & (40\%, 30\%) & -0.1542 & (-0.348, 0.040) & (-0.445, 0.137) & -0.1169 & (-0.270, 0.036) & (-0.378, 0.144) \\
    \rowcolor[rgb]{ .949,  .949,  .949} Total & (50\%, 40\%) & -0.2047 & (-0.504, 0.095) & (-0.476, 0.067) & -0.1800 & (-0.464, 0.104) & (-0.457, 0.097) \\
    \rowcolor[rgb]{ .949,  .949,  .949} Total & (50\%, 30\%) & -0.2193 & (-0.532, 0.093) & (-0.478, 0.040) & -0.1849 & (-0.461, 0.091) & (-0.434, 0.064) \\
    Overall & (30\%, 20\%) & -0.0111 & (-0.023, 0.001) & (-0.058, 0.036) & -0.0159 & (-0.032,0.0002) & (-0.067, 0.035) \\
    Overall & (40\%, 20\%) & -0.0517 & (-0.116, 0.012) & (-0.122, 0.019) & -0.0530 & (-0.120, 0.014) & (-0.131, 0.025) \\
    Overall & (50\%, 20\%) & -0.1111 & (-0.267, 0.044) & (-0.189,-0.033) & -0.1015 & (-0.242, 0.039) & (-0.188,-0.015) \\
    Overall & (40\%, 30\%) & -0.0407 & (-0.098, 0.017) & (-0.067,-0.015) & -0.0371 & (-0.089, 0.015) & (-0.067,-0.007) \\
    Overall & (50\%, 40\%) & -0.0593 & (-0.152, 0.033) & (-0.084,-0.034) & -0.0485 & (-0.121, 0.025) & (-0.072,-0.025) \\
    Overall & (50\%, 30\%) & -0.1000 & (-0.250, 0.050) & (-0.144,-0.056) & -0.0856 & (-0.210, 0.039) & (-0.132,-0.039) \\
    \bottomrule
    \end{tabular}%
  \label{tab:TRIP_noHIV}%
\end{table}%

% Table generated by Excel2LaTeX from sheet 'no_date'
\begin{table}[htbp]
  \centering
  \caption{Estimated risk differences and 95\% confidence intervals (CIs) of the effects of community alerts on HIV risk behavior at 6 months in TRIP adjusted for HIV status, shared drug equipment (e.g. syringe) in last six months, education (primary school, high school, and post high school), and employment status (employed, unemployed/looking for a work, can’t work because of health reason, and other).}
    \begin{tabular}{lc|rllrll}
    \toprule
    \multicolumn{1}{c}{Effects} & Coverage & \multicolumn{3}{c}{IPW$_1$} & \multicolumn{3}{c}{IPW$_2$} \\
         & $(\alpha, \alpha')$ & \multicolumn{1}{c}{RD} & \multicolumn{1}{c}{95\% CI (10)} & \multicolumn{1}{c}{95\% CI (20)} & \multicolumn{1}{c}{RD} & \multicolumn{1}{c}{95\% CI (10)} & \multicolumn{1}{c}{95\% CI (20)} \\
    \midrule
    \midrule
    \rowcolor[rgb]{ .949,  .949,  .949} Direct & (20\%, 20\%) & -0.0884 & (-0.195, 0.018) & (-0.276, 0.100) & 0.0077 & (-0.083, 0.099) & (-0.191, 0.206) \\
    \rowcolor[rgb]{ .949,  .949,  .949} Direct & (30\%, 30\%) & -0.1322 & (-0.332, 0.067) & (-0.340, 0.075) & -0.0874 & (-0.189, 0.014) & (-0.274, 0.099) \\
    \rowcolor[rgb]{ .949,  .949,  .949} Direct & (40\%, 40\%) & -0.1772 & (-0.478, 0.123) & (-0.391, 0.036) & -0.1588 & (-0.403, 0.086) & (-0.385, 0.067) \\
    \rowcolor[rgb]{ .949,  .949,  .949} Direct & (50\%, 50\%) & -0.2169 & (-0.610, 0.176) & (-0.440, 0.006) & -0.2083 & (-0.561, 0.145) & (-0.479, 0.062) \\
    Indirect & (30\%, 20\%) & 0.0055 & (-0.016, 0.027) & (-0.018, 0.029) & 0.0019 & (-0.012, 0.016) & (-0.045, 0.048) \\
    Indirect & (40\%, 20\%) & -0.0063 & (-0.038, 0.026) & (-0.054, 0.041) & -0.0089 & (-0.022, 0.005) & (-0.095, 0.077) \\
    Indirect & (50\%, 20\%) & -0.0266 & (-0.072, 0.019) & (-0.106, 0.053) & -0.0242 & (-0.042,-0.007) & (-0.152, 0.104) \\
    Indirect & (40\%, 30\%) & -0.0118 & (-0.028, 0.004) & (-0.039, 0.015) & -0.0108 & (-0.018,-0.003) & (-0.053, 0.031) \\
    Indirect & (50\%, 40\%) & -0.0203 & (-0.042, 0.002) & (-0.059,0.018) & -0.0153 & (-0.027,-0.003) & (-0.062, 0.031) \\
    Indirect & (50\%, 30\%) & -0.0322 & (-0.069, 0.005) & (-0.096, 0.032) & -0.0261 & (-0.045,-0.007) & (-0.113, 0.061) \\
    \rowcolor[rgb]{ .949,  .949,  .949} Total & (30\%, 20\%) & -0.1267 & (-0.308, 0.055) & (-0.347, 0.093) & -0.0854 & (-0.176, 0.005) & (-0.284, 0.113) \\
    \rowcolor[rgb]{ .949,  .949,  .949} Total & (40\%, 20\%) & -0.1836 & (-0.464, 0.096) & (-0.406, 0.039) & -0.1676 & (-0.404, 0.069) & (-0.377, 0.041) \\
    \rowcolor[rgb]{ .949,  .949,  .949} Total & (50\%, 20\%) & -0.2435 & (-0.623, 0.136) & (-0.454,-0.033) & -0.2325 & (-0.586, 0.121) & (-0.439,-0.026) \\
    \rowcolor[rgb]{ .949,  .949,  .949} Total & (40\%, 30\%) & -0.1891 & (-0.488, 0.110) & (-0.401, 0.023) & -0.1696 & (-0.418, 0.079) & (-0.378, 0.039) \\
    \rowcolor[rgb]{ .949,  .949,  .949} Total & (50\%, 40\%) & -0.2372 & (-0.637, 0.163) & (-0.443,-0.031) & -0.2236 & (-0.585, 0.138) & (-0.459, 0.012) \\
    \rowcolor[rgb]{ .949,  .949,  .949} Total & (50\%, 30\%) & -0.2491 & (-0.647, 0.149) & (-0.452,-0.047) & -0.2344 & (-0.600, 0.131) & (-0.448,-0.021) \\
    Overall & (30\%, 20\%) & -0.0165 & (-0.038, 0.005) & (-0.057, 0.024) & -0.0258 & (-0.060, 0.008) & (-0.071, 0.020) \\
    Overall & (40\%, 20\%) & -0.0595 & (-0.140, 0.021) & (-0.125, 0.006) & -0.0739 & (-0.180, 0.032) & (-0.147,-0.001) \\
    Overall & (50\%, 20\%) & -0.1174 & (-0.282, 0.047) & (-0.197,-0.038) & -0.1299 & (-0.323, 0.064) & (-0.217,-0.043) \\
    Overall & (40\%, 30\%) & -0.0431 & (-0.103,0.017) & (-0.070,-0.016) & -0.0481 & (-0.120, 0.024) & (-0.081,-0.016) \\
    Overall & (50\%, 40\%) & -0.0579 & (-0.142, 0.027) & (-0.083,-0.033) & -0.0560 & (-0.143, 0.032) & (-0.083,-0.029) \\
    Overall & (50\%, 30\%) & -0.1009 & (-0.245, 0.043) & (-0.149,-0.053) & -0.1041 & (-0.264, 0.056) & (-0.159,-0.049) \\
    \bottomrule
    \end{tabular}%
  \label{tab:TRIP_noDate}%
\end{table}%

% Table generated by Excel2LaTeX from sheet 'no_share.x'
\begin{table}[htbp]
  \centering
  \caption{Estimated risk differences and 95\% confidence intervals (CIs) of the effects of community alerts on HIV risk behavior at 6 months in TRIP adjusted for HIV status, the calendar date at first interview, education (primary school, high school, and post high school), and employment status (employed, unemployed/looking for a work, can’t work because of health reason, and other).}
    \begin{tabular}{lc|rllrll}
    \toprule
    \multicolumn{1}{c}{Effects} & Coverage & \multicolumn{3}{c}{IPW$_1$} & \multicolumn{3}{c}{IPW$_2$} \\
         & $(\alpha, \alpha')$ & \multicolumn{1}{c}{RD} & \multicolumn{1}{c}{95\% CI (10)} & \multicolumn{1}{c}{95\% CI (20)} & \multicolumn{1}{c}{RD} & \multicolumn{1}{c}{95\% CI (10)} & \multicolumn{1}{c}{95\% CI (20)} \\
    \midrule
    \midrule
    \rowcolor[rgb]{ .949,  .949,  .949} Direct & (20\%, 20\%) & -0.0691 & (-0.137,-0.001) & (-0.266, 0.128) & -0.0073 & (-0.069, 0.054) & (-0.198, 0.183) \\
    \rowcolor[rgb]{ .949,  .949,  .949} Direct & (30\%, 30\%) & -0.1053 & (-0.247, 0.036) & (-0.329, 0.118) & -0.0948 & (-0.215, 0.025) & (-0.275, 0.085) \\
    \rowcolor[rgb]{ .949,  .949,  .949} Direct & (40\%, 40\%) & -0.1490 & (-0.392, 0.094) & (-0.372, 0.074) & -0.1606 & (-0.410, 0.089) & (-0.371, 0.050) \\
    \rowcolor[rgb]{ .949,  .949,  .949} Direct & (50\%, 50\%) & -0.1919 & (-0.538, 0.155) & (-0.409, 0.025) & -0.2056 & (-0.552, 0.141) & (-0.449, 0.038) \\
    Indirect & (30\%, 20\%) & 0.0033 & (-0.016, 0.022) & (-0.017, 0.024) & -0.0020 & (-0.007, 0.003) & (-0.043, 0.039) \\
    Indirect & (40\%, 20\%) & -0.0119 & (-0.040, 0.016) & (-0.049, 0.025) & -0.0175 & (-0.033,-0.002) & (-0.088, 0.053) \\
    Indirect & (50\%, 20\%) & -0.0357 & (-0.083, 0.012) & (-0.093, 0.021) & -0.0368 & (-0.072,-0.002) & (-0.138, 0.064) \\
    Indirect & (40\%, 30\%) & -0.0153 & (-0.033, 0.002) & (-0.035, 0.004) & -0.0155 & (-0.032, 0.001) & (-0.048, 0.017) \\
    Indirect & (50\%, 40\%) & -0.0237 & (-0.050, 0.003) & (-0.051, 0.003) & -0.0194 & (-0.040, 0.001) & (-0.055, 0.016) \\
    Indirect & (50\%, 30\%) & -0.0390 & (-0.082, 0.004) & (-0.084, 0.006) & -0.0348 & (-0.072, 0.002) & (-0.102, 0.032) \\
    \rowcolor[rgb]{ .949,  .949,  .949} Total & (30\%, 20\%) & -0.1020 & (-0.230, 0.026) & (-0.336, 0.132) & -0.0969 & (-0.214, 0.020) & (-0.290, 0.097) \\
    \rowcolor[rgb]{ .949,  .949,  .949} Total & (40\%, 20\%) & -0.1610 & (-0.393, 0.071) & (-0.395, 0.073) & -0.1781 & (-0.440, 0.084) & (-0.385, 0.029) \\
    \rowcolor[rgb]{ .949,  .949,  .949} Total & (50\%, 20\%) & -0.2276 & (-0.574, 0.119) & (-0.445,-0.010) & -0.2425 & (-0.621, 0.136) & (-0.450,-0.035) \\
    \rowcolor[rgb]{ .949,  .949,  .949} Total & (40\%, 30\%) & -0.1643 & (-0.411, 0.083) & (-0.389, 0.060) & -0.1761 & (-0.441, 0.089) & (-0.379, 0.027) \\
    \rowcolor[rgb]{ .949,  .949,  .949} Total & (50\%, 40\%) & -0.2156 & (-0.573, 0.142) & (-0.425,-0.007) & -0.2250 & (-0.590, 0.140) & (-0.446,-0.004) \\
    \rowcolor[rgb]{ .949,  .949,  .949} Total & (50\%, 30\%) & -0.2309 & (-0.593, 0.131) & (-0.441,-0.021) & -0.2405 & (-0.622, 0.141) & (-0.451,-0.030) \\
    Overall & (30\%, 20\%) & -0.0145 & (-0.032, 0.003) & (-0.055, 0.026) & -0.0290 & (-0.071, 0.013) & (-0.074, 0.016) \\
    Overall & (40\%, 20\%) & -0.0577 & (-0.135, 0.019) & (-0.124, 0.008) & -0.0802 & (-0.202, 0.042) & (-0.155,-0.005) \\
    Overall & (50\%, 20\%) & -0.1178 & (-0.283, 0.048) & (-0.199,-0.037) & -0.1382 & (-0.353, 0.076) & (-0.231,-0.045) \\
    Overall & (40\%, 30\%) & -0.0433 & (-0.104, 0.017) & (-0.071,-0.015) & -0.0512 & (-0.131, 0.029) & (-0.087,-0.016) \\
    Overall & (50\%, 40\%) & -0.0600 & (-0.149, 0.029) & (-0.086,-0.034) & -0.0580 & (-0.150, 0.035) & (-0.088,-0.028) \\
    Overall & (50\%, 30\%) & -0.1033 & (-0.253, 0.046) & (-0.154,-0.053) & -0.1092 & (-0.282, 0.064) & (-0.171,-0.048) \\
    \bottomrule
    \end{tabular}%
  \label{tab:TRIP_noShare}%
\end{table}%

% Table generated by Excel2LaTeX from sheet 'no_edu'
\begin{table}[htbp]
  \centering
  \caption{Estimated risk differences and 95\% confidence intervals (CIs) of the effects of community alerts on HIV risk behavior at 6 months in TRIP adjusted for HIV status, shared drug equipment (e.g. syringe) in last six months, the calendar date at first interview, and employment status (employed, unemployed/looking for a work, can’t work because of health reason, and other).}
    \begin{tabular}{lc|rllrll}
    \toprule
    \multicolumn{1}{c}{Effects} & Coverage & \multicolumn{3}{c}{IPW$_1$} & \multicolumn{3}{c}{IPW$_2$} \\
         & $(\alpha, \alpha')$ & \multicolumn{1}{c}{RD} & \multicolumn{1}{c}{95\% CI (10)} & \multicolumn{1}{c}{95\% CI (20)} & \multicolumn{1}{c}{RD} & \multicolumn{1}{c}{95\% CI (10)} & \multicolumn{1}{c}{95\% CI (20)} \\
    \midrule
    \midrule
    \rowcolor[rgb]{ .949,  .949,  .949} Direct & (20\%, 20\%) & -0.0621 & (-0.130, 0.006) & (-0.341, 0.216) & 0.0073 & (-0.082, 0.096) & (-0.189, 0.203) \\
    \rowcolor[rgb]{ .949,  .949,  .949} Direct & (30\%, 30\%) & -0.0933 & (-0.206, 0.019) & (-0.411, 0.225) & -0.0820 & (-0.171, 0.007) & (-0.276, 0.112) \\
    \rowcolor[rgb]{ .949,  .949,  .949} Direct & (40\%, 40\%) & -0.1338 & (-0.325, 0.057) & (-0.449, 0.181) & -0.1440 & (-0.353, 0.065) & (-0.363, 0.075) \\
    \rowcolor[rgb]{ .949,  .949,  .949} Direct & (50\%, 50\%) & -0.1738 & (-0.453, 0.105) & (-0.473, 0.125) & -0.1838 & (-0.477, 0.110) & (-0.426, 0.058) \\
    Indirect & (30\%, 20\%) & 0.0055 & (-0.014, 0.025) & (-0.029, 0.040) & -0.0019 & (-0.008, 0.004) & (-0.046, 0.042) \\
    Indirect & (40\%, 20\%) & -0.0094 & (-0.028, 0.009) & (-0.074, 0.056) & -0.0220 & (-0.049, 0.005) & (-0.097, 0.053) \\
    Indirect & (50\%, 20\%) & -0.0357 & (-0.076, 0.004) & (-0.130, 0.059) & -0.0483 & (-0.109, 0.012) & (-0.152, 0.056) \\
    Indirect & (40\%, 30\%) & -0.0149 & (-0.031, 0.002) & (-0.048, 0.018) & -0.0200 & (-0.047, 0.007) & (-0.054, 0.014) \\
    Indirect & (50\%, 40\%) & -0.0263 & (-0.060, 0.007) & (-0.062, 0.009) & -0.0264 & (-0.061, 0.008) & (-0.062, 0.009) \\
    Indirect & (50\%, 30\%) & -0.0412 & (-0.091, 0.008) & (-0.108, 0.026) & -0.0464 & (-0.108, 0.015) & (-0.114, 0.021) \\
    \rowcolor[rgb]{ .949,  .949,  .949} Total & (30\%, 20\%) & -0.0878 & (-0.192, 0.016) & (-0.419, 0.243) & -0.0839 & (-0.172, 0.004) & (-0.289, 0.121) \\
    \rowcolor[rgb]{ .949,  .949,  .949} Total & (40\%, 20\%) & -0.1432 & (-0.333, 0.047) & (-0.464, 0.177) & -0.1660 & (-0.400, 0.068) & (-0.380, 0.048) \\
    \rowcolor[rgb]{ .949,  .949,  .949} Total & (50\%, 20\%) & -0.2095 & (-0.512, 0.093) & (-0.489, 0.070) & -0.2321 & (-0.585, 0.121) & (-0.440,-0.024) \\
    \rowcolor[rgb]{ .949,  .949,  .949} Total & (40\%, 30\%) & -0.1487 & (-0.352, 0.055) & (-0.457, 0.160) & -0.1640 & (-0.399, 0.071) & (-0.375, 0.047) \\
    \rowcolor[rgb]{ .949,  .949,  .949} Total & (50\%, 40\%) & -0.2001 & (-0.506, 0.106) & (-0.479, 0.079) & -0.2101 & (-0.538, 0.118) & (-0.432, 0.012) \\
    \rowcolor[rgb]{ .949,  .949,  .949} Total & (50\%, 30\%) & -0.2150 & (-0.533, 0.103) & (-0.485, 0.055) & -0.2302 & (-0.585, 0.124) & (-0.442,-0.019) \\
    Overall & (30\%, 20\%) & -0.0101 & (-0.026, 0.006) & (-0.069, 0.049) & -0.0280 & (-0.069, 0.013) & (-0.075, 0.019) \\
    Overall & (40\%, 20\%) & -0.0505 & (-0.117,0 .016) & (-0.140, 0.039) & -0.0810 & (-0.206, 0.044) & (-0.160,-0.002) \\
    Overall & (50\%, 20\%) & -0.1102 & (-0.268, 0.048) & (-0.207,-0.013) & -0.1417 & (-0.364, 0.081) & (-0.239,-0.044) \\
    Overall & (40\%, 30\%) & -0.0404 & (-0.098, 0.018) & (-0.073,-0.008) & -0.0530 & (-0.137, 0.031) & (-0.091,-0.016) \\
    Overall & (50\%, 40\%) & -0.0597 & (-0.153, 0.033) & (-0.084,-0.035) & -0.0606 & (-0.159, 0.037) & (-0.092,-0.029) \\
    Overall & (50\%, 30\%) & -0.1001 & (-0.251, 0.051) & (-0.149,-0.051) & -0.1137 & (-0.296, 0.068) & (-0.179,-0.049) \\
    \bottomrule
    \end{tabular}%
  \label{tab:TRIP_noEdu}%
\end{table}%

% Table generated by Excel2LaTeX from sheet 'no_employ'
\begin{table}[htbp]
  \centering
  \caption{Estimated risk differences and 95\% confidence intervals (CIs) of the effects of community alerts on HIV risk behavior at 6 months in TRIP adjusted for full set of confounding variables, HIV status, shared drug equipment (e.g.  syringe) in last six months, the calendar date at first interview, and education (primary school, high school, and post high school).}
    \begin{tabular}{lc|rllrll}
    \toprule
    \multicolumn{1}{c}{Effects} & Coverage & \multicolumn{3}{c}{IPW$_1$} & \multicolumn{3}{c}{IPW$_2$} \\
         & $(\alpha, \alpha')$ & \multicolumn{1}{c}{RD} & \multicolumn{1}{c}{95\% CI (10)} & \multicolumn{1}{c}{95\% CI (20)} & \multicolumn{1}{c}{RD} & \multicolumn{1}{c}{95\% CI (10)} & \multicolumn{1}{c}{95\% CI (20)} \\
    \midrule
    \midrule
    \rowcolor[rgb]{ .949,  .949,  .949} Direct & (20\%, 20\%) & -0.0653 & (-0.151, 0.020) & (-0.366, 0.235) & -0.0079 & (-0.076, 0.060) & (-0.210, 0.194) \\
    \rowcolor[rgb]{ .949,  .949,  .949} Direct & (30\%, 30\%) & -0.0983 & (-0.250, 0.053) & (-0.458, 0.262) & -0.1122 & (-0.254, 0.030) & (-0.304, 0.080) \\
    \rowcolor[rgb]{ .949,  .949,  .949} Direct & (40\%, 40\%) & -0.1397 & (-0.381, 0.102) & (-0.492, 0.213) & -0.1938 & (-0.498, 0.110) & (-0.437, 0.050) \\
    \rowcolor[rgb]{ .949,  .949,  .949} Direct & (50\%, 50\%) & -0.1795 & (-0.512, 0.153) & (-0.493, 0.135) & -0.2500 & (-0.675, 0.175) & (-0.549, 0.049) \\
    Indirect & (30\%, 20\%) & 0.0059 & (-0.017, 0.029) & (-0.024, 0.036) & 0.0122 & (-0.019, 0.043) & (-0.048, 0.072) \\
    Indirect & (40\%, 20\%) & -0.0078 & (-0.036, 0.020) & (-0.053, 0.037) & 0.0113 & (-0.033, 0.056) & (-0.101, 0.124) \\
    Indirect & (50\%, 20\%) & -0.0331 & (-0.073, 0.007) & (-0.096, 0.029) & 0.0028 & (-0.048, 0.053) & (-0.159, 0.165) \\
    Indirect & (40\%, 30\%) & -0.0138 & (-0.028, 0.001) & (-0.035, 0.008) & -0.0008 & (-0.015, 0.014) & (-0.055, 0.053) \\
    Indirect & (50\%, 40\%) & -0.0253 & (-0.053, 0.002) & (-0.055, 0.004) & -0.0085 & (-0.018, 0.001) & (-0.062, 0.045) \\
    Indirect & (50\%, 30\%) & -0.0391 & (-0.079, 0.001) & (-0.088, 0.010) & -0.0094 & (-0.031, 0.013) & (-0.115, 0.097) \\
    \rowcolor[rgb]{ .949,  .949,  .949} Total & (30\%, 20\%) & -0.0924 & (-0.228, 0.043) & (-0.468, 0.283) & -0.1000 & (-0.213, 0.013) & (-0.300, 0.100) \\
    \rowcolor[rgb]{ .949,  .949,  .949} Total & (40\%, 20\%) & -0.1475 & (-0.371, 0.076) & (-0.514, 0.219) & -0.1825 & (-0.443, 0.078) & (-0.390, 0.025) \\
    \rowcolor[rgb]{ .949,  .949,  .949} Total & (50\%, 20\%) & -0.2126 & (-0.540, 0.115) & (-0.525, 0.100) & -0.2472 & (-0.625, 0.131) & (-0.451,-0.044) \\
    \rowcolor[rgb]{ .949,  .949,  .949} Total & (40\%, 30\%) & -0.1535 & (-0.396, 0.089) & (-0.505, 0.198) & -0.1946 & (-0.486, 0.096) & (-0.409, 0.020) \\
    \rowcolor[rgb]{ .949,  .949,  .949} Total & (50\%, 40\%) & -0.2048 & (-0.551, 0.141) & (-0.507, 0.098) & -0.2585 & (-0.680, 0.163) & (-0.515,-0.002) \\
    \rowcolor[rgb]{ .949,  .949,  .949} Total & (50\%, 30\%) & -0.2185 & (-0.566, 0.129) & (-0.519, 0.082) & -0.2593 & (-0.667, 0.149) & (-0.481,-0.038) \\
    Overall & (30\%, 20\%) & -0.0105 & (-0.026, 0.005) & (-0.077, 0.056) & -0.0199 & (-0.044, 0.004) & (-0.068, 0.029) \\
    Overall & (40\%, 20\%) & -0.0506 & (-0.116, 0.014) & (-0.150, 0.049) & -0.0646 & (-0.154, 0.025) & (-0.139, 0.009) \\
    Overall & (50\%, 20\%) & -0.1098 & (-0.260, 0.040) & (-0.214,-0.005) & -0.1206 & (-0.297, 0.056) & (-0.205,-0.036) \\
    Overall & (40\%, 30\%) & -0.0401 & (-0.095, 0.014) & (-0.076,-0.005) & -0.0447 & (-0.111, 0.022) & (-0.075,-0.014) \\
    Overall & (50\%, 40\%) & -0.0592 & (-0.146, 0.027) & (-0.087,-0.031) & -0.0560 & (-0.143, 0.031) & (-0.082,-0.030) \\
    Overall & (50\%, 30\%) & -0.0993 & (-0.240, 0.041) & (-0.152,-0.047) & -0.1007 & (-0.254, 0.053) & (-0.151,-0.050) \\
    \bottomrule
    \end{tabular}%
  \label{tab:TRIP_noemploy}%
\end{table}%

% Table generated by Excel2LaTeX from sheet 'no_adj'
\begin{table}[htbp]
  \centering
  \caption{Estimated risk differences and 95\% confidence intervals (CIs) of the effects of community alerts on HIV risk behavior at 6 months in TRIP not adjusted for any covariates.}
    \begin{tabular}{lc|rllrll}
    \toprule
    \multicolumn{1}{c}{Effects} & Coverage & \multicolumn{3}{c}{IPW$_1$} & \multicolumn{3}{c}{IPW$_2$} \\
         & $(\alpha, \alpha')$ & \multicolumn{1}{c}{RD} & \multicolumn{1}{c}{95\% CI (10)} & \multicolumn{1}{c}{95\% CI (20)} & \multicolumn{1}{c}{RD} & \multicolumn{1}{c}{95\% CI (10)} & \multicolumn{1}{c}{95\% CI (20)} \\
    \midrule
    \midrule
    \rowcolor[rgb]{ .949,  .949,  .949} Direct & (20\%, 20\%) & -0.0802 & (-0.154,-0.006) & (-0.279, 0.119) & 0.0063 & (-0.102, 0.114) & (-0.218, 0.231) \\
    \rowcolor[rgb]{ .949,  .949,  .949} Direct & (30\%, 30\%) & -0.1299 & (-0.276, 0.017) & (-0.385, 0.126) & -0.0748 & (-0.145,-0.004) & (-0.319, 0.169) \\
    \rowcolor[rgb]{ .949,  .949,  .949} Direct & (40\%, 40\%) & -0.1806 & (-0.427, 0.066) & (-0.475, 0.113) & -0.1492 & (-0.353, 0.055) & (-0.433, 0.135) \\
    \rowcolor[rgb]{ .949,  .949,  .949} Direct & (50\%, 50\%) & -0.2254 & (-0.573, 0.123) & (-0.549, 0.098) & -0.2108 & (-0.542, 0.121) & (-0.540, 0.118) \\
    Indirect & (30\%, 20\%) & 0.0055 & (-0.010, 0.022) & (-0.021, 0.032) & 0.0060 & (-0.010, 0.022) & (-0.047, 0.059) \\
    Indirect & (40\%, 20\%) & -0.0057 & (-0.024, 0.013) & (-0.068, 0.057) & 0.0005 & (-0.019, 0.020) & (-0.102, 0.103) \\
    Indirect & (50\%, 20\%) & -0.0241 & (-0.056, 0.008) & (-0.140, 0.092) & -0.0096 & (-0.033, 0.013) & (-0.165, 0.146) \\
    Indirect & (40\%, 30\%) & -0.0112 & (-0.024, 0.002) & (-0.050, 0.027) & -0.0055 & (-0.012, 0.001) & (-0.057, 0.046) \\
    Indirect & (50\%, 40\%) & -0.0184 & (-0.039, 0.002) & (-0.075, 0.038) & -0.0101 & (-0.019,-0.001) & (-0.068, 0.048) \\
    Indirect & (50\%, 30\%) & -0.0296 & (-0.062, 0.003) & (-0.124, 0.065) & -0.0156 & (-0.030,-0.001) & (-0.123, 0.092) \\
    \rowcolor[rgb]{ .949,  .949,  .949} Total & (30\%, 20\%) & -0.1244 & (-0.261, 0.012) & (-0.377, 0.128) & -0.0688 & (-0.128,-0.009) & (-0.318, 0.181) \\
    \rowcolor[rgb]{ .949,  .949,  .949} Total & (40\%, 20\%) & -0.1863 & (-0.429, 0.057) & (-0.451, 0.078) & -0.1488 & (-0.337, 0.040) & (-0.402, 0.105) \\
    \rowcolor[rgb]{ .949,  .949,  .949} Total & (50\%, 20\%) & -0.2495 & (-0.607, 0.109) & (-0.498,-0.001) & -0.2205 & (-0.538, 0.097) & (-0.457, 0.016) \\
    \rowcolor[rgb]{ .949,  .949,  .949} Total & (40\%, 30\%) & -0.1918 & (-0.447, 0.064) & (-0.461, 0.078) & -0.1548 & (-0.358, 0.048) & (-0.412, 0.102) \\
    \rowcolor[rgb]{ .949,  .949,  .949} Total & (50\%, 40\%) & -0.2438 & (-0.607, 0.119) & (-0.522, 0.035) & -0.2209 & (-0.554, 0.112) & (-0.502, 0.060) \\
    \rowcolor[rgb]{ .949,  .949,  .949} Total & (50\%, 30\%) & -0.2550 & (-0.626, 0.116) & (-0.509,-0.0005) & -0.2265 & (-0.558, 0.106) & (-0.475, 0.022) \\
    Overall & (30\%, 20\%) & -0.0174 & (-0.039, 0.004) & (-0.055, 0.020) & -0.0177 & (-0.039, 0.004) & (-0.069, 0.033) \\
    Overall & (40\%, 20\%) & -0.0619 & (-0.147, 0.023) & (-0.123,-0.001) & -0.0605 & (-0.144, 0.023) & (-0.137, 0.016) \\
    Overall & (50\%, 20\%) & -0.1207 & (-0.296, 0.054) & (-0.196,-0.045) & -0.1163 & (-0.286, 0.054) & (-0.200,-0.033) \\
    Overall & (40\%, 30\%) & -0.0445 & (-0.109, 0.020) & (-0.070,-0.019) & -0.0428 & (-0.106, 0.020) & (-0.072,-0.014) \\
    Overall & (50\%, 40\%) & -0.0588 & (-0.150, 0.032) & (-0.086,-0.031) & -0.0558 & (-0.142, 0.030) & (-0.082,-0.029) \\
    Overall & (50\%, 30\%) & -0.1033 & (-0.259, 0.052) & (-0.152,-0.055) & -0.0986 & (-0.248, 0.051) & (-0.146,-0.051) \\
    \bottomrule
    \end{tabular}%
  \label{tab:TRIP_noadj}%
\end{table}%

%\label{lastpage}

\end{document}